\documentclass[twocolumn,showpacs,preprintnumbers,amsmath,amssymb]{revtex4}

\usepackage{graphicx}
\usepackage{amssymb}
\usepackage{amsmath}
\usepackage{color}
\usepackage{ wasysym }
\usepackage{hyperref}
 \usepackage{tabu}

\def\barray{\begin{array}}
\def\earray{\end{array}}
\def\be{\begin{equation}}
\def\ee{\end{equation}}
\def\ben{\begin{equation} \nonumber}
\def\een{\end{equation}}
\def\ban{\begin{eqnarray*}}
\def\ean{\end{eqnarray*}}
\def\ba{\begin{eqnarray}}
\def\ea{\end{eqnarray}}

\def\({\left(}
\def\){\right)}

\def\apj{ApJ}

\def\mnras{MNRAS}

\def\aj{AJ}

\def\apjs{ApJS}

\def\prd{Phys. Rev. D}

\newcommand{\CC}{\Lambda}

\graphicspath{{./fig/}}

\begin{document}

\title{CAN DARK ENERGY BE EXPRESSED AS A POWER SERIES\\  OF THE HUBBLE PARAMETER?}

\author{Mehdi Rezaei}
\email{rezaei@irimo.ir}
\affiliation{Research Institute for Astronomy and Astrophysics of Maragha (RIAAM), Maragha, Iran, P.O.Box:55134-441}
\affiliation{Iran meteorological organization, Hamedan Research Center for Applied
Meteorology, Hamedan 65199, 99711, Iran}

\author{Mohammad Malekjani}
\email{malekjani@basu.ac.ir}
\affiliation{Department of Physics, Bu-Ali Sina University, Hamedan
65178, 016016, Iran}

\author{Joan Sol\`a Peracaula}
\email{sola@fqa.ub.edu}
\affiliation{Departament de F\'\i sica Qu\`antica i Astrof\'\i sica, and Institute of Cosmos Sciences (ICCUB),Univ. de Barcelona, Av. Diagonal 647 E-08028 Barcelona, Catalonia, Spain}

\begin{abstract}
In this work, we examine the possibility that the dark energy (DE) density, $\rho_{\rm de}$, can be dynamical and appear as a power series expansion of the Hubble rate (and its time derivatives), i.e. $\rho_{\rm de}(H,\dot{H},...)$. For the present universe, however, only the terms $H$, $\dot{H}$ and $H^2$ can be relevant, together with an additive constant term. We fit these models to the current cosmological data on the main observables SnIa+$H(z)$+BAO+LSS+CMB+BBN. Our analysis involves both the background as well as the cosmic perturbation equations. The latter include, apart from the matter density perturbations, also the DE density perturbations. We assume that matter and dynamical DE are separately self-conserved. As a result the equation of state  of the DE becomes a nontrivial function of the cosmological redshift, $w_D=w_D(z)$. The particular subset of DE models of this type having no additive constant term in $\rho_{\rm de}$ include the so-called ``entropic-force'' and ``QCD-ghost'' DE models, as well as the pure linear model $\rho_{\rm de}\sim H$, all of which are strongly disfavored in our fitting analysis. In contrast, the models that include the additive term plus one or both of the dynamical components $\dot{H}$ and $H^2$ appear more favored than the $\Lambda$CDM. In particular, the dynamical DE models provide a value of $\sigma_8\simeq 0.74-0.77$ which is substantially lower than that of the $\Lambda$CDM and hence more in accordance with the observations. This helps to significantly reduce the $\sigma_8$-tension in the structure formation data. At the same time the predicted value for $H_0$ is in between the local and Planck measurements, thus helping to alleviate this tension as well.

\end{abstract}
\maketitle

\section{Introduction}\label{sect:intro}

The supernovae (SnIa) observations \citep{Riess1998,Perlmutter1999} reveal that our universe experiences an accelerated expansion during recent cosmological times. Furthermore, the other independent observations including cosmic microwave background (CMB)
\citep{Komatsu2009,Jarosik:2010iu,Komatsu2011,Ade:2015yua},
large scale structure (LSS), baryonic acoustic oscillation (BAO)
\citep{Tegmark:2003ud,Cole:2005sx,Eisenstein:2005su,Percival2010,Blake:2011rj,Reid:2012sw},
high redshift galaxies \citep{Alcaniz:2003qy}, high redshift galaxy
clusters \citep{Wang1998,Allen:2004cd} and weak gravitational
lensing \citep{Benjamin:2007ys,Amendola:2007rr,Fu:2007qq} have confirmed the discovery of SnIa observations. These observations suggest that the universe in the context of standard gravity is dominated by an unknown form of energy with negative pressure, the so-called dark energy (DE). The latter can explain the current accelerated expansion of the Universe. The simplest candidate for DE is the cosmological
constant (CC), denoted $\Lambda$, with constant equation of state (EoS) parameter $w_{\rm \Lambda}=-1$  \citep{Peebles2003}. However there are difficult fine-tuning and coincidence problems associated with explaining
why $\Lambda$ should have today's energy scale \citep{Weinberg1989,Sahni:1999gb,Sola:2013gha}. These
problems led the cosmologists to suggest a time-evolving energy density with negative pressure as an alternative
to $\Lambda$. Some of these dynamical dark energy (DDE) models are constructed on the basis of quantum gravity theories. Models such as holographic \citep{Horava:2000tb,Thomas:2002pq} and agegraphic dark energy models \citep{Wei:2007ty} are derived in the framework of quantum gravity, by introducing a new degree of freedom. However, in recent works such as \citep{Akhlaghi:2018knk,Malekjani:2018qcz}, the authors show that some of holographic DE models are not consistent with cosmological observations, see also the previous studies\,\citep{Basilakos:2012ra,Basilakos:2014tha}. In this work we focus on two scenarios for DE whose energy density can be expressed as a power series expansion of the Hubble rate (and its time derivatives):  $\rho_{\rm de}=\rho_{\rm de}(H,\dot{H},...)$. At the present epoch, however, the relevant terms can only be of order $H^2$ at most (this includes $\dot{H}$), whereas the higher orders  $H^n (n>2)$ can be used in the early Universe to successfully implement inflation, see e.g.\,\citep{Lima:2012mu,Perico:2013mna,Sola:2015rra,Sola:2015csa}.

As a first type of scenario we have the ghost DE which was considered in\, \citep{Schutzhold:2002pr,Urban2010a} and was proposed without
introducing new degrees of freedom\,\citep{Schutzhold:2002pr,Urban2010a}. In such context, it is claimed that the CC arises from the contribution of the ghost fields which are supposed to be present in the low-energy effective theory of QCD and to find a solution to the $U(1)$ problem \citep{Witten1979, Veneziano1979,Rosenzweig1980}. Although the ghost make no contribution to the vacuum energy density in a flat Minkowski space time, in the case of curved spacetime it gives rise to a small vacuum energy density $\rho\sim H \Lambda^3_{QCD}$ where $H$ is the Hubble parameter and $\Lambda_{QCD}$ is the QCD mass scale of order $100 MeV$. The (approximate) right order of magnitude $\sim (3\times10^{-3}eV)^4$ of the DE density is obtained on using the current value of the Hubble parameter ($H_0\sim 10^{-33}eV$).  In another conceptual vein, the entropic-force idea\,\cite{Verlinde:2010hp} with its cosmological implications \,\cite{Easson:2010av} leads to a form of DE which is akin to the previous ones. Let us also mention the generalized models of this kind based on the holographic principle, see e.g.\,\cite{Komatsu:2018meb} and references therein. Our second main focus is the class of DDE models (with self-conserved DE density)  proposed in \citep{Gomez-Valent:2015pia}, in which the effective form of the DE energy density in Quantum Field Theory (QFT) in curved spacetime can be expressed as a generic power series of $H$ and $\dot{H}$. The same sort of models, but considered as dynamical vacuum models (in some cases interacting with matter) were previously studied in \citep{Gomez-Valent:2014rxa,Gomez-Valent:2014fda,Sola:2015wwa} and have been further investigated in \citep{Sola:2016jky,Sola:2016ecz,Sola:2017jbl,Sola:2017znb}. From the theoretical point of view, this kind of scenarios can be motivated from the renormalization group approach in QFT in curved spacetime \citep{Sola:2013gha,Sola:2015rra,Sola:2015csa}.

Comparing a model with observational data is one of the most important tools to study its validity. Previously, the ghost DE (hereafter GDE) model has been tested by different observational data sets in the literature. The authors of \cite{Cai:2010uf} have fitted the GDE model to observational data, including SnIa, BAO, CMB, BBN
and $H(z)$ data points. Their fitting results indicate that the observational data used do not favor the GDE option in comparison with the concordance $\Lambda$CDM cosmology. The viscous GDE variant was considered in \cite{Feng:2012gr}, in which the presence of bulk viscosity effects is introduced. Using the Markov Chain Monte Carlo (MCMC) method, they could not distinguish between GDE models with and without viscosity. In \cite{Khurshudyan:2013oba} some generalizations of GDE were studied which involved interaction of GDE with matter.  In the work \cite{Alavirad:2014kqa} the cosmological constraints on the parameters of the GDE are revisited in the framework of Brans-Dicke theory and using the MCMC method. Their results showed that the best fit values of the free parameters in the GDE model are compatible with the results of the $\Lambda$ cosmology.

In general, the GDE models are not only phenomenologically problematic (as we will reconfirm here) but are also theoretically troublesome since they all involve linear terms in $H$ which cannot be generated from a fully covariant effective action. For this reason the models containing $\dot{H}$ and $H^2$ are  preferred since they have an even number of derivatives of the scale factor and hence are compatible with general covariance, see \citep{Sola:2013gha,Sola:2015rra} for further discussion.

Dynamical DE models, specially if they are well motivated and are not incompatible with general covariance, are all the most interesting and welcome if we take into account that a number of persisting tensions with the data suggest  that the standard $\CC$CDM model, with rigid $\Lambda$-term, might be performing insufficiently at the observational level.  One of the tensions concerns the large scale structure  formation data\,\citep{Macaulay:2013swa}, which are in conflict with the too large value of $\sigma_8$ predicted by the $\CC$CDM. Another acute tension concerns the discrepancy between the local (distance ladder) determination of the current Hubble parameter $H_0$\,\citep{Riess:2018uxu} against the Planck determination based on the CMB anisotropies\, \citep{Aghanim:2018eyx}. In this work we shall show that both such tensions can be significantly relaxed within the main models of the DDE class.

In the three comprehensive studies\,\citep{Gomez-Valent:2014rxa,Gomez-Valent:2014fda,Sola:2016jky} the authors analyzed some dynamical vacuum energy models whose vacuum density consists of a constant term and a series of powers of the Hubble rate. These models were also analyzed as self-conserved DE models with a dynamical EoS in \citep{Gomez-Valent:2015pia}. In the last study the models are fitted to  SnIa+ CMB+ BAO+ $f\sigma_8$ data sets and the results show that the pure linear model $\rho_{de}\sim H$ as well as models without a constant additive term are strongly disfavored. In this work we revisit these scenarios but from a more general point of view. We do not only consider a more complete and updated set of data, but we perform the analysis with two important new ingredients. On the first place,  we use the MCMC method as a more systematic way to explore the parameter space; and another novelty is that  we take the DE perturbations fully into account (in the clustered DE scenario).
Our main goal in this work is to assess to which extent the DDE models based on a power series of the Hubble rate are consistent with observations. We will analyze both the models of this type without any additive constant term and those including it. To implement our analysis, we fit the models to a large number of SnIa+$H(z)$+BAO+LSS+CMB data, which we describe in detail in the paper, including the Big Bang Nucleosynthesis (BBN) bound. These data can reveal the role of DE in the accelerating of the universe expansion.
Moreover, since we deal with DDE with time varying EoS parameter $w_{\rm de}\neq -1$ the growth rate of cosmic structures can also be affected by perturbations of DE  \citep{Erickson:2001bq,Bean:2003fb,Mehrabi:2015kta,Malekjani:2016edh,Abramo:2008ip,Grande:2008re}. We use the background expansion data  in conjunction  with the growth rate data of large scale structures in order to put constraints on the parameters of cosmology and DE models. This combination of data have been used in a variety of DE models and studies, see e.g. \citep[see][]{mota6,Mehrabi:2015kta,Malekjani:2016edh,Rezaei:2017yyj,Malekjani:2018qcz,Rezaei:2019roe,Sola:2018sjf,Sola:2019lnw}.

The summary of our paper is as follows. We start by defining the DDE models under study in Sect.\,\ref{sect:back}. In Sect.\,\ref{sect:DATA} we introduce different data sets that we use in our analysis and describe the statistical methods as well as the procedure to select the best models. In Sect.\,\ref{sect:RS} we report on the numerical results in two different steps: first, we use expansion data to run MCMC at background level; and second, we combine all of the data sets (viz. expansion data + growth rate data) to run an overall likelihood analysis. Finally, in Sect.\,\ref{sect:conclusion} we summarize our findings and expose the main conclusions.

\section{DDE models and cosmological parameters}\label{sect:back}
In this section, we investigate the cosmological equations of  the different DDE models under study both at background and perturbation levels. In all cases the models involve powers of the expansion rate and/or its first derivative.  We restrict to powers that can be of relevance for the post-inflationary universe, hence  $H$, $H^2$ and $\dot{H}$ only.  Any other higher power has negligible influence for the evolution of the universe since the radiation-dominated epoch till now.  See \,\citep{Lima:2012mu,Sola:2013gha,Perico:2013mna,Sola:2015rra,Sola:2015csa} for a detailed discussion of this aspect and for the impact of the higher powers of $H$ in the very early universe, where they play a role to trigger inflation and graceful exit. They can even help to resolve the entropy problem of the $\CC$CDM model\,\cite{Sola:2015csa}.

We divide the DE models under consideration into two basic classes: 1)  Those that do {\emph not} include a constant additive term in the series  and involve heterogeneous powers of the Hubble term with different dimensions, such as $H$ and/or $H^2$. They are inspired in the context of QCD and are usually called the ``ghost DE  class'' (GDE); and 2) those which involve dimensionally homogeneous powers of the Hubble term, such as $H^2$ and/or $\dot{H}$ (both of energy dimension $+2$ in natural units $\hbar=c=1$). They may include  a non vanishing constant additive term or not.  Notice that only those having such additive constant have a well-defined $\CC$CDM limit. The models in the two classes generalize the vacuum class\citep{Sola:2016jky,Sola:2016ecz,Sola:2017jbl} in the sense that their EoS is not $-1$ but a nontrivial function of the scale factor, $w(a)$.  More specifically, the models which we analyze in this study are the following:
\begin{itemize}
\item GDE models whose DE density is linear in  $H$ or  its generalized form involving also the power $H^2$:
\begin{eqnarray}\label{gde1}
Model(1): \rho_{\rm de}(z)&=&\alpha\, H(z)\;\\
Model(2): \rho_{\rm de}(z)&=&\alpha H(z)+ \beta H^2(z).\label{gde2}
\end{eqnarray}
Model(2) is a more general form of the Veneziano ghost field in QCD theory which was proposed in\,\citep{Zhitnitsky:2011tr}. In this generalized form, the term $H^2$  could play a significant role in the evolution of the early universe. There  is also the particular realization of this model with $\alpha=0$, i.e. $\rho_{\rm de}(z)=\beta H^2$, which was discussed as vacuum model in \cite{Basilakos:2009wi}. Notice that the $H^2$ component in Model(2) is not necessarily sub leading as compared to the linear term in $H$ since the latter has the coefficient $\alpha$ of dimension $+3$ in natural units, whereas the former has coefficient $\beta$ of dimension $+2$. In the GDE context, coefficient $\alpha$ is of order $\Lambda^3_{QCD}$, whereas  $\beta$ is of order $m_{Pl}^2=1/G$, with $G$ the Newton's constant. Here $m_{Pl}\sim 10^{19}$ GeV is the  Planck mass. Thus, the ratio of the two coefficients, $\alpha/\beta\sim \Lambda^3_{QCD}/m_{Pl}^2$,  is roughly of order of the current value  of the Hubble parameter, i.e.  $ H_0\sim 10^{-42}$ GeV. Consequently, the two terms involved in Model (2) are of order $m_{Pl}^2H_0^2$ for the present universe, and hence both are close to the current value of the vacuum energy density, $\rho_{\Lambda 0}\sim 10^{-47}$ GeV$^4$. This is at least the idea behind the GDE models proposed in the literature.

\item The second class of models was proposed in \citep{Gomez-Valent:2015pia} an define the   "D-class" of dynamical DE models. They  are formally similar to the vacuum class of DE models, which was studied in \citep{Gomez-Valent:2014rxa,Gomez-Valent:2014fda,Basilakos:2009wi}. There is, however, an important difference in that the vacuum class was interacting with matter whereas the D-class is not, and hence the DE density is covariantly  self-conserved, together with matter.
    The various types of D-class models (all of them involving dimensionally homogeneous dynamical terms) read as follows:
\begin{eqnarray}\label{gde3}
Model(3): \rho_{\rm de}(z)&=&\dfrac{3}{\kappa^2}\left[ c_0+\nu H^2(z)\right] \,,\\
Model(4): \rho_{\rm de}(z)&=&\dfrac{3}{\kappa^2}\left[c_0+\dfrac{2}{3}\mu \dot{H}(z)\right]\,,\label{gde4}\\
Model(5): \rho_{\rm de}(z)&=&\dfrac{3}{\kappa^2}\left[\dfrac{2}{3}\mu \dot{H}(z)+\nu H^2(z)\right]\,,\label{gde5}\\
Model(6): \rho_{\rm de}(z)&=&\dfrac{3}{\kappa^2}\left[c_0+\dfrac{2}{3}\mu \dot{H}(z)+\nu H^2(z)\right]\label{gde6}
\end{eqnarray}
where $\kappa^2=8\pi G$.
Notice that in the above equations, the constant parameter $c_0$ has dimension $+2$ in natural units. In the case of parameter $\mu$ we have extracted an explicit factor of $2/3$ for convenience. Obviously, models (3) ,(4) and  (5) are  particular cases of Model(6). In addition,  for $\nu,\mu\to 0$ models models (3), (4) and (6) smoothly reduce to the $\CC$CDM. These models are actually the most promising ones, as we shall see. In contrast, Model (5) has $c_0=0$ and hence it does not have a well-defined $\CC$CDM limit, a feature which is shared with Models (1) and (2). Model (5) is of the kind used to discuss the so-called entropic-force cosmological scenario\,\cite{Easson:2010av}. As we shall see, it does not provide a comparably good fit to the data as models (3), (4) and (6), and is problematic in the radiation-dominated epoch.

\end{itemize}

\subsection{Cosmology at background level}

In an isotropic and homogeneous spatially flat Friedmann-Robertson-Walker (FRW) cosmology, filled by radiation, pressure less matter and a DE component with equation of state
$w_{\rm de}=p_{\rm de}/\rho_{\rm de}$, the first Friedmann equation reads:
\begin{eqnarray}\label{frid1}
H^2=\frac{8\pi G}{3}(\rho_{\rm r}+\rho_{\rm m}+\rho_{\rm de})\;,
\end{eqnarray}
where $H\equiv {\dot a}/a$ is the Hubble parameter, $\rho_{\rm r}$, $\rho_{\rm m}$ and $\rho_{\rm de}$ are the energy densities of radiation, pressure less matter and DE, respectively. All these densities are assumed to be covariantly conserved. Inserting $\rho_{\rm de}$ from Eq.(\ref{gde2}) into Eq.(\ref{frid1}) we can easily obtain the Hubble function for Model(2):

\begin{equation}\label{h1}
H(z)=\tilde\kappa \alpha \pm \left[(\tilde\kappa \alpha)^2+2\tilde\kappa (\rho_{\rm m,0}(1+z)^3+ \rho_{\rm r,0}(1+z)^4) \right] ^{1/2}\;.
\end{equation}
where we have defined $\tilde\kappa=\frac{4\pi G}{3 \gamma}$ and $\gamma=1-8\pi G \beta /3$. It is easy to see that by setting $\gamma=1$ leads to $\beta=0$, which means that Model(2) reduces to Model(1). So, by studying the  cosmology of Model(2) we can also investigate Model(1) as a special case of the generalized form. Taking the time derivative of Eq.(\ref{frid1}) and using the conservation equations for radiation, DM and DE, we get

\begin{eqnarray}\label{continuity}
&& \dot{\rho_{\rm r}}+4H\rho_{\rm r}=0\;,\label{radiation}\\
&&\dot{\rho_{\rm m}}+3H\rho_{\rm m}=0\;,\label{matter}\\
&&\dot{\rho_{\rm de}}+3H(1+w_{\rm de})\rho_{\rm de}=0\;\label{de}.
\end{eqnarray}
Let us note that there are other, alternative, DDE scenarios in which there is an interaction of the DE with matter, see e.g. \citep{Gomez-Valent:2014rxa,Gomez-Valent:2014fda,Sola:2015wwa,Sola:2016jky,Sola:2016ecz,Sola:2017jbl}. But these will not be considered here.
By computing the time derivative of Eq. (\ref{frid1}) and introducing the dimensionless cosmological parameters  $\Omega_{\rm i}=8\pi G \rho_{\rm i}/3H^2$, it is easy to see that
\begin{equation}\label{gde10}
\frac{\dot{H}}{H^2}=-\frac{3}{2}(1+w_{\rm de}\Omega_{\rm de}+\dfrac{\Omega_{\rm r}}{3})\;,
\end{equation}
where $w_{\rm de}$ and $\Omega_{\rm de}$ are, respectively, the EoS and the dimensionless density parameter of the DE and $\Omega_r$ is the dimensionless energy density of radiation. In order to calculate the EoS of Model(2) we take the time derivative from Eq.(\ref{gde2}) and insert the result in Eq.(\ref{de}). Finally, using Eq.(\ref{gde10}) the desired EoS parameter for Model(2) ensues:
\begin{equation}\label{gde12}
w_{\rm de}=\frac{(1-\Omega_{\rm de}-\gamma)+\dfrac{\Omega_{\rm r}}{3}(1+\Omega_{\rm de}-\gamma)}{\Omega_{\rm de}(1-\Omega_{\rm de}+\gamma)}\;.
\end{equation}
Upon setting $\gamma=1$ the above equation reduces to the EoS parameter for Model(1), which reads

\begin{equation}\label{gde11}
w_{\rm de}=\frac{\dfrac{\Omega_{\rm r}}{3}-1}{2-\Omega_{\rm de}}\;.
\end{equation}

Differentiating from $\Omega_{\rm de}=\rho_{\rm de}/\rho_{\rm c}$ with respect to time and using Eqs.(\ref{gde2} and \ref{gde10}) and finally  using the relation  $dz=-(1+z)\,H dt$  to trade the time derivative for the redshift derivative,  we obtain \citep[see also][]{Malekjani:2015pza}
\begin{equation}\label{gde13}
\frac{d\Omega_{\rm de}(z)}{dz}=-\frac{3 (\Omega_{\rm de}+\gamma-1)}{2 (1+z)}\left[ 1+w_{\rm de}\Omega_{\rm de}+\dfrac{\Omega_{\rm r}}{3}\right] \;.
\end{equation}
Now by inserting Eq.(\ref{gde12}) in Eq.(\ref{gde13}) and solving the latter we can find the evolution of DE density in Model(2). We set the initial condition as $\Omega_{\rm de}(z=0)=1- \Omega_{\rm m,0}-\Omega_{\rm r,0}$, where $\Omega_{\rm m,0}=\Omega_{\rm DM,0}+\Omega_{\rm b,0}$, $\Omega_{\rm r0}=2.469\times 10^{-5}h^{-2}(1.6903)$ and $h=H_{0}/100$ \citep{Hinshaw:2012aka}.
Inserting the results in Eq.(\ref{gde12}), we can also obtain the EoS parameter for Model(2). In a spatially flat universe we obtain the corresponding dimensionless Hubble parameter $(E=H/H_0)$ as follows:
\begin{equation}\label{gde11}
E(z)=\sqrt{\frac{\Omega_{\rm m,0}(1+z)^3 +\Omega_{\rm r,0}(1+z)^4}{1-\Omega_{\rm de}}}\;.
\end{equation}
Replacing $\Omega_{\rm de}(z)$ with the result we obtained from solving Eq.(\ref{gde13}) we can calculate the evolution of Hubble parameter in Model(2) and Model(1) as a particular case of it. We do not show the explicit result.

Let us now turn to models (3)-(6). By solving Eq.(\ref{de}) we find:
\begin{equation}\label{28}
\rho_{\rm de}(a)=\rho_{\rm de,0}\exp \left[  -3\int^a_1 (1+w_{\rm de}(a'))\dfrac{da'}{a'} \right]  \;.
\end{equation}
Now by differentiating from Eq.(\ref{28}) with respct to time we find:
\begin{equation}\label{31}
w_{\rm de}(a)=-1-\dfrac{a}{3\rho_{\rm de}(a)}\dfrac{d\rho_{\rm de}(a)}{da}\;.
\end{equation}
On inserting $\rho_{\rm de}$ from Eq.(\ref{gde6}) into Eq.(\ref{frid1}) and resorting once more to the redshift variable, we obtain:
\begin{equation}\label{34}
\mu a \dfrac{dH^2}{da}-3(1-\nu)H^2+3H^2_0(\Omega_{\rm m,0}a^{-3}+\Omega_{\rm r,0}a^{-4})+3c_0=0\;.
\end{equation}
Re-arranging Eq.(\ref{gde6}) for the present epoch and using Eq.(\ref{gde10}) we can find the constant parameter $c_0$ in terms of the current values of the cosmological parameters:
\begin{equation}\label{35}
c_0=H^2_0[\Omega_{\rm de,0}-\nu+\mu(1+w_{\rm de,0}\Omega_{\rm de,0}+\dfrac{\Omega_{\rm r,0}}{3})]\;.
\end{equation}
Finally by integrating  Eq.(\ref{34}) we can find the evolution of dimensionless Hubble parameter:
\begin{eqnarray}\label{37}
E^2(a)=a^{3\eta}&+&\dfrac{c_0(1-a^{3\eta})}{(1-\nu)H^2_0}+\dfrac{\Omega_{\rm m,0}(a^{-3}-a^{3\eta})}{1+\mu-\nu}\nonumber\\
&+&\dfrac{\Omega_{\rm r,0}(a^{-4}-a^{3\eta})}{1-\nu + 4 \mu /3}\;.
\end{eqnarray}
where $\eta=\dfrac{1-\nu}{\mu}$. Solving these equations we can find the evolution of main cosmological parameters of Model(6) and its limiting cases (models (3), (4) \& (5)). Notice that, in the case of Model (6), $\mu$ must be positive and small in absolute value ($0<\mu\ll 1$) in order to recover the solution for Model (3) in the limit $\mu\to 0^+$.  For this reason, the term $a^{3\eta}\to 0$ for virtually any $a<1$. As for $\nu$, it can have any sign provided $|\nu|\ll1$.

Using equations (\ref{31}) and (\ref{37}) and expanding for small redshift $z$, i.e. around our current epoch, we can determine the effective EoS for the the general Model (6):
\begin{eqnarray}
\label{eq:EoSmod3}
w_{\rm de}(z)&\simeq&-1+\frac{H_0^2(1-\nu)}{c_0}\,\Omega_{m,0}\,(\nu-\mu)\,(1+z)^3\nonumber\\
&\simeq& -1+\frac{\Omega_{m,0}}{1-\Omega_{m,0}}\,(\nu-\mu)\,(1+z)^3\,,
\end{eqnarray}
where $c_0$ is given by (\ref{35}). In the expansion we have neglected terms beyond linear order in $\nu$ and $\mu$ since these are small for Model (6).
The above equation is particularly illustrative to assess the effective quintessence or phantom-like behavior shown by these models. We shall come back to it in Sect. IV A corresponding to the numerical analysis. Needles to say, the EoS for models (3) and (4) are recovered from Eq.\,(\ref{eq:EoSmod3}) in the limits $\mu=0$ and $\nu=0$ respectively. The EoS for Model (5) cannot be expressed analytically within the same approximation since for that model $c_0=0$ and this enforces the parameters $\mu$ and $\nu$ be of order one, see e.g. Table \ref{tab:bestfit}.

Finally, let us emphasize that we must consider some limitation for the model parameters $\mu, \nu$ and $c_0$ as below:
\begin{itemize}
\item In all models we must have $\mu \geqslant 0$, since otherwise the term $\dfrac{dH^2}{da}$ in Eq.(\ref{34}) could become arbitrarily large and negative in $a\rightarrow0$, which  leads to negativity of $H^2$ term.
As indicated, this also warrants the retrieve of the solution for Model (3) in the limit $\mu\to 0^+$ of  Model (6).
\item Parameter $c_0$ is not independent,  as it is determined from the fitted values of  $\mu$ and $\nu$ and the other conventional parameters , see Eq.(\ref{35}). For  Model (5), in contrast, $c_0=0$ and this imposes a constraint on the remaining parameters.
\item In order to break degeneracies among the parameters in model(6),  we set $\mu=-\nu$\,\citep{Gomez-Valent:2015pia}.
\end{itemize}

\subsection{Cosmology at perturbation level}
In this part of the paper we investigate the linear evolution of matter perturbations in DDE cosmology. In order to identify the effects of dynamical DE on the linear growth of matter fluctuations we will introduce two distinct approaches, which usually have been considered in the literature: first, ``homogeneous DE'' in which the DDE component remain homogeneous and hence unperturbed ($\delta_{\rm de}\equiv 0$) and only dark matter is allowed to cluster
($\delta_{\rm m}\ne 0$); and, second, ``clustered DE'' in which the DDE component can cluster together with dark matter ($\delta_{\rm m}\ne 0$ and $\delta_{\rm de}\ne 0$). \citep{ArmendarizPicon:2000dh,Bean:2003fb,Abramo2007,Ballesteros:2008qk,Pace2010,Basilakos2010,Basilakos:2010rs,Batista:2013oca,Pace:2014taa,Mehrabi:2015hva,Malekjani:2016edh,Rezaei:2017yyj,Rezaei:2017hon,Malekjani:2018qcz,Rezaei:2019roz}.

The main equations which control the evolution of perturbations in matter and DE components within the Newtonian gauge are given by \citep{Abramo:2008ip,Grande:2008re}
 \begin{eqnarray}
 &&\dot{\delta_{\rm m}}+\frac{\theta_{\rm m}}{a}=0\;,\label{grwoth1}\\
&& \dot{\delta_{\rm de}}+(1+w_{\rm de})\frac{\theta_{\rm de}}{a}+3H(c_{\rm eff} ^2 -w_{\rm de})\delta_{\rm de}=0\;,\label{grwoth2}\\
&& \dot{\theta_{\rm m}}+H \theta_{\rm m} - \frac{k^2 \phi}{a}=0\;,\label{grwoth3}\\
 &&\dot{\theta_{\rm de}}+H \theta_{\rm de} - \frac{k^2 c_{\rm eff} ^2 \theta_{\rm de}}{(1+w_{\rm de})a} - \frac{k^2 \phi}{a} =0\;,\label{grwoth4}
 \end{eqnarray}
where $k$ and $c_{\rm eff}$ are the wave number and the effective sound speed of perturbations respectively, $\phi$ is the potential in the Newtonian gauge (assuming that there is no anisotropic stress) and $\theta_i$ is the divergence of the peculiar velocity for each component (matter and DE).

\begin{table*}
	\centering
	
	\caption{The best fit values of the free parameters  for the different DDE models and the $\CC$CDM using expansion data sets.
	}
	\begin{tabular}{c  c  c c c c c c}
		\hline \hline
		Model & Model(1) & Model(2) & Model(3) & Model(4) & Model(5) & Model(6) &$\Lambda$CDM\\
		\hline
		$\Omega_{\rm m}^{(0)}$ & $0.2818^{+0.0075}_{-0.0072}$ & $0.2826\pm 0.0070$ & $0.2874\pm 0.0074$ & $0.2872\pm 0.0057$ & $0.2889\pm 0.0060$   & $0.2856\pm 0.0075$ &$0.281^{+0.0096}_{-0.0089}$ \\
		\hline
		$ h $ & $0.6622^{+0.0049}_{-0.0058}$ & $0.7143\pm 0.0086$ & $0.7144\pm 0.0080$ & $0.7152\pm 0.0051$  & $0.7128^{+0.0074}_{-0.0078}$ & $0.7166\pm 0.0066$ & $0.7117^{+0.0082}_{-0.0096}$ \\
		\hline
		$\gamma $ & $1.0$ & $1.141\pm 0.017$ & $-$ & $-$ & $-$ & $-$ & $-$ \\
		\hline
		$ \mu $ & $-$ & $-$ & $0.0$ & $0.092^{+0.0071}_{-0.0078}$ & $0.921\pm 0.029$  & $0.0057\pm 0.0024$  & $-$ \\
	    \hline
		$ \nu $ & $-$ & $-$ & $-0.0095\pm 0.0112$ & $0.0$  & $0.928\pm 0.024$ & $-0.0057\pm 0.0024$  &  $-$ \\
        \hline
		$ w_{\rm de}(z=0) $ & $-0.780\pm 0.023$ & $-0.840\pm 0.0167$ & $-1.0037\pm 0.0031$ & $-1.0003^{+0.0022}_{-0.0023}$  & $-1.0748\pm 0.030$ & $-1.002\pm 0.0017$  & $-1.0$\\
		\hline
		$ \Omega_{\rm de}(z=0) $ & $0.7182$ & $0.7174$ & $0.7126$ & $0.7128$  & $0.7111$ & $0.7144$ & $0.719$ \\
		\hline \hline
	\end{tabular}\label{tab:bestfit}
\end{table*}

Next we consider  the perturbed Poisson equation
  \begin{equation}
\label{Possnew}
 -\frac{k^2}{a^2}\phi=\frac{3}{2} H^2[\Omega_{\rm m} \delta_{\rm m} + (1+3 c_{\rm eff} ^2)\Omega_{\rm de} \delta_{\rm de}]\;,
\end{equation}
 and combine it with Eqs. (\ref{grwoth3} \& \ref{grwoth4}). Since we mainly consider perturbations within the subhorizon scales ($k^2 >> H^2$), as usual we can ignore the terms  $H\dot{\phi}$ and $H^2\phi$ in the above equations\,\citep{Gomez-Valent:2018nib}. Upon removing $\theta_{\rm m}$ and $\theta_{\rm de}$ and trading the cosmic time derivative for the derivative with respect to the scale factor $a$, one can obtain the following system
of differential equations \citep[see also][]{Malekjani:2016edh,Rezaei:2017yyj}:
\begin{eqnarray}
&&\delta_{\rm m}''+{\cal A}_{\rm m} \delta_{\rm m}'= {\cal C}_{\rm m}[\Omega_{\rm m} \delta_{\rm m} +\Omega_{\rm de}  (1+3 c_{\rm eff} ^2)\delta_{\rm de}]\;, \label{growth5}\\
&&\delta_{\rm de}''+{\cal A}_{\rm de} \delta_{\rm de}'+{\cal B}_{\rm de} \delta_{\rm de}={\cal C}_{\rm de}[\Omega_{\rm m} \delta_{\rm m} +\Omega_{\rm de}  (1+3 c_{\rm eff} ^2)\delta_{\rm de}]\,.\nonumber
 \end{eqnarray}
 where ${\cal A}_{\rm m}$ and ${\cal C}_{\rm m}$ are given by

\begin{eqnarray}
&&{\cal A}_{\rm m}=\frac{3}{2 a}(1-w_{\rm de} \Omega_{\rm de})\;,\nonumber \\
&&{\cal C}_{\rm m}=\frac{3}{2a^2}\;,\nonumber
\end{eqnarray}

and ${\cal A}_{\rm de}$, ${\cal B}_{\rm de}$ and ${\cal C}_{\rm de}$ have the form
\begin{eqnarray}
&&{\cal A}_{\rm de}=\frac{1}{a}[-3 w_{\rm de} -\frac{a w_{\rm de}'}{1+w_{\rm de}}+\frac{3}{2}(1-w_{\rm de} \Omega_{\rm de})],\nonumber \\
&&{\cal B}_{\rm de}=\frac{1}{a^2}[-a w_{\rm de}' +\frac{a w_{\rm de}' w_{\rm de}}{1+w_{\rm de}}-\frac{1}{2}w_{\rm de}(1-3 w_{\rm de} \Omega_{\rm de})],\nonumber \\
&&{\cal C}_{\rm de}=\frac{3}{2a^2}(1+w_{\rm de}).\nonumber
\end{eqnarray}

As the  initial condition, we set the initial scale factor $a_{\rm i}=10^{-4}$ and $\delta_{\rm m}(a_{\rm i})=10^{-5}$. By these choices we verify that matter perturbations always remain in the linear regime.
The other appropriate initial conditions which we need in order to solve the above system are obtained as follows \citep{Batista:2013oca,Mehrabi:2015kta,Malekjani:2016edh}:

  \begin{eqnarray}\label{initialcondition}
&& \delta_{\rm m}'(a_{\rm i})=\frac{\delta_{\rm m}(a_{\rm i})}{a_{\rm i}}\;,\nonumber \\
&& \delta_{\rm de}(a_{\rm i})=\frac{1+w_{\rm dei}}{1-3w_{\rm dei}}\delta_{\rm m}(a_{\rm i})\;,\nonumber \\
&&\delta_{\rm de}'(a_{\rm i})=\frac{4 w_{\rm dei}'}{(1-3w_{\rm dei})^2}\delta_{\rm m}(a_{\rm i})+\frac{1+w_{\rm dei}}{1-3w_{\rm dei}}\delta_{\rm m}'(a_{\rm i})\;,
\end{eqnarray}
where $w_{\rm dei}$ is the value of $w_{\rm de}(a)$ at $a=a_{\rm i}$. The two scenarios for DE perturbations that we are going to consider are defined as follows.  By setting $c_{\rm eff}\equiv 1$ leads to the homogeneous DE option, whereas by setting  $c_{\rm eff}\equiv 0$ we explore the fully clustered DE results.   In the first case the sound horizon is equal or larger than the Hubble horizon, which means that DE perturbations are occurring only at very large scales;  in the second case, instead, the sound horizon is much smaller than the Hubble radius and thus DE perturbations can grow  in a similar manner to matter perturbations.
These two scenarios are at opposite ends and therefore if clustered DE has any impact it should show up in the second scenario.

The combined system of matter and DE perturbations are treated as follows. After solving the coupled  system of equations (\ref{growth5})
we obtain the evolution of matter fluctuation $(\delta_{\rm m})$ and from it we can compute the growth function using
\begin{eqnarray}\label{fz}
f(z)=\dfrac{d\ln{\delta_{\rm m}}}{d\ln{a}}\;.
\end{eqnarray}

Subsequently, we obtain the rms mass variance for spheres of $R=8h^{-1}$ Mp  following the procedure of \citep{Gomez-Valent:2014rxa,Gomez-Valent:2014fda,Sola:2016jky,Sola:2015wwa,Sola:2016ecz,Sola:2017jbl}:
\begin{eqnarray} \label{s8z}
\sigma_8(z)=\sigma^\Lambda_{8}(z=0)\frac{\delta_{\rm m}(z)}{\delta^\Lambda_{\rm m}(z=0)}\times \; \nonumber \\
\left[\dfrac{\int^\infty_0 k^{n_s+2}T^2(\Omega_{\rm m,0},k)W^2(kR_8)dk}{\int^\infty_0 k^{n_s+2}T^2(\Omega^\Lambda_{\rm m,0},k)W^2(kR_8)dk} \right]^{1/2} \;.
\end{eqnarray}
In the above equation, we use a spherical top-hat window function, whose Fourier transform reads as follows:  $W(k,R_8)=(3\sin kR_8-3kR_8\cos kR_8)(kR_8)^{-3}$.   For the transfer function $T(\Omega_{\rm m,0},k)$,  we use the BBKS  form \,\cite{Bardeen:1985tr,Liddle:2000cg}:
\begin{eqnarray}
T(\Omega_{\rm m,0},k)=\dfrac{\ln{(1+2.34q)}}{2.34q}\times \;\nonumber \\
\left[ 1+3.89q+(16.1q)^2+(5.46q)^3+(6.71q)^4\right]^{-1/4} \,,
\end{eqnarray}
where
\begin{eqnarray}
q(k)=\dfrac{k.Mpc}{\Omega_{\rm m,0}h^2}\exp (\Omega_{\rm b,0}+\sqrt{2h}\dfrac{\Omega_{\rm b,0}}{\Omega_{\rm m,0}})\,.
\end{eqnarray}

\begin{figure*}
	\centering
	\includegraphics[width=8cm]{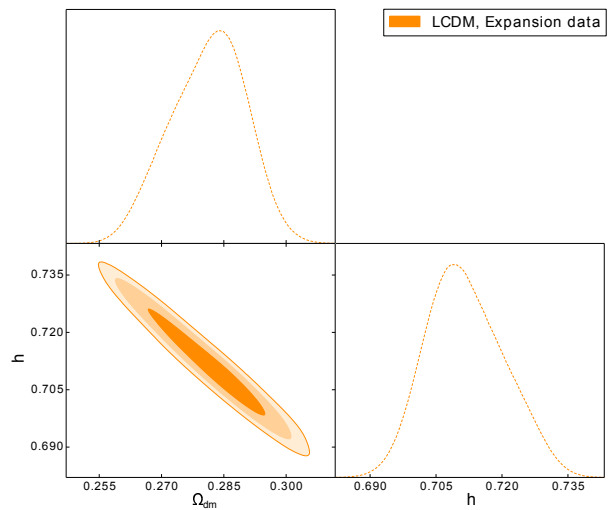}
		\caption{ Contours of $1\sigma$, $2\sigma$ and $3\sigma$ confidence level
for the free parameters of the $\Lambda$CDM model obtained from the expansion data.}
	\label{fig:contour1}
\end{figure*}

We use here the  BBKS transfer function as a sufficient approximation to reach the main  results of our study. One could use a variety of MCMC Boltzmann codes to improve
accuracy.  However, our main aim in this case is to correctly identify the order of magnitude of the results so as to insure the main conclusions of our study insofar as concerns the possible relaxation of the main tensions of the $\CC$CDM, see our discussion in Sect. V  (Conclusion). For more details on related models and comparison of different levels of treatment of the structure formation,  see e.g. the studies   in Refs. \citep{Sola:2016jky,Sola:2016ecz,Sola:2017jbl} and  \citep{Sola:2018sjf}. Thus, so long as we deal with the order of magnitude of the results, using BBKS is a good approximation to probe the ability of the dynamical DE models under study to deal with the mentioned tensions.
In this work we set the scalar spectral index $n_s=0.965$ and the matter fluctuation amplitude $\sigma^\Lambda_8(z=0)=0.811\pm 0.006$ from the base-$\Lambda$CDM results of Planck 2018 \citep{Aghanim:2018eyx}.
 By multiplying the results of Eqs.(\ref{fz} \& \ref{s8z}) we can calculate the important observable quantity $f(z)\sigma_8(z)$, which is a crucial ingredient in our analysis insofar as concerns the structure formation data.

\begin{figure*}
	\centering
\includegraphics[width=8cm]{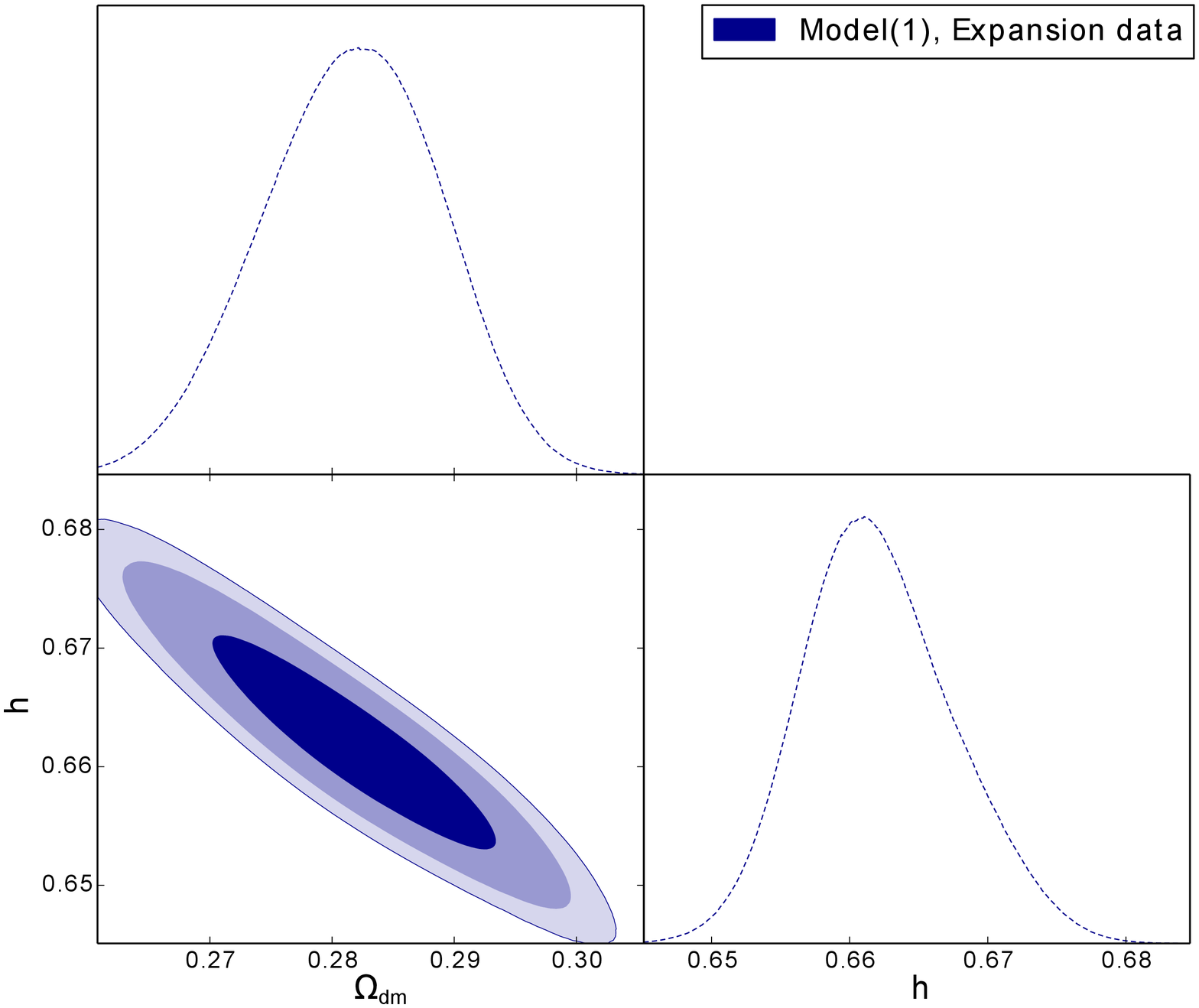}
\includegraphics[width=8cm]{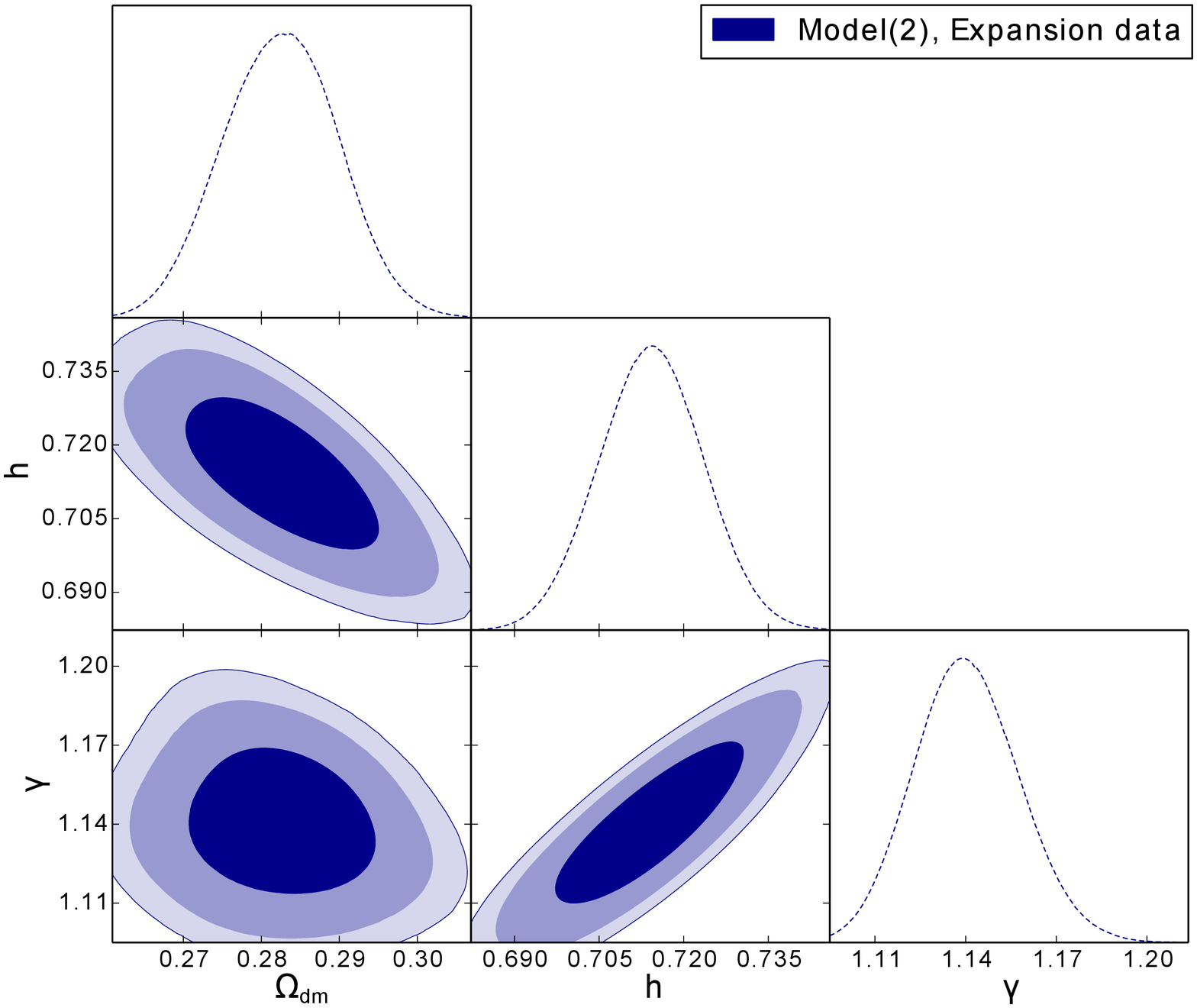}
\includegraphics[width=8cm]{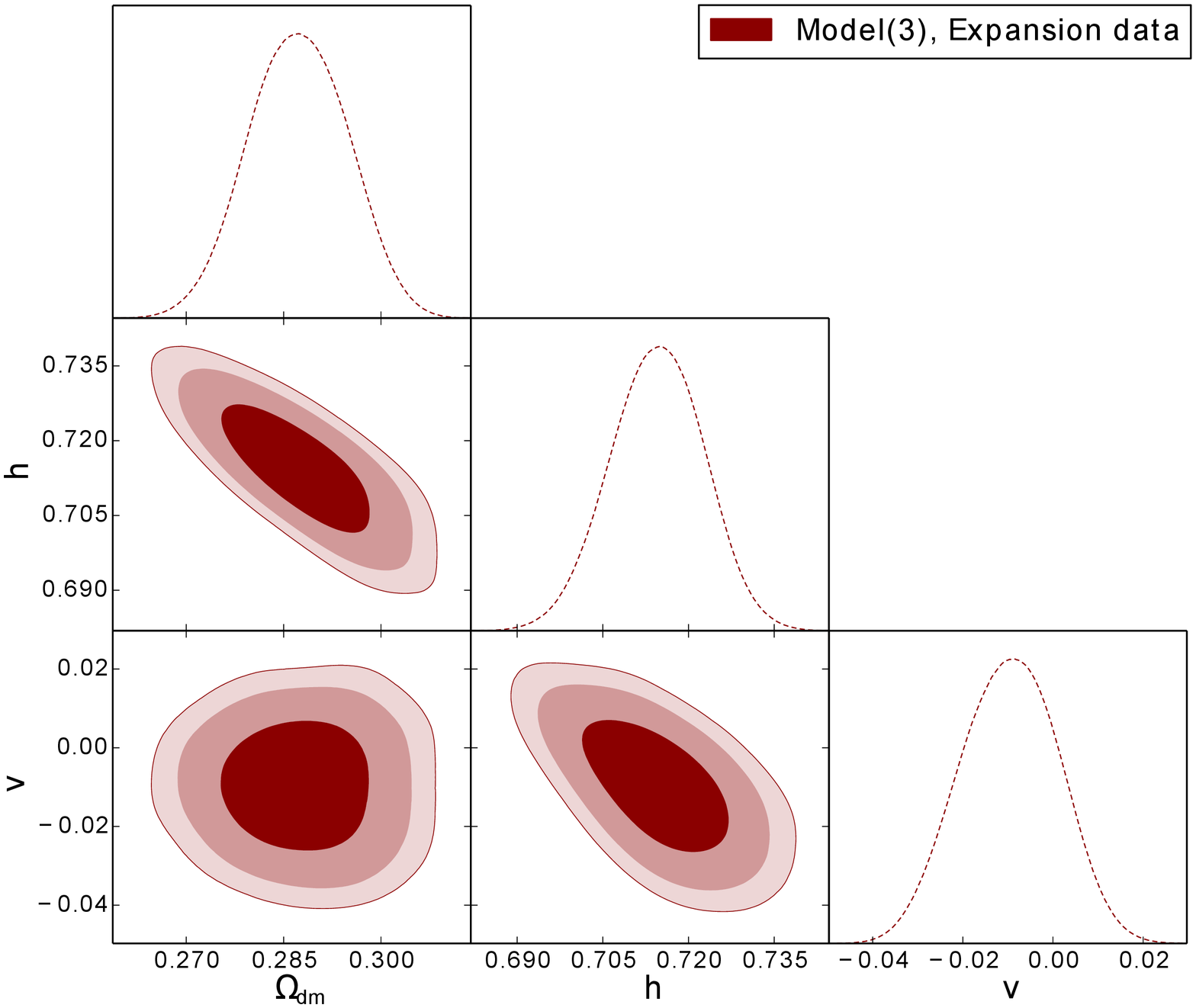}
\includegraphics[width=8cm]{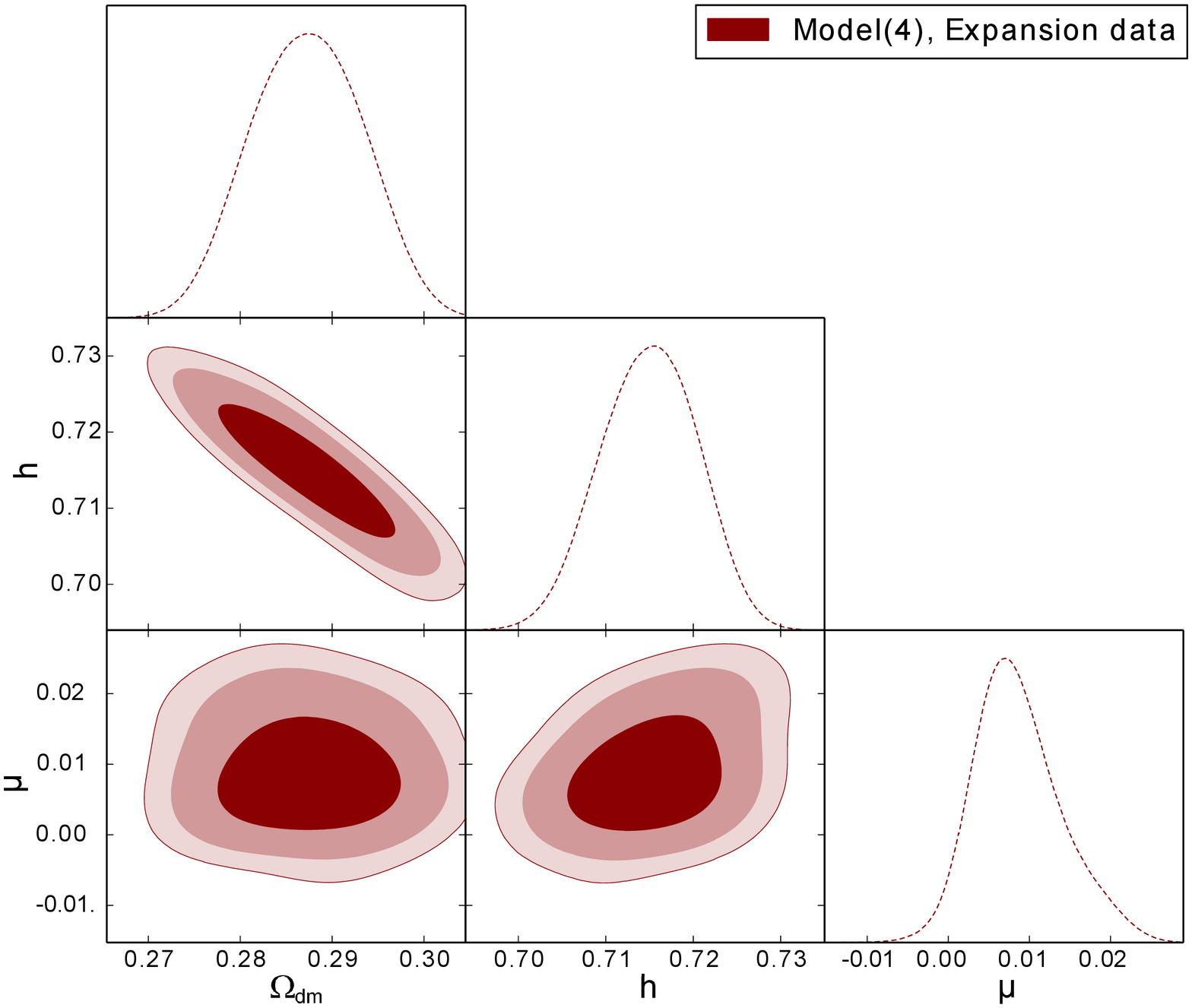}
\includegraphics[width=8cm]{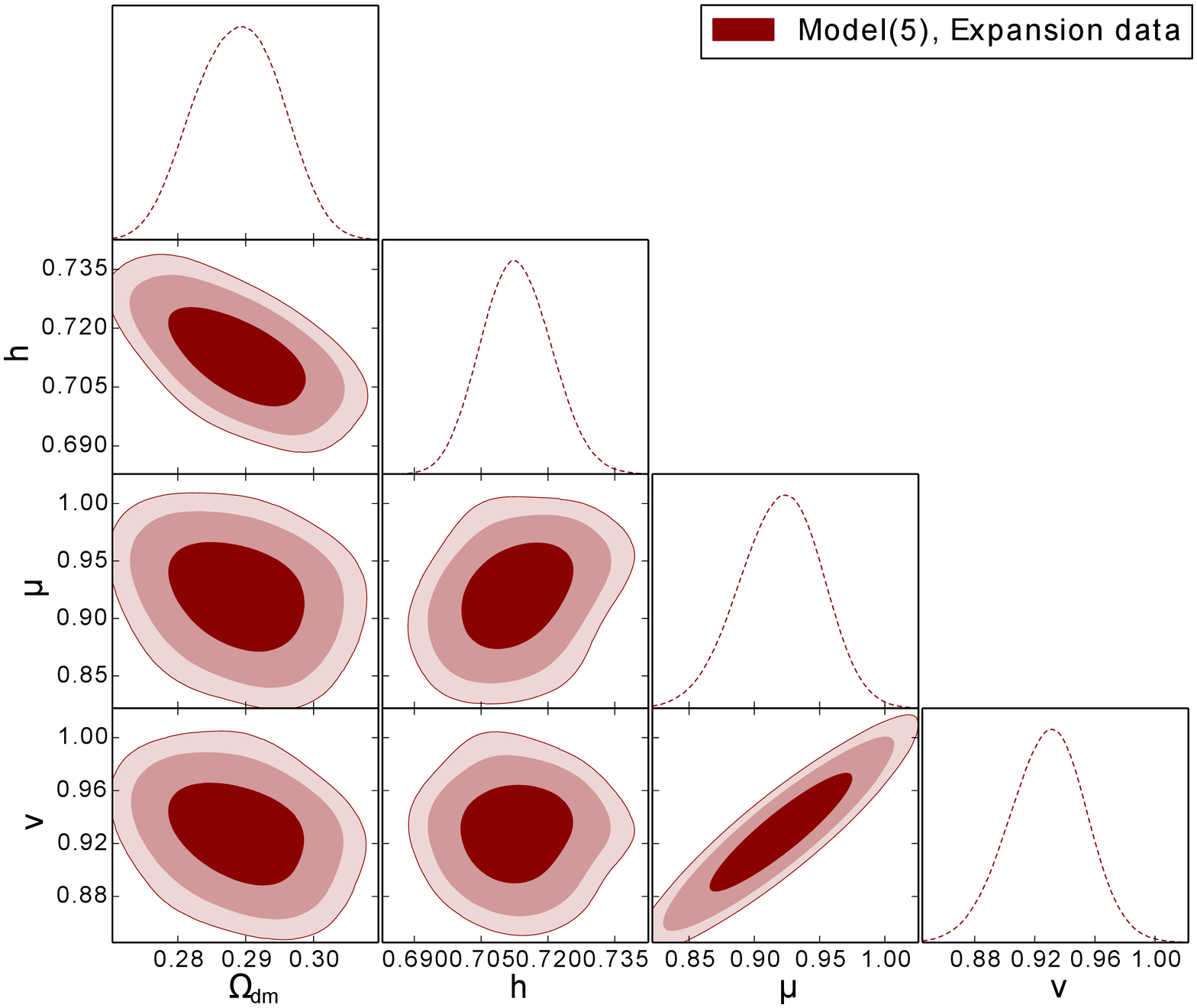}
\includegraphics[width=8cm]{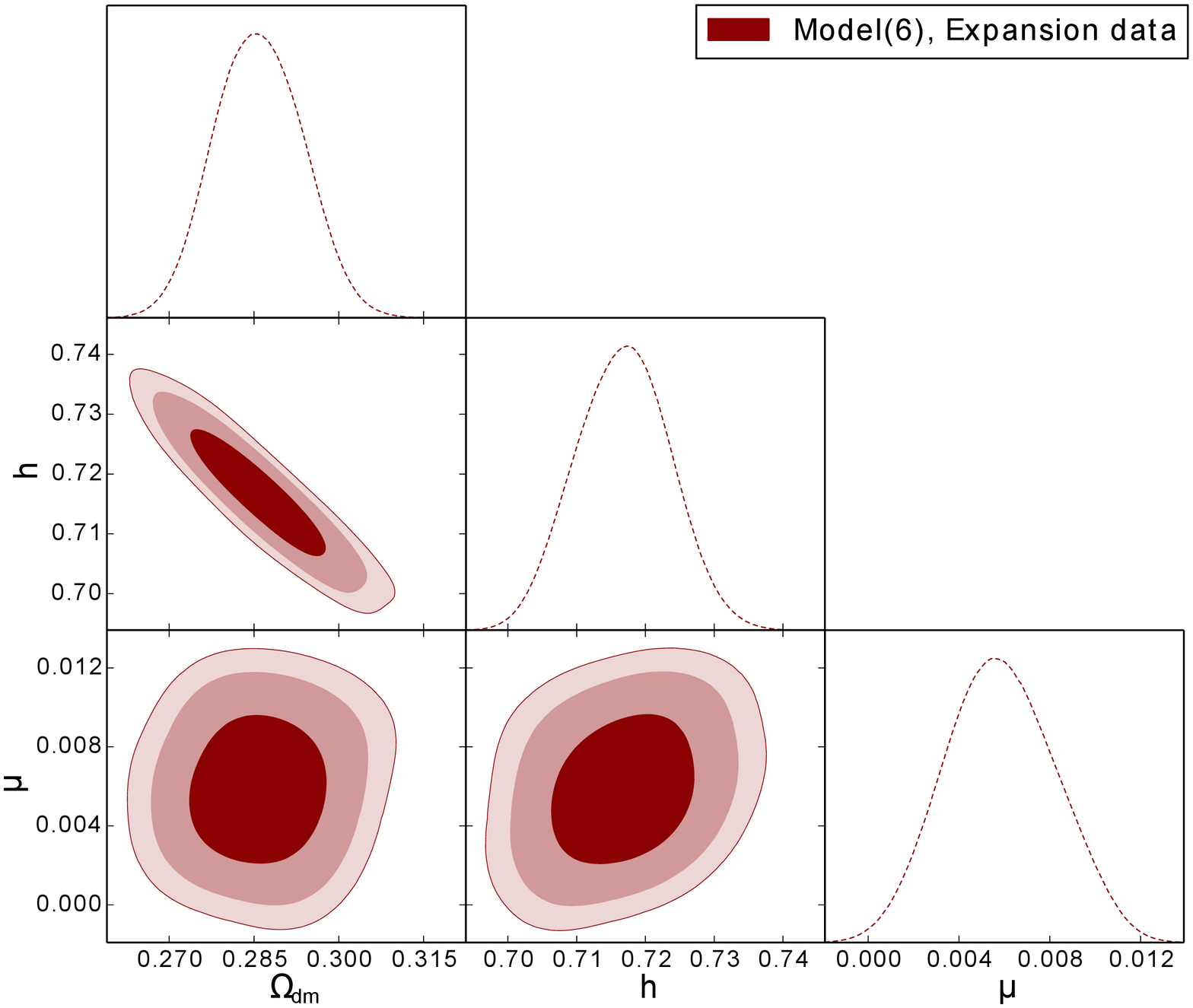}

	\caption{ Contours of $1\sigma$, $2\sigma$ and $3\sigma$ confidence level
for  the free parameters of the different DDE models obtained from the expansion data. The model names are shown in the legend of each panel.}
	\label{fig:contour2}
\end{figure*}

\section{data analysis}\label{sect:DATA}
In the following, we describe the observational samples and statistical data analysis method that will be adopted to constrain the free parameters of the different DE models under study. We devote the rest of this section to explain the statistical methods and to introduce our criteria for selecting the best models.

\subsection{Data samples}
First of all, we use the expansion data at background level. Using these data sets we can investigate the evolution of the universe in the presence of DE. In this case the presence of the DE (and its possible dynamics)  just affects the expansion rate of the universe owing to its negative pressure. Pantheon Sample, a set of latest 1048 type Ia supernovae of \citep{Scolnic:2017caz}, is the largest sample of data points we use in this study to constrain the cosmological parameters through the comparison  of  their  apparent  luminosities  over  a  range of redshifts. Furthermore, a deep geometrical probe of dark energy is the position of the CMB acoustic peak which provides accurate data to constrain dark energy models. In this paper we deal with the CMB data through the method of distance priors, which encodes in a sufficiently precise way the compressed likelihood as a substitute for the
full CMB power spectrum analysis \citep[see][]{Bond:1997wr,Efstathiou:1998xx,Wang:2007mza,Chen:2018dbv}.  As we shall see, it is enough to capture the main traits of the dynamical DE features and to exemplify the effectiveness of dynamical DE as an alternative to a rigid CC term. In the distance prior method the  CMB data are incorporated by using constraints on the parameters $(R,l_a,z_{\star})$, where $R$ is the scale distance to recombination \citep[see][]{Basilakos:2009wi,Mehrabi:2015hva} and references therein]. It has been shown that measuring the parameters$(R,l_a,z_{\star})$  provides an efficient and intuitive summary of CMB data as far as dark energy constraints are concerned.  Recently, the results of the distance prior method were compared with the results of full CMB power spectra analysis in \citep{Chen:2018dbv}. This study shows that the results obtained from both methods are in very good agreement.  Furthermore, we use the position of the acoustic peak from the Planck measurements, in which we have $(R=1.7499,l_a=301.65,z_{\star}=1090.41)$\citep{Shafer:2013pxa}. We also use the BAO scale originated in the last scattering surface by the contest between gravity and the pressure of the coupled photon-baryon fluid. The generated acoustic waves deposit an overdensity signature at certain length scales of the matter distribution. They become visible as continual periodic matter fluctuations in large scale structure resulting from sound waves spreading in the early Universe. Different studies prove that measurements of BAO scale is useful as a standard ruler that we can use in order to constrain DE models. As mentioned in Sect.\ref{sect:intro}, we use a compilation which includes 11 distinct measurements of the baryon acoustic scale from \citep{Beutler:2011hx,Font-Ribera:2013wce,Ross:2014qpa,Alam:2016hwk,Bautista:2017zgn,Ata:2017dya} given in Table 1 of \citep{Park:2018tgj}. Moreover,  Big Bang Nucleosynthesis(BBN)  provides a single data point which constrains mostly the present value of baryon density parameter $\Omega_{\rm b,0}$\citep{Serra:2009yp,Burles:2000zk}. A further data set used in our analysis is the Hubble data from the redshift evolution of cosmic chronometers. This dataset includes $23$ data points from the redshift interval between $z=0.07$ and $z=1.75$. We start from the list of 28 data points quoted from a variety of sources in the literature in\citep{Farooq:2013hq}.  However, three data points of that collection were obtained from \citep{Blake:2012pj}, and are correlated with other data which were used in this work. Additionally, two other data points from \citep{Chuang:2011fy} and \citep{Busca:2012bu} are not fully model independent. Therefore we ignore these five data points in our analysis and only use the remaining  $23$  points (see also Table 2 in \citep{Rezaei:2019roe}).  The final data point on $H$ used at background level in our study is the recently measured local value of the Hubble parameter  $H_0$  given in\,\citep{Riess:2018uxu}.

All of the above data sets are significant for investigating the effect of DE on the evolution of the universe at background level. Thus we use the expression  "expansion data" to refer to them.  However, in order to investigate the effects of DE on the growth of matter fluctuations in linear regime, we need to resort to other data sets which reflect the role of DE in the formation of large scale structures.
Dark energy and its possible evolution can affect the formation of cosmic structures through three different mechanisms. First, the DE increases the expansion rate of the universe, so it suppresses the formation of structures. Furthermore,  because the DE becomes the dominating component of the universe, it slows down the growth of large scale overdensities, and the process of gravitational structure formation will reduce at scales comparable to the Hubble distance. These two mechanisms do affect the formation of structures through changing the Hubble expansion rate.  Notice that the dynamical character of the DE already affects the background evolution owing to the corresponding change of the Hubble function induced by a variable DE. But a third effect that should not be forgotten is that the DE density can  fluctuate. If such fluctuation were not negligible, the DE would not only  feel the gravitational pull of dark matter structures, but it should tend to form structures itself. An important data source which provides precious information about the effect of DE on the growth of matter perturbations is the LSS data obtained from redshift space distortions (RSD) in different galaxy surveys.  Such data set includes 18 independent $f\sigma_8$ data points collected by \citep{Nesseris:2017vor} from different references e.g.,\citep{Song:2008qt,Samushia:2011cs,Blake:2012pj,Hudson:2012gt,Blake:2013nif,Sanchez:2013tga, Chuang:2013wga,Howlett:2014opa,Feix:2015dla,Okumura:2015lvp,Pezzotta:2016gbo,Huterer:2016uyq}. Different authors use different  sets of independent points, see e.g. \cite{Sola:2018sjf}, \cite{Park:2018fxx}, \cite{Martinelli:2019dau} and references therein.  If the points in each set are not correlated, they should be essentially equivalent. These LSS data points provide the values of the quantity $f\sigma_8(z)$ at redshifts in between $z=0.02$ and $z=1.4$ and must be confronted with the theoretically predicted value from the DDE models under study, see  Sect. \ref{sect:back}.

\subsection{Statistical methods}\label{subsect:method}
In the previous part, we introduced all of the data sets which we use in this work. In what follows and for the sake of completeness we will briefly introduce the procedure under which we constraint our models with observational data.  Given a model with a set of free parameters and a set of observational data points, we should define a merit function in order to quantify the agreement between our model and observations. In this way by maximizing the degree of agreement, we can obtain the best values of the free parameters.  Therefore, any useful fitting procedure should provide: \textit{(i)} the best fit values of the parameters \textit{(ii)}  an estimate of the error on each of the parameters, and \textit{(iii)}  a reasonable measure of the goodness of the fit. We should bear in mind that if the model can't fit the observations, then the obtained best fit values of the free parameters are obviously meaningless. In our analysis we use the minimum chi-squared  ($\chi^2$) test for model fitting. When we have a set of data points $D$ and a theoretical model for these data , $M(x,\textbf{p})$, which depends on set of parameters $\textbf{p}$, the $\chi^2$-test in its simplest form becomes

\begin{eqnarray}\label{lsq}
\chi^2=\sum_{i}\frac{1}{\sigma^2_i}[D_{i}-M(x_i\mid\textbf{p})]^2\;,
\end{eqnarray}
where $\sigma_i$ is the error on data point $i$.  The best values for the collection of free parameters $\textbf{p}$ are those that minimize the  $\chi^2$- function.
If we deal with some data points which are correlated, we can not use the above equation for computing $\chi^2$. In this case we should apply the $\chi^2$-test  for correlated data points, instead of Eq.(\ref{lsq}):

\begin{eqnarray}\label{lsq2}
\chi^2=\sum_{i,j}[D_{i}-M(x_i\mid\textbf{p})]{\cal Q}_{ij}[D_{j}-M(x_j\mid\textbf{p})]\;.
\end{eqnarray}
 where ${\cal Q}_{ij}$ is the inverse of the covariance matrix. This matrix describes the covariance
between the data points. Among  the various data sets we are using in this work, the CMB data points are correlated. Also 11 data points in BAO sample have correlation among themselves. Therefore, in the case of these two data sets we use Eq.(\ref{lsq2}) for calculating the value of $\chi^2$.

\begin{table}
 \centering
 \caption{Relevance of $\Delta{\rm AIC}$ for support to a given model}
\begin{tabular}{c  c }
\hline \hline
$\Delta{\rm AIC}$  & Level of support to model\\
 \hline
$<2$ (including $<0$) &  Significant support \\
 \hline
Between 4 and 7 & Considerably less support for the model\\
 \hline
Greater than 10 & Essentially no support for the model\\
 \hline \hline
\end{tabular}\label{daic}
\end{table}


\begin{table}
 \centering
 \caption{Relevance of $\Delta{\rm BIC}$ as evidence \emph{against} a model}
\begin{tabular}{c  c }
\hline \hline
$\Delta{\rm BIC}$  & Evidence against\\
 \hline
Less than 2 &   No evidence \\
 \hline
Between 2 and 6 & Mild to positive \\
 \hline
Between 6 and 10 & Strong  \\
 \hline
Greater than 10 & Very Strong \\
 \hline \hline
\end{tabular}\label{dbic}
\end{table}

When we want to use several data sets with different $\chi_n^2$ functions, we should  totalize all of the $\chi_n^2$ values and finally minimize the result. In order to test a wide range of values for each of the parameters, we perform a Markov chain Monte Carlo (MCMC) analysis. For a wide range of cases the probability distribution for different values of the $\chi^2$-square function, Eq.(\ref{lsq}), around it's minimum is the distribution of $\chi^2$ for $N-k$ degrees of freedom where $k$ and $N$  are the number of free parameters and data points respectively. For more details concerning the $\chi^2(\textbf{p})$ function and the MCMC analysis we refer the reader to \cite{Mehrabi:2015hva} \citep[see also][]{Hinshaw:2012aka,Mehrabi:2015kta,Malekjani:2016edh}. The authors of \cite{Capozziello:2011tj} investigated the statistical performance of the MCMC procedure. Their results indicated that when we deal with a multi-dimensional space of the cosmological parameters, the MCMC algorithm provides better constraints compared to other popular fitting skills. In this paper we perform our analysis in two steps. Firstly, we limit our study to background level and just use expansion data to constraint DE model. In this step the total $\chi^2_{\rm T, exp}$ function becomes

\begin{equation}\label{eq:like-tot_chi}
 \chi^2_{\rm T, exp}({\bf p})=\chi^2_{\rm SN}+\chi^2_{\rm BAO}+\chi^2_{\rm CMB}+\chi^2_{\rm BBN}+\chi^2_{\rm H}+\chi^2_{\rm H_0}\;,
\end{equation}
where in view of the considerations made in the previous section, the statistical vector  of free parameters, ${\bf p}$, for each model reads as follows:
\begin{itemize}
\item Model(1): ${\bf p}=\{\Omega_{\rm DM0}, \Omega_{\rm b0}, h\}$
\item Model(2): ${\bf p}=\{\Omega_{\rm DM0}, \Omega_{\rm b0}, h, \gamma\}$
\item Model(3): ${\bf p}=\{\Omega_{\rm DM0}, \Omega_{\rm b0}, h, \nu\}$
\item Model(4): ${\bf p}=\{\Omega_{\rm DM0}, \Omega_{\rm b0}, h, \mu\}$
\item Model(5): ${\bf p}=\{\Omega_{\rm DM0}, \Omega_{\rm b0}, h, \mu, \nu\}$
\item Model(6): ${\bf p}=\{\Omega_{\rm DM0}, \Omega_{\rm b0}, h, \mu\}$
\item $\Lambda$CDM: ${\bf p}=\{\Omega_{\rm DM0}, \Omega_{\rm b0}, h\}$
\end{itemize}
In the next step we extend our investigation from background to perturbations level and combine the expansion data with the growth rate data to perform a joint statistical analysis. This means that the total $\chi^2$-square will now take the form:
\begin{equation}\label{eq:like-tot_chi2}
 \chi^2_{\rm T}({\bf p})=\chi^2_{\rm T, exp}+\chi^2_{\rm growth}\;,
\end{equation}
The results of these analyses for different DDE models and $\Lambda$CDM are discussed in Sect.\ref{sect:RS}

\begin{figure}
	\centering
	\includegraphics[width=8cm]{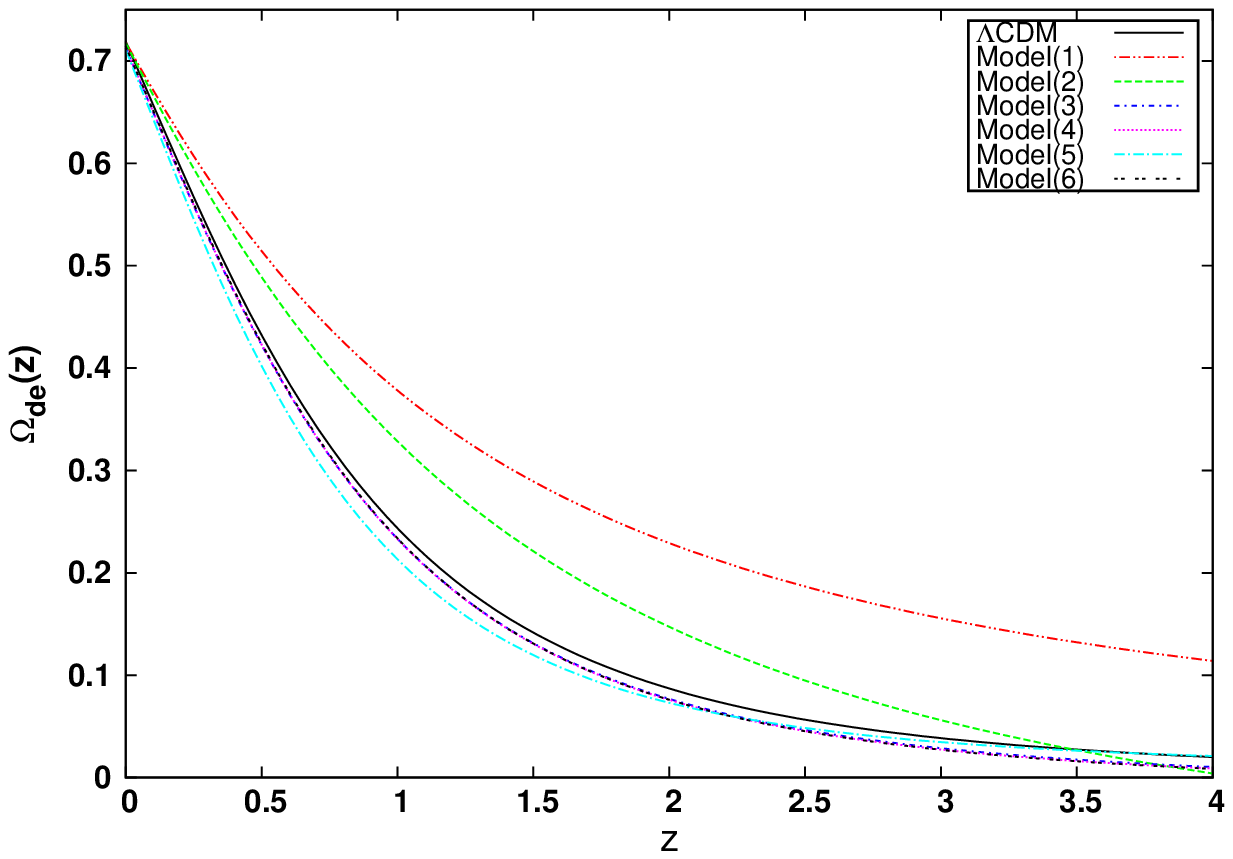}
	\includegraphics[width=8cm]{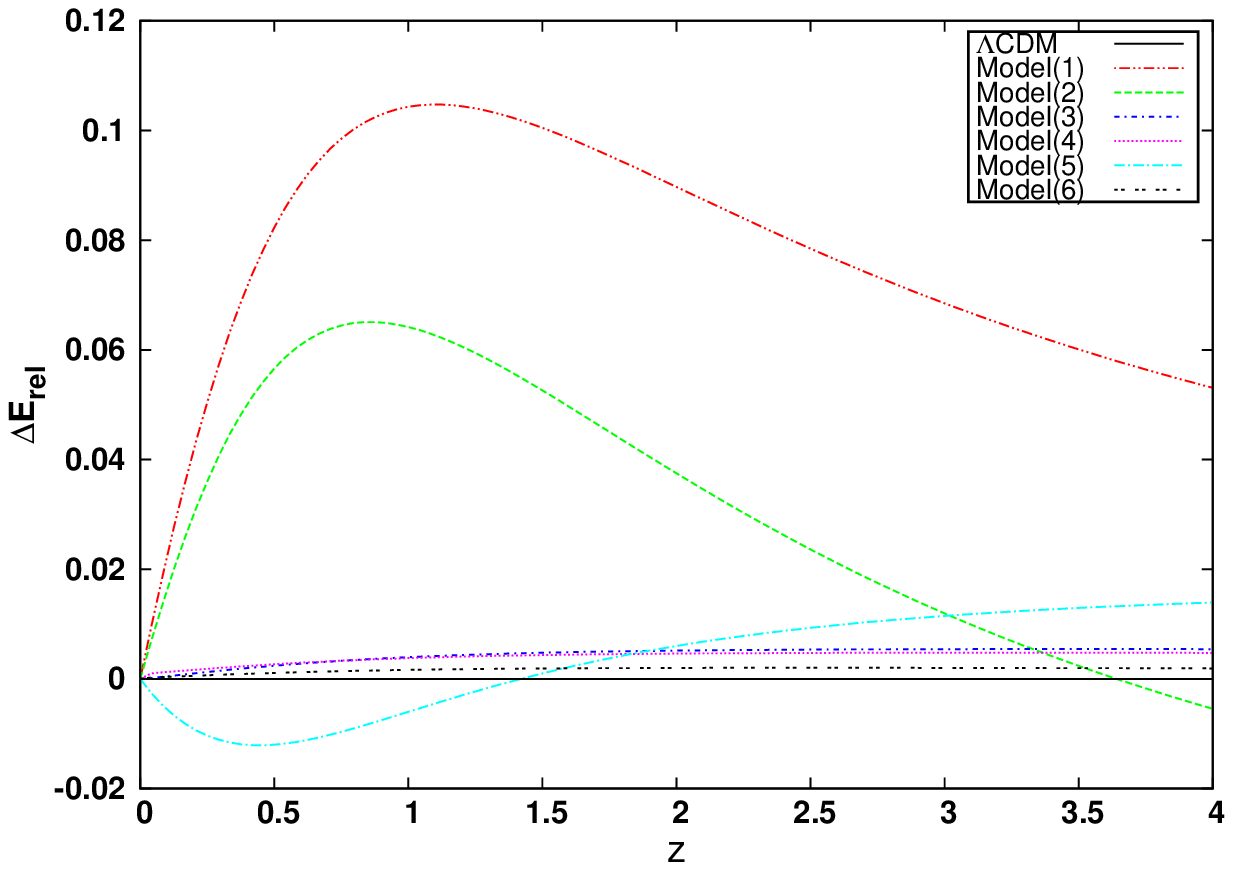}
	\caption{Evolution of the main cosmological quantities with the redshift,  based on the best fit values (reported in Table \ref{tab:bestfit}).
For each model, the density parameter $\Omega_{\rm de}(z)$ is shown in the upper panel, and the relative difference of the dimensionless Hubble parameter
$\Delta E_{rel}=[(E_{\rm model}-E_{\rm \Lambda})/E_{\rm \Lambda}]$ is plotted in the bottom panel.}
	\label{fig:back}
\end{figure}

\begin{figure*}
	\centering
	\includegraphics[width=5cm]{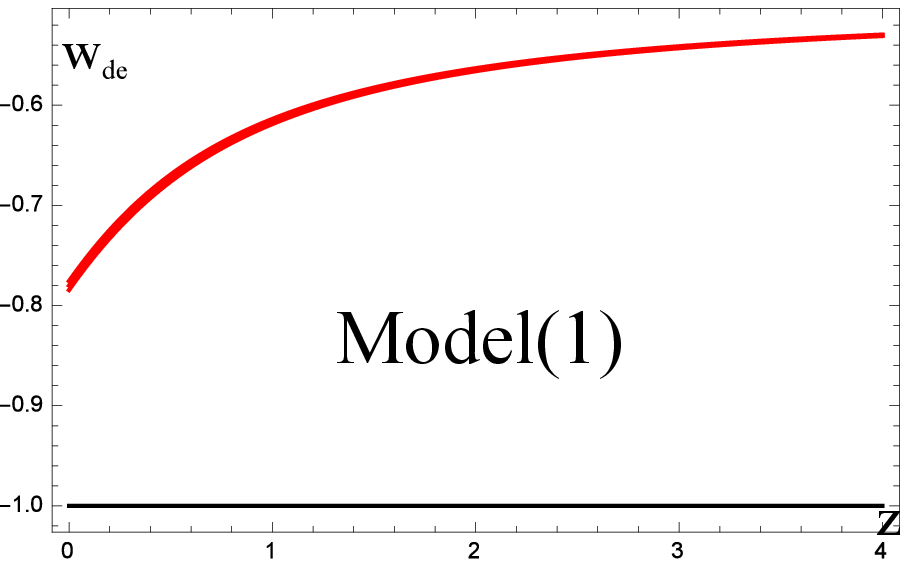}
	\includegraphics[width=5cm]{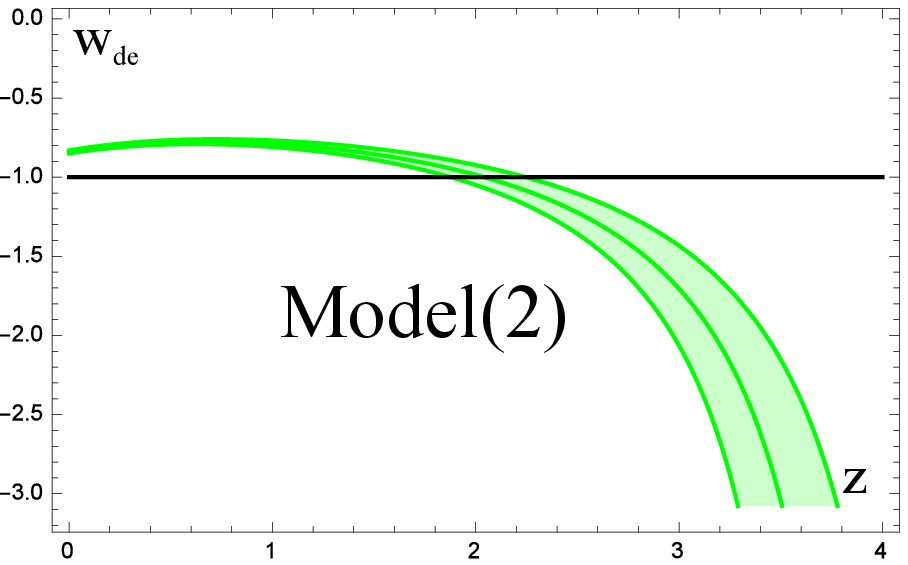}
	\includegraphics[width=5cm]{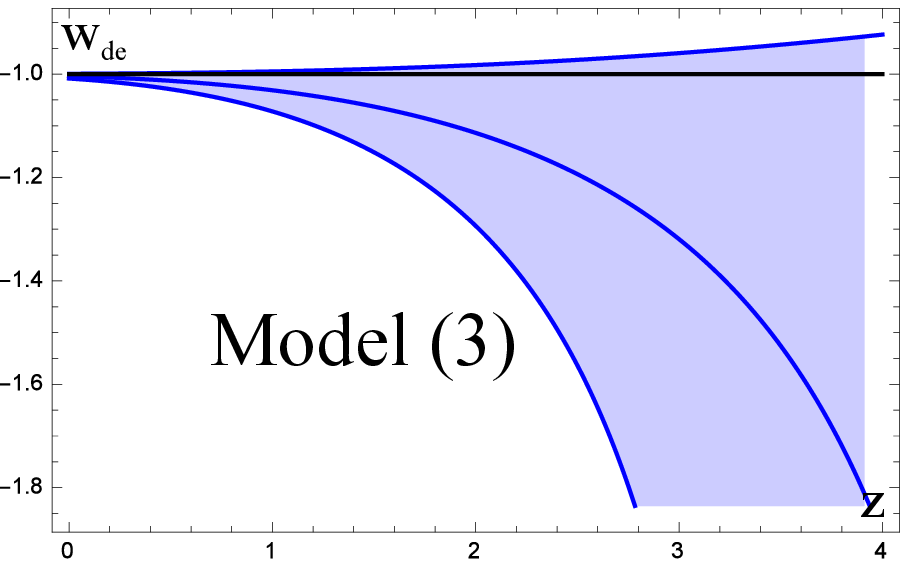}
	\includegraphics[width=5cm]{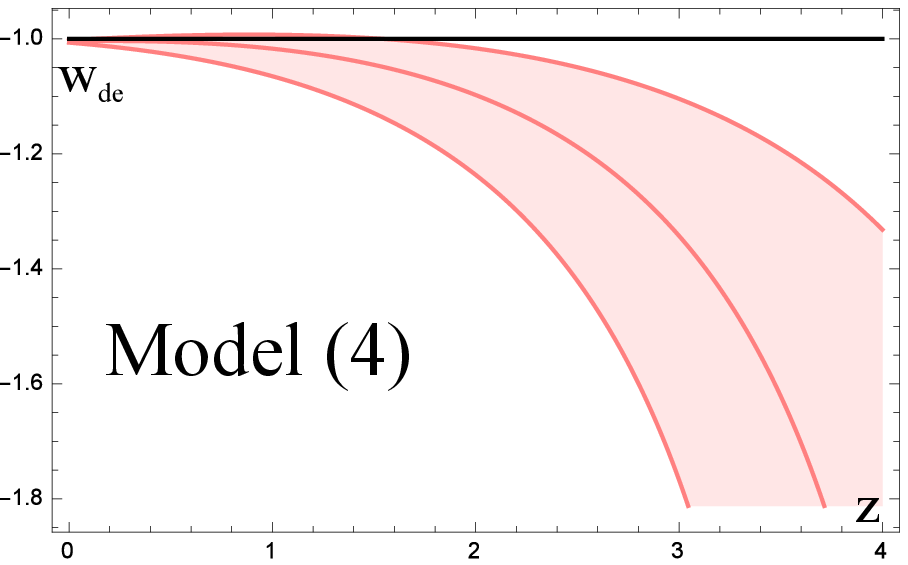}
	\includegraphics[width=5cm]{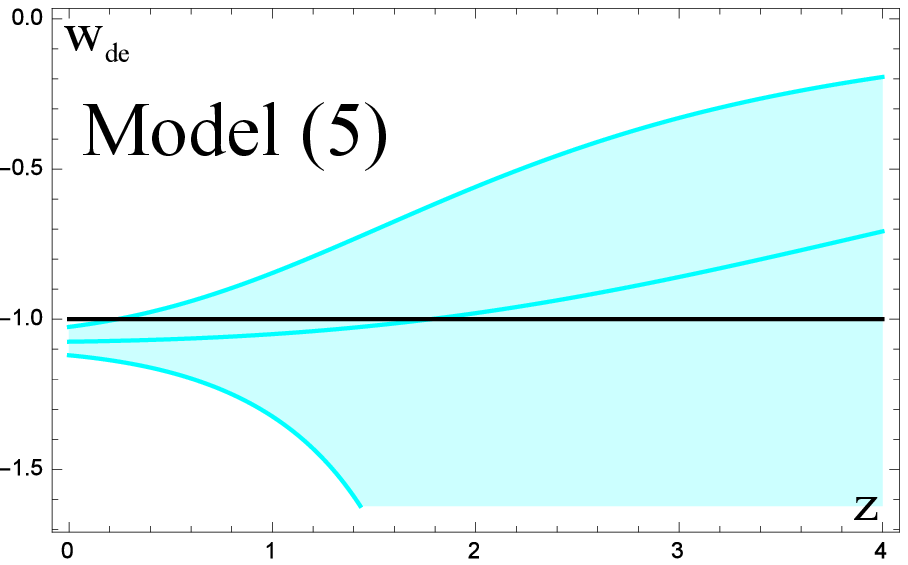}
	\includegraphics[width=5cm]{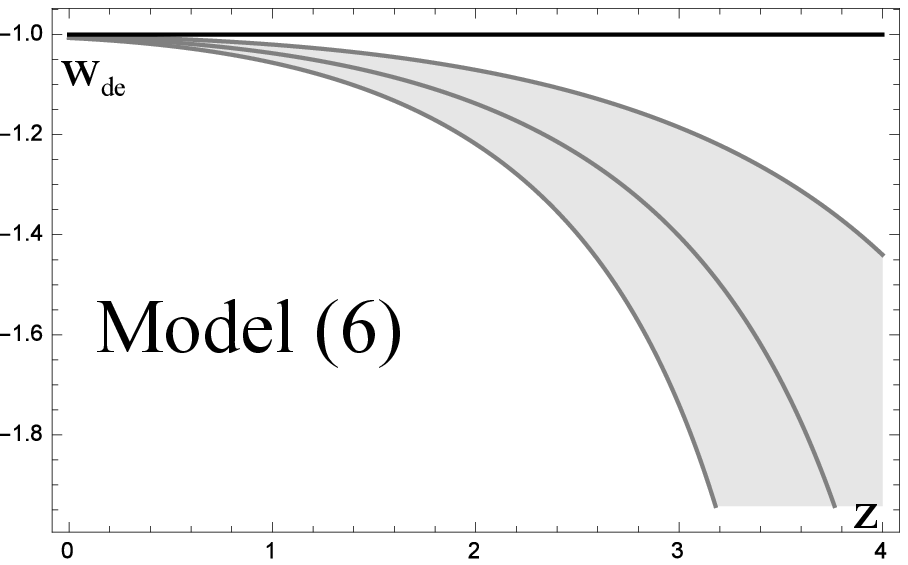}
	\caption{Evolution of the EoS parameter $w_{\rm de}(z)$ with the redshift, based on the best fit values of the free parameters and their $1-\sigma$ error bars (reported in Table \ref{tab:bestfit}). For each model in its particular panel, we show the $1-\sigma$ region of $w_{\rm de}(z)$ alongside with $w_{\rm \Lambda}=-1$. }
	\label{fig:w}
\end{figure*}

\begin{figure}
	\centering
	\includegraphics[width=8cm]{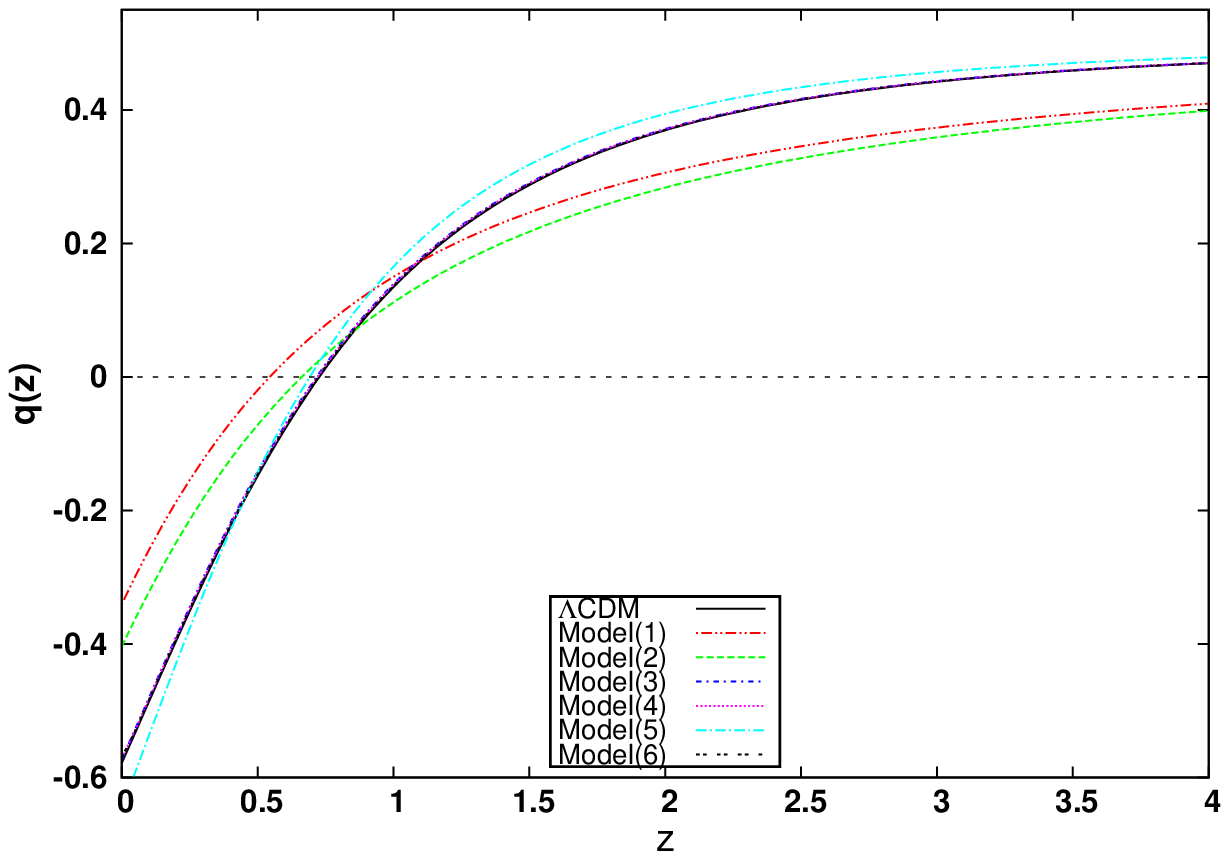}
    \includegraphics[width=8cm]{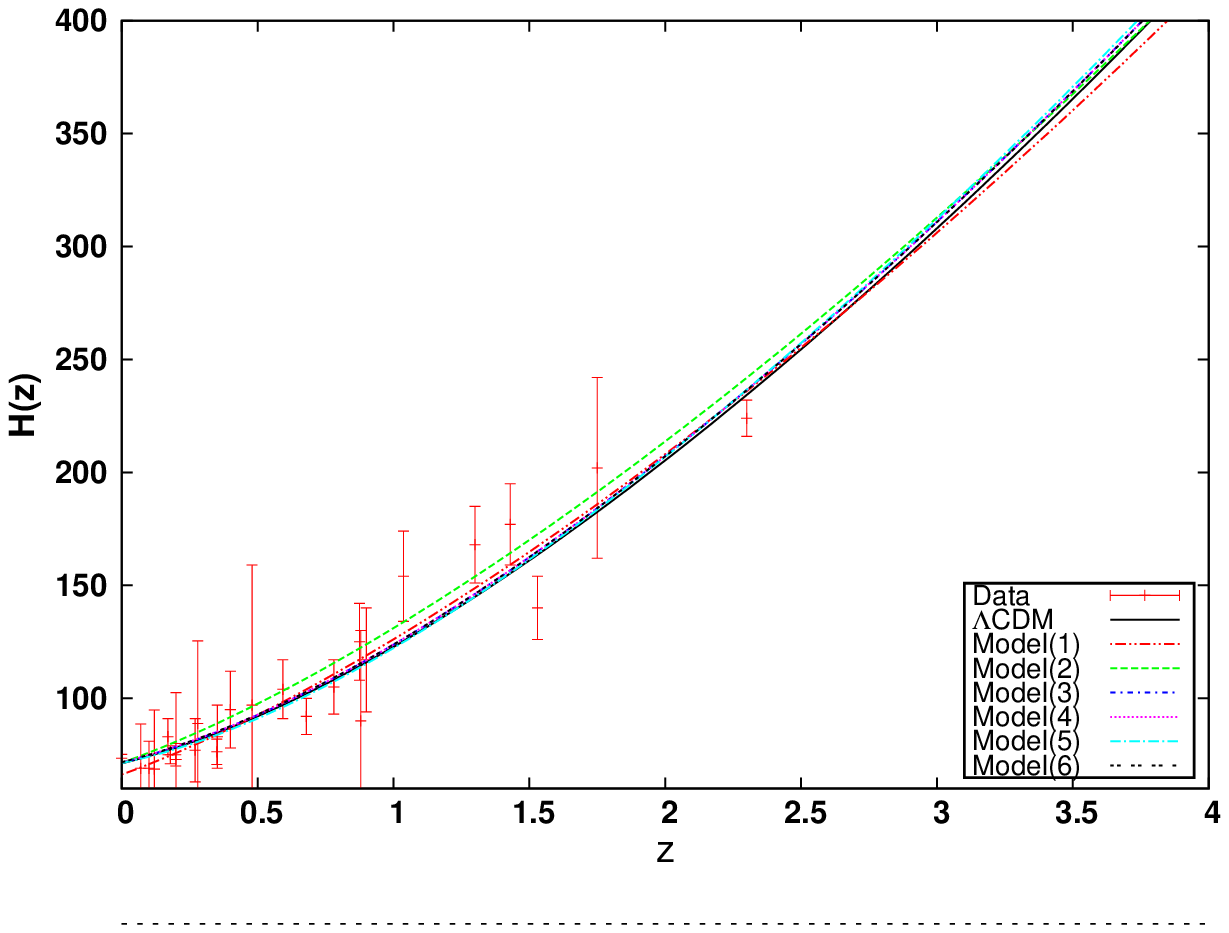}
    \includegraphics[width=8cm]{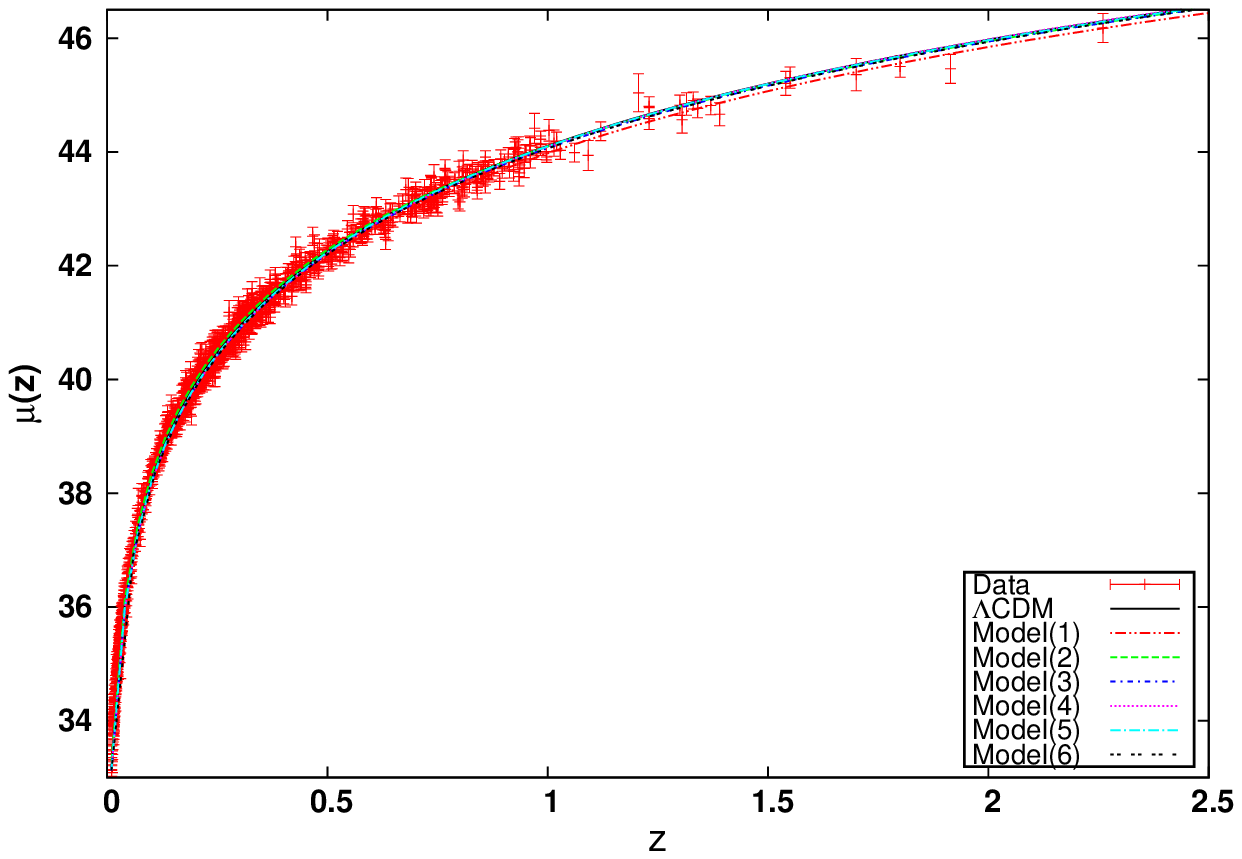}
	\caption{Evolution of $q(z)$  for the DE models considered in this work (upper panel) and comparison between the observed values of the
		Hubble parameter $H(z)$ and the theoretical predictions as a function of the redshift $z$ (middle panel). The evolution of the distance modulus $\mu(z)$ for different models and its observed values can be seen in the bottom panel. }
	\label{fig:q2}
\end{figure}

\subsection{Model selection}

As previously indicated, the simplest tool for comparing different models from the viewpoint of the fit quality  is the $\chi^2_{\rm min}$-test. That is to say, when different models compete to fit the same data,  the model which has smaller value of $\chi^2_{\rm min}$ is assumed to be the best.  However, to be fairer we should also take into account `Occam's razor' criterion, namely the idea that the simplest model  (the one with the smallest number of parameters) could perhaps be the best despite  its $\chi^2_{\rm min}$ may not be the minimum.  In order to balance the two criteria, which may well point towards opposite directions,  it is natural to penalize  appropriately the model having more parameters by adding an extra contribution to its  $\chi^2_{\rm min}$.  The penalty may depend on the total number of free parameters and/or the total number of data points entering the fit. To remove the advantage provided by the extra number of parameters (and therefore to better balance Occam'ss razor) it is conventional to use two well known information criteria, namely the Akaike and Bayesian criteria, AIC  and BIC\,\citep{Kass:1995loi, Burnham}. Both criteria attempt to restore the balance in the model competition by introducing a penalty term as follows:
\begin{eqnarray}
{\rm AIC} = \chi^2_{\rm min}+2k\;,\nonumber\\
{\rm BIC} = \chi^2_{\rm min}+k\ln N\;,
\end{eqnarray}
\begin{table*}
 \centering
 \caption{Comparison of the statistical results for the different DDE models  and the $\CC$CDM at background level using expansion data sets.
 }
\begin{tabular}{c  c  c c c c c c}
\hline \hline
 Model & Model(1) & Model(2) & Model(3) & Model(4) & Model(5) & Model(6) &$\Lambda$CDM\\
 \hline
$\chi^2_{\rm min}$ & 1273.6 & 1126.8 & 1067.0 & 1067.1 & 1068.0& 1066.9 & 1071.1  \\
 \hline
 AIC & 1279.6 & 1134.8 & 1075.0 & 1075.1 & 1078.0 & 1074.9 & 1077.1\\
 \hline
 $\Delta$AIC & 202.5 & 57.7 & -2.1 & -2.0 & 0.9 & -2.2 & 0.0  \\
 \hline
 ${\cal W}_{m}$ & 1.1 E-45 & 2.9 E-14 & 0.276 & 0.262 & 0.062 & 0.290 & 0.097  \\
 \hline
 BIC & 1294.6 & 1154.7  & 1095.0  & 1095.1  & 1103.0 & 1094.9 & 1092.1  \\
 \hline
 $\Delta$BIC & 202.5  & 62.6  & 2.9  & 3.0 & 10.9 & 2.8 & 0.0  \\
 \hline \hline
\end{tabular}\label{tab:best}
\end{table*}
where $k$ and $N$ are the number of free parameters and the number of data points, respectively. The above formula for $\Delta$AIC assumes $N\gg k$ (which is indeed the case here). With these information criteria the rule is now the following:  given a set of candidate models competing for the description of the same observational data, the preferred model is the one which has the minimum value of AIC and BIC. Hence when comparing  one candidate  DDE model versus the $\CC$CDM we can use the model differences $\Delta{\rm AIC}$ and $\Delta{\rm BIC}$. These are defined between the AIC and BIC values of the given  DDE model and the corresponding  $\CC$CDM values, taken as  reference.  The resulting  $\Delta{\rm AIC}$ is then used to determine the level of support for each model,  as indicated in Table \ref{daic} \citep{Burnham}. Small values  of $\Delta{\rm AIC}$ below $2$, and specially negative values, denote significant support to a given DDE versus the  $\CC$CDM.  As a complementary information criterion we use $\Delta{\rm BIC}$ to gauge the evidence \emph{against} a given DDE model as compared to the $\CC$CDM, see Table (\ref{dbic}).

The mentioned values for  $\Delta{\rm AIC}$ and $\Delta{\rm BIC}$ in Tables (\ref{daic} \& \ref{dbic}) are merely general rules of thumb.
Notice that for ${\rm BIC}$  we use the corresponding difference $\Delta{\rm BIC}$  to characterize the strength of the evidence, but in this case \text{\it against} the given DDE model, if the difference is positive. The  higher are the (positive) values of $\Delta{\rm BIC}$ the highest is the level of rejection of the given DDE model with respect to the $\CC$CDM, see Table \ref{dbic}  \citep{Kass:1995loi}.

In order to measure the strength of evidence in favor of each candidate model there is another interesting parameter dubbed Akaike weights ${\cal W}_m$ \citep{Burnham}. By normalizing the relative likelihood values one can obtain this parameter.  The relative likelihood for each of the candidate DDE models is obtained as follows:

\begin{eqnarray}\label{rellik}
{\cal L}_{rel}=\exp (-\dfrac{1}{2}\Delta{\rm AIC})\;.
\end{eqnarray}

Now we can compute the Akaike weight for the \textit{ith} candidate model among $n$ different models as follows:

\begin{eqnarray}\label{akaweit}
{\cal W}_{m i}=\frac{{\cal L}_{rel,i}}{\sum_{j=1}^n{\cal L}_{rel,j}}\;.
\end{eqnarray}

Hence one can say that the calculated Akaike weight for model \textit{ith} is the probability that this model is the best among the set of $n$ candidate models. Despite one can use full Bayesian Evidence to better judge the fit quality differences between the different models,  here  we use the AIC and BIC criteria as a first estimate. In our analysis we have used a MCMC algorithm which takes large-enough uninformative priors for each of the free parameters, and hence the AIC and BIC can be sufficient. Let us note that it is not our main aim to identify the small differences among the main models selected in this analysis, namely Models (3), (4) and (6)  --- see the discussion in the next sections --- but to emphasize that each one of them  has approximately  the same ability to partially alleviate the tensions under consideration.  The calculation of the full Bayesian evidence can be rather cumbersome, see e.g. \citep{Sola:2018sjf}  and references therein.  It would however not alter the main conclusion as to the fact that the mentined DDE models have the capacity to amelliorate the theoretical description as compared to the $\CC$CDM.  Depending on the evolution of the mentioned tensions in the future one may consider if further precision is necessary, but at the moment we are expectant to the possible impact of further and better observational data and in general on collecting more statistics on the different sources.

\section{NUMERICAL RESULTS}\label{sect:RS}

In this section we will report on the numerical results of our analysis for different DDE models, together with the concordance $\Lambda$CDM model, based on the statistical methods presented in Sec.\ref{subsect:method}.

\subsection{Expansion data}\label{subsect:bg}

Here we apply expansion data sets in order to put constraints on DE model parameters. In Table \ref{tab:best} we present the statistical results for comparing the fit quality of the different DDE models and the $\Lambda$CDM cosmology in the light of the expansion data..
In Table (\ref{tab:bestfit}) we reported the best fit values of free parameters for these models obtained using expansion data sets. In this table we also presented the error bars related to each of parameters. The $1\sigma$, $2\sigma$ and $3\sigma$ confidence levels of these parameters can be seen in the related contours in Fig.\. (\ref{fig:contour1} (for the $\CC$CDM) and  Fig.\,\ref{fig:contour2})  (for the various DDE models) respectively.  The name of the models can be seen in the legend.   Considering the first row panels of Fig.\, (\ref{fig:contour2}), we observe that the value of $\gamma=1.0$ (which represent the Model (1))  is not even within $3\sigma$ confidence level of the model parameter $\gamma$ for Model (2). This means that the the expansion data do not favor Model (1) (the GDE model) in comparison to Model (2). This result can be verified from the statistical results reported in the Table (\ref{tab:best}). Comparing the $\Delta$AIC values in this table and using the information of Table (\ref{daic}) we conclude that there is {\it essentially no support at all} for Model (1) and Model (2). In contrast,  there are {\it substantial evidences} to support the other DDE models, except Model (5) which lags behind models (3), (4) and (6), and in addition it suffers from another problem to be discussed later on connected with radiation.  Furthermore,  we can confirm from the Akaike weight values that there are negligible chances of $10^{-47}$ and $10^{-14}$ respectively for Model (1) and Model (2) to be the best model among all the considered DDE models in this work.   On the other hand, the probability for Model (6) is about $0.29$, which means that this model is the best model in our list. From this viewpoint, we can verify that Models (3) and (4) also have greater chance in comparison to the $\Lambda$CDM. In addition, from the point of view of $\Delta$BIC we can say that there is {\it very strong evidence against} Models (1), (2) and (5) while there is no compelling evidence against Models (3), (4) and (6). Summarizing these results, we conclude  that at background level:
\begin{itemize}
\item From the $\Delta$AIC analysis alone: Model (6) is the best model and Models (1) and  (2) are completely unacceptable.
\item From the point of view of the Akaike weight criteria: Model(6) is the best model and Models (1) and  (2) are once more completely unacceptable.
\item From the $\Delta$BIC analysis alone: $\Lambda$CDM remains the best model, having no significant objection against  Models (3),  (4) and  (6), whilst  Models (1),  (2) and  (5) are judged as completely unacceptable.
\end{itemize}

\begin{table*}
	\centering
	\caption{The best fit values of the free parameters for the different DDE models and the $\CC$CDM using combined data in Homogeneous DE scenario.
}
		\begin{tabular}{c c c c c c c c}
			\hline \hline
		Model  & Model(1) & Model(2) & Model(3) & Model(4) & Model(5) & Model(6) & $\Lambda$CDM\\
		\hline
		$\Omega_{\rm m}^{(0)}$ & $0.2843\pm 0.0072$ & $0.2793\pm 0.0068$ & $0.2857\pm 0.0067$ & $0.2845\pm 0.0080$ & $0.2853\pm 0.0053$  & $0.2878\pm 0.0069$ &$0.2953\pm 0.0107$ \\
		\hline
		$ h $ & $0.6599 \pm 0.0061$ & $0.7174\pm 0.0089$ & $0.7142\pm 0.0081$ & $0.7154\pm 0.0050$  & $0.7152\pm 0.0072$ & $0.7162\pm 0.0063$ & $0.7013\pm 0.0080$ \\
		\hline
		$\gamma $ & $1.0$ & $1.143\pm 0.0163$ & $-$ & $-$ & $-$ & $-$ & $-$ \\
		\hline
		$ \mu $ & $-$ & $-$ & $0.0$ & $0.0085^{+0.0065}_{-0.0067}$ & $0.955\pm 0.032$  & $0.00589\pm 0.0022$  & $-$ \\
	    \hline
		$ \nu $ & $-$ & $-$ & $-0.0071\pm 0.0097$ & $0.0$  & $0.956\pm 0.023$ & $-0.00589\pm 0.0022$  &  $-$ \\
        \hline
		$ \sigma_8 $ & $0.741\pm 0.024$ & $0.748^{+0.022}_{-0.024}$ & $0.766\pm 0.021$ & $0.766\pm 0.021$ & $0.766^{+0.023}_{-0.022}$ & $0.761\pm 0.021$ & $0.801\pm 0.019$\\
		\hline \hline
		\end{tabular}\label{tab:bestfit2h}
\end{table*}

\begin{table*}
	\centering
	\caption{The best fit values of the free parameters for the different DDE models and the $\CC$CDM using combined data in Clustered DE scenario.
}
		\begin{tabular}{c c c c c c c }
			\hline \hline
		Model  & Model(1) & Model(2) & Model(3) & Model(4) & Model(5) & Model(6) \\
		\hline
		$\Omega_{\rm m}^{(0)}$ & $0.284\pm 0.0075$ & $0.2812\pm 0.0060$ & $0.2849\pm 0.0068$ & $0.2836\pm 0.0062$ & $0.2858\pm 0.0051$  & $0.2867\pm 0.0069$ \\
		\hline
		$ h $ & $0.6616\pm 0.0065$ & $0.7164\pm 0.0070$ & $0.7163\pm 0.0079$ & $0.7149\pm 0.0072$  & $0.7155\pm 0.0072$ & $0.7134\pm 0.0070$ \\
		\hline
		$\gamma $ & $1.0$ & $1.1417\pm 0.016$ & $-$ & $-$ & $-$ & $-$ \\
		\hline
		$ \mu $ & $-$ & $-$ & $0.0$ & $0.0086^{+0.0076}_{-0.0078}$ & $0.958\pm 0.022$  & $0.00587\pm 0.0021$  \\
	    \hline
		$ \nu $ & $-$ & $-$ & $-0.0073\pm 0.0096$ & $0.0$  & $0.957\pm 0.0229$ & $-0.00587\pm 0.0021$   \\
        \hline
		$ \sigma_8 $ & $0.741^{+0.024}_{-0.023}$ & $0.751\pm 0.021$ & $0.767 \pm 0.022$ & $0.767^{+0.021}_{-0.023}$ & $0.766^{+0.023}_{-0.022}$ & $0.763^{+0.021}_{-0.022}$ \\
		\hline \hline
		\end{tabular}\label{tab:bestfit2c}
\end{table*}

Taking into account that the AIC criterion is in favor whereas the BIC one is not against, from the above analysis of the expansion data we can assert that Model (6) and $\Lambda$CDM are the best models, while Models (1) and (2) can not fit the expansion data at all.  On the other hande Models (3) and (4) are acceptable, but Model (5) is in trouble. In a nutshell, this is the verdict of Table (\ref{tab:best}).

Based on the best fit values of free parameters in Table (\ref{tab:bestfit}), we plot the evolution of the main cosmological quantities for the investigated models in Figs.(\ref{fig:back}, \ref{fig:w} \& \ref{fig:q2}). In the upper panel of Fig.(\ref{fig:back}) we show the redshift evolution of density parameter $\Omega_{\rm de}$. For all of the models,  $\Omega_{\rm de}$ falls down at  high redshift, where the role of dark matter is more prominent as compared to that of  DE. In the case of Model (1), $\Omega_{\rm de}$ reduces more slowly in comparison with the other models.  In the bottom panel one can see the evolution of the relative difference of the dimensionless Hubble parameter $E(z)=H(z)/H_0$ of the DDE models versus that of the standard $\CC$-model ($\CC$CDM), i.e. we plot  $\Delta E_{rel}\equiv [(E_{\rm model}-E_{\rm \Lambda})/E_{\rm \Lambda}]$. We observe that  the expansion rate for Models (1) and (2) deviates quite significantly from the standard one, which is in accordance with the anomalous character of these two models and their highly unfavored status within our analysis. In the case of Model(5) we observe smaller but still significant deviations, and in particular at low redshifts $\Delta E_{rel}$ for this becomes negative.  Models (3), (4) and (6), instead, show small departures with respect to the $\CC$CDM concerning their expansion rates.

In different panels of Fig.\ref{fig:w} we show the evolution of the EoS parameter of the various DDE models upon best fit parameters and their $1-\sigma$ error bars. While Model (1) evolves within quintessence regime ($-1<w_{\rm de}<-1/3$) at all redshifts, Model (2) crosses the CC divide  $w_{\rm de}=-1$ and enters  the phantom regime ($w_{\rm de} <-1$) at $z\sim 2.1$. However, the EoS values of these two models do not approach to $-1$ sufficiently at present (cf. Table \ref{tab:bestfit} ), and therefore they depart significantly from the $\CC$CDM behavior. This is again a reflect that the quality of their fits to the overall set of observations is substantially poorer as compared to the concordance model. The EoS evolution of Model(5), on the other hand, is quite different since it evolves into the phantom region from the quintessence region and approaches better the CC divide at present. As for the remaining Models, (3), (4) and (6), the most remarkable feature is that their EoS parameter is very close to $w_{\rm de}=-1$ near $z=0$ from below (i.e. $w_{\rm de}(0)\lesssim-1$ ). These models, as we pointed out before, have a well-defined continuous limit towards the $\CC$CDM for $\mu,\nu\to 0$ and they are actually the most favored DDE models in our fits. We can see that these models exhibit an ``effective phantom behavior'', which is however very small near the present time, i.e. $w_{\rm de}\lesssim-1$. This is perfectly compatible with the current Planck 2018 data, which yields $w_{\rm de}=-1.03\pm 0.03$\,\citep{Aghanim:2018eyx}.  The fact that the phenomenological  EoS range permits a non-negligible phantom window  is similar to previous Planck data (2015, 2013).  Specifically,  the following (approximate) current values of the EoS for Models (3), (4)  and (6) are obtained from their fitting data,  i.e. their values at $z=0$:  $w^{(3)}_{{\rm de} 0}=  -1.004 \pm 0.003$, $w^{(4)}_{{\rm de} 0}=   -1.0003  \pm 0.002$ and  $w^{(6)}_{{\rm de} 0}=  -1.002 \pm 0.002$, respectively, which are very close to the CC divide from the phantom region. That kind of behavior was indeed expected for these models, see Eq.(\ref{eq:EoSmod3}). For example, Model (3) has $\mu=0$ and $\nu<0$ with  $|\nu|\ll1$ (see e.g. Tables \ref{tab:bestfit2h} and {tab:bestfit2c} with combined data); therefore we expect $w^{(3)}_{{\rm de} 0}\lesssim-1$ near $z=0$. Similarly, for Model (4) we have $\nu=0$, with $0<\mu\ll1$, so again Eq.(\ref{eq:EoSmod3}) predicts  $w^{(4)}_{{\rm de} 0}\lesssim-1$ near $z=0$. Finally, for Model (6) the two parameters are small, $|\nu,\mu|\ll1$, and we have broken degeneracies by setting $\mu=-\nu$ (see Sect. II A),  and hence $\nu-\mu=2\nu<0$ from the tables, so once more Eq.(\ref{eq:EoSmod3}) entails $w^{(6)}_{{\rm de} 0}\lesssim-1$ near our time. In all cases we find a mild phantom behavior, which is of course merely effective since there are no fundamental phantom fields here.  As indicated, this behavior is fully compatible with the current data and therefore the latter can find a natural explanation in this kind of dynamical DE models, specifically models (3), (4) and (6).
See Fig.(\ref{fig:w}) for the corresponding plots of the EoS as a function of $z$, and where the behavior $w_{{\rm de} 0}\lesssim-1$ at $z=0$ can also be appraised. The only two models departing significantly from it are models (1) and (2). These models are also the ones providing the less favorable fit to the overall data.

The following  comment on Model (5) is now in order.  In fifth panel of Fig.(\ref{fig:w}) we can see that such model also approaches an EoS behavior close to the CC similarly to the previous three, although  it starts  first quintessence-like and subsequently moves to phantom-like regime in the last stretch near the present ($w^{(5)}_{{\rm de} 0}=   -1.075  \pm 0.003$). Notwithstanding, Model (5) has an insurmountable pitfall in the radiation-dominated epoch.  Owing to the fact that  $c_0=0$ for such model, this enforces $\mu\simeq\nu\simeq 1$, as can be seen in Table I using expansion data sets, or in Tables \ref{tab:bestfit2h} and \ref{tab:bestfit2c} when perturbations are included. As a result, in the radiation dominated epoch, where  $a\gg1$, the  relevant term of the Hubble function in Eq.\,(\ref{37}) behaves as $\Omega_{r,0} a^{-4}/(1-\nu+4\alpha/3)\simeq 3\Omega_{r,0} a^{-4}/4$ rather than just $\Omega_{r,0} a^{-4}$. Such departure from the $\CC$CDM in the radiation epoch is not acceptable as it implies an effective ``renormalization'' of the standard $\Omega_{r,0}$ parameter by roughly $-25\%$. Therefore, this model cannot be considered viable for the description of the standard cosmic history.  Having rejected also Models (1) and (2), we are left with Models (3), (4) and (6) as the only ones that can  successfully pass all tests. Not only so, they are even capable to significantly improve the description of the data as compared to the $\CC$CDM.  Let us note in particular that the fitting values of $\nu$ and $\mu$ for Model (6) are non-vanishing at roughly $2.8\sigma$ (cf. Tables \ref{tab:bestfit2h} and \ref{tab:bestfit2c}).  The successfulness of these models is further corroborated by the information criteria, as shown in Table \ref{tab:best2}.

As a complementary information, we calculate the deceleration parameter $q=-1-\dot{H}/H^2$.  Recall that $q=0$ indicates the position of the  transition point from early decelerated expansion to current accelerated expansion in the Universe.  Recalling  Eq.(\ref{gde10}) we find
		\begin{eqnarray}\label{eq:q2}
		q=\frac{1}{2}+\frac{3}{2} w_{\rm de}\Omega_{\rm de}+\frac{\Omega_{\rm r}}{2}\,.
		\end{eqnarray}
In the upper panel of Fig.(\ref{fig:q2}) we plot the deceleration parameter as a function of the redshift,  $q(z)$,  using the best fit values of the parameters in Table (\ref{tab:bestfit}). The transition redshift, $z_{\rm tr}$, i.e. the value for which we have $q(z_{\rm tr})=0$,  for each model reads: $z_{\rm tr}=0.55$ for Model (1), $z_{\rm tr}=0.67 $ for Model (2), $z_{\rm tr}=0.71 $ for Model (3), $z_{\rm tr}=0.72 $ for Model (4), $z_{\rm tr}=0.69 $ for Model (5),  $z_{\rm tr}=0.72 $ for Model(6) and $z_{\rm tr}=0.72$ for the $\Lambda$CDM model.  The first two models deviate as always very significantly, whereas the other remain close to the $\CC$CDM. These results are  in agreement with those reported in \citep{Farooq:2016zwm}, which were obtained from observations.  In the middle panel of Fig.(\ref{fig:q2}) we superimpose the  predicted value of the Hubble parameter at different redshifts for the various DE models and the observational data points. Finally, in the bottom panel we have plotted the evolution of theoretical distance modulus  $\mu(z)=5\log \left[(1+z)\int_0^z \dfrac{dx}{E(x)} \right]+\mu_0$,  where $\mu_0=42.384-5\log h$ for different DE models and its observational values from SnIa sample.

\subsection{Combined data}\label{subsect:TOT}
The verdict on the various DE models expressed in Table (\ref{tab:best}) is based exclusively  on the expansion data. In this section we combine the expansion data and the growth data to reconsider our constraints on cosmological parameters from a more complete standpoint. This enforces us to examine the performance of the DE models from both the expansion and perturbations perspectives. The extended expression that describes the $\chi^2$-function for combined data was introduced in Eq.(\ref{eq:like-tot_chi2}).  Upon minimizing  such  function  in the context of the MCMC analysis we display the fitting results for the different DE models under scrutiny in Tables \ref{tab:bestfit2h} and \ref{tab:bestfit2c}. In them we show our results for both homogeneous and clustered DDE scenarios, respectively. Furthermore,  in Table  (\ref{tab:best2})  we report on the statistical return (including the AIC and BIC information criteria) of the various DDE models as compared to the $\CC$CDM in the respective homogeneous and clustered DDE realizations. As we can see, the   $\Delta$AIC values of the extended analysis show once more that there is essentially no support at all  for Models (1) and (2). Furthermore, we find that Models (3),(4) and (6)  attain now  an even more comfortable advantage position versus the concordance  $\Lambda$CDM model (cf.  Tables (\ref{tab:best}) and  (\ref{tab:best2})). In the case of Akaike weight criteria, we can assert once more that Models (1) and (2) can be outright  rejected.  From the AIC perspective alone, Model (5) is equally good as the concordance $\Lambda$CDM model.  In contradistinction, Models (3), (4) and(6) have superior chances ranging between three to four times greater than the $\Lambda$CDM. On the other hand,  using $\Delta$BIC we find consistent conclusions.
To summarize, in the case of combined data sets, we can conclude as follows:
\begin{itemize}
\item From the point of view of $\Delta$AIC: Model(6) is the best, Model (5) and $\CC$CDM are comparable, whilst Models (1) \& (2) are highly unfavorable.
\item  Using the Akaike weights: Model(6) is the best model.  Models  (1) \& (2) are completely unfavorable.
\item  Using $\Delta$BIC results in isolation,  $\Lambda$CDM is the best model, while we find  very strong evidence against Models (1), (2) and (5), and  only mild objection against Models (3), (4) and (6), which is minimum for Model (6).
\end{itemize}
Overall, the joint verdict of the various criteria tends to  favor the DDE models (3), (4) and (6), specially the latter,  over the $\CC$CDM. These results are in good agreement with those obtained early with the expansion data sets.

\begin{figure}
	\centering
	\includegraphics[width=8cm]{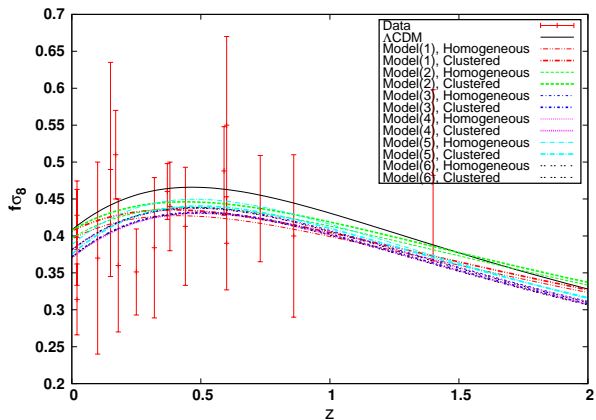}
	\caption{The predicted redshift evolution of $f (z)\sigma_8 (z)$ for different DDE models and the $\CC$CDM compared to the observed values of $f (z)\sigma_8 (z)$.}
	\label{fig:fsigma8}
\end{figure}

Finally, in Fig. (\ref{fig:fsigma8}), we plot the theoretical prediction of the growth rate weighted function $f(z)\sigma_8(z)$ for the different DE scenarios studied in this work, including the $\CC$CDM. The theoretical evolution of  $f(z)\sigma_8(z)$ is plotted using the best fit parameters of Tables (\ref{tab:bestfit2h} \& \ref{tab:bestfit2c}). We observe that in all of the DDE models,  whether with clustered or homogeneous DE, the results are very similar.
Therefore, we find that it is not possible to clearly distinguish between homogeneous and clustered DE scenarios.   The numerical results for these two options are very close to each other and for the time being we cannot project a preferred  scenario concerning the DE perturbations.  This is confirmed from comparing the statistical results and information criteria of Tables (\ref{tab:best}) and (\ref{tab:best2}).

We summarize the numerical results of this section as follows:

\begin{itemize}
\item  Using AIC alone,  there is no support at all for Models (1) \&  (2) neither at background nor at  perturbations levels.

\item Using Akaike weights, we find once more that  at background Models (1) \&  (2)  are completely unsuited, while Model (5) and $\Lambda$CDM have similar same chance to be the best models (viz.  $\sim7 \%$ and $\sim9 \%$, respectively). In contrast, Model(3), (4) \& (6) have the greatest chance to be the best models. At perturbations level and using combined data we find that the results are essentially the same as for the background level.

\item On the basis of BIC alone, there is "very strong" evidence against model(1), (2) \& (5) whether using expansion data or combining it with perturbations.

\item No significant difference can be presently appraised  between homogeneous and clustered DE scenarios.

\item Although adding growth data to the expansion data does not produce very significant changes in the numerical results,  Models (3),(4) and (6) improve their advantage position versus the concordance  $\Lambda$CDM model.  The small improvement of Model (5), however, cannot compensate for the troublesome behavior of this model in the radiation epoch.

\end{itemize}

\begin{table*}
	\centering
	\caption{Comparison of the statistical results for the different DDE models and the $\CC$CDM using combined data in Homogeneous (Clustered) DE scenario.}
	\begin{tabular}{c  c c c c c c c }
		\hline \hline
		Model  & Model(1) & Model(2) & Model(3) & Model(4) & Model(5) & Model(6) & $\Lambda$CDM\\
		\hline
		$\chi^2_{\rm min}$ & 1283.6(1283.4) & 1136.8(1136.5) & 1077.0(1077.2) & 1077.1(1077.1) & 1078.0(1078.2) & 1076.9(1076.8) & 1081.1\\
		\hline
		AIC & 1289.6(1289.4) & 1144.8(1144.5) & 1085.0(1085.2) & 1085.1(1085.1) & 1088.0(1088.2) & 1084.9(1084.8) & 1087.1\\
		\hline
		$\Delta$AIC  & 202.5(202.3) & 57.7(57.4) & -2.1(-1.9) & -2.0(-2.0) & 0.9(1.1) & -2.2(-2.3) & 0.0  \\
		\hline
		${\cal W}_{m}$ & 1.1 E-45(1.2 E-45) & 2.9 E-14(3.4 E-14) & 0.280(0.258) & 0.266(0.271) & 0.062(0.057) & 0.294(0.315) & 0.099\\
		\hline
		BIC & 1304.6(1304.4) & 1164.8(1164.5) & 1105.0(1105.2) & 1105.1(1105.1) & 1113.0(1113.2) & 1104.9(1104.8) & 1102.1\\
	    \hline
		$\Delta$BIC & 202.5(202.3) & 62.7(62.4) & 2.9(3.1) & 3.0(3.0) & 10.9(11.1) & 2.8(2.7) & 0.0\\
		\hline \hline
	\end{tabular}\label{tab:best2}
\end{table*}

The following comment is in order. We do understand that by adding the local $H_0$ value to our analysis, the results should tend to better agree with that prior. However, it may be useful so as to test explicitly the influence that this has on  our results on $\sigma_8$ and the $H_0$ tension. Therefore,  we next proceed to remove the local value of the Hubble parameter $H_0$  given in\,\citep{Riess:2018uxu} from our data sets. The new results are displayed in Table (\ref{tabr18}).  For  this part, we just report on the main cosmological parameters $\Omega_m, h$ and $\sigma_8$.  On comparing these results with those obtained in the presence of the local $H_0$ input (Tables (V) \& (VI)), we can see that the changes in the best fit parameters  are not significant and remain compatible within error bars. In addition, for  the various models under study, the value of $\sigma_8$ still remains low,  which is a welcome result and indicates that this is the natural tendency of this kind of models.   We can see that the trend of  our results is similar to the one obtained in the similar test carried out in \citep{Sola:2017znb}, but in contrast to the current results for $H_0$ the ones obtained in the last work (in which the vacuum is in interaction with matter)  tend to remain more in accordance with the Planck $H_0$ value obtained from the CMB.  The upshot of the current test is that the addition or not of the local $H_0$ input is not essential to obtain a best fit value for $H_0$ substantially higher than  the current Planck 2018 value. With or without the inclusion of the local $H_0$ input in our analysis  we confirm that the main DDE models under study can be in better concordance with \citep{Riess:2018uxu}. Thus, for these models we find as a rewarding novelty that $H_0$ tends to remain higher alongside  with low values of  $\sigma_8$, which is tantamount to  saying that we can better alleviate the two tensions at the same time  See the next section for further discussion. 

\begin{table*}
	\centering
	\caption{The results of analysis using combined data after removing the local $H_0$ value from data sets.}
	\begin{tabular}{lc c c | c c c |} \hline \hline
 & \multicolumn{3}{|c}{Homogeneous DE}&\multicolumn{3}{|c}{Clustered DE} \\
 Model & \{$\Omega_{\rm m}^{(0)}$ & $h$ & $\sigma_8$\} & \{$\Omega_{\rm m}^{(0)}$ & $h$ & $\sigma_8$\}\\
		\hline
Model(1) &$0.2921\pm 0.0073$ & $0.652 \pm 0.0063$ & $0.743\pm 0.022$ & $0.2916\pm 0.0076$ &$0.6698\pm 0.0069$ &$0.742 \pm 0.027$ \\
		\hline
Model(2) &$0.2865\pm 0.0060$ & $0.7112 \pm 0.0077$ & $0.749\pm 0.024$ & $0.2886\pm 0.0069$ &$0.7089\pm 0.0073$ &$0.751 \pm 0.023$ \\
		\hline
Model(3) &$0.2932\pm 0.0063$ & $0.7067 \pm 0.0072$ & $0.767\pm 0.021$ & $0.2933\pm 0.0074$ &$0.7089\pm 0.0074$ &$0.767 \pm 0.024$ \\
\hline
Model(4) &$0.2922\pm 0.0077$ & $0.7072 \pm 0.0059$ & $0.765\pm 0.022$ & $0.2911\pm 0.0066$ &$0.7068\pm 0.0069$ &$0.766 \pm 0.024$ \\
		\hline
Model(5) &$0.2936\pm 0.0061$ & $0.7077 \pm 0.0071$ & $0.767\pm 0.022$ & $0.2936\pm 0.0056$ &$0.7088\pm 0.0074$ &$0.766 \pm 0.023$ \\
		\hline
Model(6) &$0.2951\pm 0.0068$ & $0.7082 \pm 0.0064$ & $0.762\pm 0.022$ & $0.2946\pm 0.0070$ &$0.7058\pm 0.0068$ &$0.743 \pm 0.021$ \\
		\hline
$\CC$CDM &$0.3024\pm 0.0094$ & $0.6945 \pm 0.0078$ & $0.802\pm 0.020$ & $--$ &$--$ &$--$ \\
\hline \hline
	\end{tabular}\label{tabr18}
\end{table*}

\section{Discussion and conclusions}\label{sect:conclusion}

In this paper, we have studied the behavior of the cosmic fluid in the presence of dynamical dark energy (DDE) whose density appears in the form of powers of the Hubble rate, $H$, and/or  its cosmic time derivative, $\dot{H}$.  These are the models indicated in Eqs.\,(\ref{gde1})-\ref{gde6}).  The first two models  had previously  been considered from the point of view of ghost dark energy and the others as running vacuum models, except Model (5)  which  had been dealt with as entropic-force model.  Using this ample spectrum of models we have studied the corresponding cosmological equations at  both background and perturbations levels.   We have taken into account not only the matter perturbations, but also the perturbations on the DE component in each case. This feature had not been considered in previous studies. Thanks to it we have been able to compare the homogeneous DDE versus the clustered DDE scenarios.  Initially, using the latest observational data at background level in the context of the Markov Chain Monte Carlo (MCMC)  method we have performed a likelihood analysis. Our results clearly indicate  that Models (1) and (2)  can not fit the expansion data;  and as a matter of fact we find that there is  strong compelling evidence against them on the basis of expansion and growth data. At the same time we find that Model (5) is in trouble for the correct description of the radiation epoch. The common source of problems for Models (1), (2) and (5) is the lack of an additive constant term in the structure of their DE density, what prevents them from having a smooth $\CC$CDM limit.  In contradistinction to them, Models (3), (4)  \& (6) do have such an additive term and well defined limit, and these models do pass successfully the test with the expansion data.
Subsequently, using combined (expansion+growth rate) data and the theoretical equations at perturbations level we have extended our analysis by considering both homogeneous DE and clustered DE. These, more complete, results confirmed  that Models (3), (4) \& (6) keep on providing better fitting results in comparison to the $\Lambda$CDM.  The differences between the homogeneous and clustered DE scenarios, though,  are not acute enough at present as to be able to distinguish them with clarity.  Notwithstanding this fact, the presence of the growth data actually enhances the fitting quality and hence the privileged position of the main Models (3), (4) \& (6) as compared to the concordance model. We remain hopeful to obtain improved future results concerning the  clustering properties of DE in the light of new observational data which will be obtained from the next generation of high precision surveys, such as those based on Euclid \citep[see][]{Taddei:2014wqa}.

The analysis presented here has extended that of \citep{Gomez-Valent:2015pia} by taking into account the DE perturbations as well as a more complete and updated set of observations. Not only so, for the first time we find that these DDE models with self-conserved DE and no interaction with matter do offer a possible solution to  well-known existing tensions of the $\CC$CDM with the data, as we further comment below. This in contradistinction to the situation when the same type of models are treated as dynamical vacuum models in interaction with matter, in which the $\sigma_8$ tension can be improved but the $H_0$ one is not amelliorated\,\cite{Sola:2017jbl,Sola:2017znb}.
Before extending this important remark, let us also note that the main DDE models under study, namely Models (3), (4)  \& (6),  exhibit an `effective phantom behavior' very close to a cosmological constant term near our time, i.e. with an equation of state parameter which satisfies $w_{\rm de}\lesssim-1$. Such behavior is not associated to any fundamental scalar field and is entirely caused by the dependence of the DE density on $H^2$ and/or $\dot{H}$.  It could explain why the current Planck data  is perfectly compatible with a narrow phantom window very close (from below)  to the cosmological constant divide  $w_{\rm de}=-1$.  Notice that such effective phantom behavior  would not show up when we allow interaction with matter, as in such case the effective behavior is quintessence-like\,\cite{Sola:2017jbl}.

Finally, we should not gloss over the important fact that the main DDE models under study could provide a significant alleviation of two of the most intriguing tensions of the $\CC$CDM with the observational data throughout the last few years. Let us further extend our preliminary comments on this issue.  To start with, these DDE models  lead to a global decrease of the rms mass fluctuation parameter $\sigma_8$  (associated to spheres of $8 h^{-1}$Mpc at $z=0$)   by about $2.2\,\sigma$ as compared to  the prediction from the  `Base-$\CC$CDM'  from Planck 2018 ($\sigma_8=0.811\pm 0.006$ )\,\citep{Aghanim:2018eyx}.  If we compare e.g. the corresponding $\sigma_8$ result from our most favored Model (6) with the  value that we find for our  fitted $\CC$CDM model, the discrepancy is of course smaller, of about  $1.4\sigma$ (cf. Tables \ref{tab:bestfit2h} and \ref{tab:bestfit2c}) since the Base-$\CC$CDM uses  TT,TE,EE+lowE+lensing data only.  Taking into account that the $\sigma_8$-tension of the LSS data with such Base-$\CC$CDM cosmology is of $\sim 3.3\sigma$,  the best DDE under study  brings the remaining $\sigma_8$ tension with the data down to $\sim 1\sigma$ level and therefore renders it essentially irrelevant. In addition, we can see in the same tables that the corresponding values of  $H_0$ remain relatively higher as compared to the $\CC$CDM, what also contributes to mitigate the tension between the Planck value and the local value of $H_0$ by Riess et al. \cite{Riess:2018uxu}.  In point of fact, a recent analysis by this team\,\cite{Riess:2019cxk} has further aggravated the tension between the local measurement of $H_0$ and the one based on Planck data. They find $H_0 = 74.03 \pm 1.42 $ km s$^{-1}$ Mpc$^{-1}$, which raises the discrepancy with respect to the current Planck 2018 value $H_0 = 67.4 \pm 0.5 $ km s$^{-1}$ Mpc$^{-1}$  up to $4.4\sigma$.   We have not used the local value of $H_0$ from the last paper,  which is very recent. However, if we compare it with the value that we find from our best DDE models (3), (4)  \& (6) in Tables \ref{tab:bestfit2h} and \ref{tab:bestfit2c}, we see that in the three cases the tension is lowered to only  $1.5-1.7\sigma$  (after taking into account the respective errors in quadrature). The small differences depend on the model and the DE scenario, homogeneous or clustered, the former being slightly more favored, but the basic result is that there is no significant $H_0$-tension as compared to the situation of the  $\CC$CDM with a strict cosmological constant. As previously noted, the inclusion of the local $H_0$ input in the analysis naturally drags the resulting best fit value toward it.  Thus,  in order to better assess the ability of these DDE models to deal with the tensions, it is natural to remove the $H_0$  input from our analysis and observe what is the outcome. The results are recorded in Table VIII  and they can be considered the genuinely unbiased results. Remarkably enough, we find that the new best-fit values of $H_0$  still remain relatively large.  Indeed, for the main DDE models under consideration it shows a moderate discrepancy in between $1.9-2.2\sigma$ at most (which varies slightly from the clustered to the non-clustered DE scenario)  with respect to the local measurement from \,\cite{Riess:2019cxk}, in stark contrast with the appaling  $4.4\sigma$ tension generated with  the Base-$\CC$CDM value from Planck 2018.  Taking into account that the local value cannot be completely ignored from any analysis since it is a real measurement in the context of current observations, we can assert that the level of $H_0$-tension found in the main DDE models considered here remains roughly at a tolerable level of  $2\sigma$ at most, and this is certainly a substantial improvement as compared to many other options in the literature. This fact is all the more noticeable if we take into account that the $\sigma_8$ value remains sufficiently small in all the cases (cf. Table VIII), and therefore a possible simultaneous solution or alleviation of the two main tensions of the $\CC$CDM is on the horizon. The upshot  is that Models (3), (4)  \& (6)  provide a better global fit to the data as compared to the $\CC$CDM and are able to significantly alleviate both of the two intriguing  tensions, $\sigma_8$ and $H_0$, currently afflicting the concordance model. We conclude that if the dark energy is dynamical and can be expressed as a power series of the Hubble rate, including also a nonvanishing additive constant (which insures a smooth limit with the $\CC$CDM), it is possible to achieve a better compatibility of the global cosmological data with the theoretical predictions than with just a rigid $\CC$-term. Maybe the dynamical DDE models that we have examined here could be an illustrative approach (in the absence of  the perfect scenario) of the kind of sought-for  refinement of the standard $\CC$CDM invoked by Riess et al \,\cite{Riess:2019cxk}. In the words of these authors:  `While it is difficult and perhaps debatable
to identify the precise threshold at which a tension passes the point of being attributable to a fluke, the one presently involving $H_0$  appears to have passed that point';  and they continue with the following suggestive words:  ``With multiple, independent corroborations now demonstrated at both ends of cosmic history, we may need to seek resolution in a refinement of the model that joins them, (Vanilla) $\CC$CDM'.

\vspace{0.5cm}

\section{Acknowledgements}
The work of MR has been supported financially by Research Institute for Astronomy \& Astrophysics of Maragha (RIAAM) under research project No. 1/6025-32. The work of JSP has
been partially supported by projects  FPA2016-76005-C2-1-P (MINECO), 2017-SGR-929 (Generalitat de Catalunya) and MDM-2014-0369 (ICCUB).  We are grateful to A. G\'omez-Valent for useful discussions. We would like to thank the anonymous Referee for the detailed revision of our work and for providing a number of useful suggestions which have helped to improve the presentation of our results.


 \bibliographystyle{apsrev4-1}
  \bibliography{ref}

\begin{thebibliography}{124}%
\makeatletter
\providecommand \@ifxundefined [1]{%
 \@ifx{#1\undefined}
}%
\providecommand \@ifnum [1]{%
 \ifnum #1\expandafter \@firstoftwo
 \else \expandafter \@secondoftwo
 \fi
}%
\providecommand \@ifx [1]{%
 \ifx #1\expandafter \@firstoftwo
 \else \expandafter \@secondoftwo
 \fi
}%
\providecommand \natexlab [1]{#1}%
\providecommand \enquote  [1]{``#1''}%
\providecommand \bibnamefont  [1]{#1}%
\providecommand \bibfnamefont [1]{#1}%
\providecommand \citenamefont [1]{#1}%
\providecommand \href@noop [0]{\@secondoftwo}%
\providecommand \href [0]{\begingroup \@sanitize@url \@href}%
\providecommand \@href[1]{\@@startlink{#1}\@@href}%
\providecommand \@@href[1]{\endgroup#1\@@endlink}%
\providecommand \@sanitize@url [0]{\catcode `\\12\catcode `\$12\catcode
  `\&12\catcode `\#12\catcode `\^12\catcode `\_12\catcode `\%12\relax}%
\providecommand \@@startlink[1]{}%
\providecommand \@@endlink[0]{}%
\providecommand \url  [0]{\begingroup\@sanitize@url \@url }%
\providecommand \@url [1]{\endgroup\@href {#1}{\urlprefix }}%
\providecommand \urlprefix  [0]{URL }%
\providecommand \Eprint [0]{\href }%
\providecommand \doibase [0]{http://dx.doi.org/}%
\providecommand \selectlanguage [0]{\@gobble}%
\providecommand \bibinfo  [0]{\@secondoftwo}%
\providecommand \bibfield  [0]{\@secondoftwo}%
\providecommand \translation [1]{[#1]}%
\providecommand \BibitemOpen [0]{}%
\providecommand \bibitemStop [0]{}%
\providecommand \bibitemNoStop [0]{.\EOS\space}%
\providecommand \EOS [0]{\spacefactor3000\relax}%
\providecommand \BibitemShut  [1]{\csname bibitem#1\endcsname}%
\let\auto@bib@innerbib\@empty
\bibitem [{\citenamefont {Riess}\ \emph {et~al.}(1998)\citenamefont {Riess},
  \citenamefont {Filippenko}, \citenamefont {Challis},\ and\ \citenamefont
  {et~al.}}]{Riess1998}%
  \BibitemOpen
  \bibfield  {author} {\bibinfo {author} {\bibfnamefont {A.~G.}\ \bibnamefont
  {Riess}}, \bibinfo {author} {\bibfnamefont {A.~V.}\ \bibnamefont
  {Filippenko}}, \bibinfo {author} {\bibfnamefont {P.}~\bibnamefont {Challis}},
  \ and\ \bibinfo {author} {\bibnamefont {et~al.}},\ }\href@noop {} {\bibfield
  {journal} {\bibinfo  {journal} {\aj}\ }\textbf {\bibinfo {volume} {116}},\
  \bibinfo {pages} {1009} (\bibinfo {year} {1998})}\BibitemShut {NoStop}%
\bibitem [{\citenamefont {Perlmutter}\ \emph {et~al.}(1999)\citenamefont
  {Perlmutter}, \citenamefont {Aldering}, \citenamefont {Goldhaber},\ and\
  \citenamefont {et~al.}}]{Perlmutter1999}%
  \BibitemOpen
  \bibfield  {author} {\bibinfo {author} {\bibfnamefont {S.}~\bibnamefont
  {Perlmutter}}, \bibinfo {author} {\bibfnamefont {G.}~\bibnamefont
  {Aldering}}, \bibinfo {author} {\bibfnamefont {G.}~\bibnamefont {Goldhaber}},
  \ and\ \bibinfo {author} {\bibnamefont {et~al.}},\ }\href@noop {} {\bibfield
  {journal} {\bibinfo  {journal} {\apj}\ }\textbf {\bibinfo {volume} {517}},\
  \bibinfo {pages} {565} (\bibinfo {year} {1999})}\BibitemShut {NoStop}%
\bibitem [{\citenamefont {Komatsu}\ \emph {et~al.}(2009)\citenamefont
  {Komatsu}, \citenamefont {Dunkley}, \citenamefont {Nolta},\ and\
  \citenamefont {et~al.}}]{Komatsu2009}%
  \BibitemOpen
  \bibfield  {author} {\bibinfo {author} {\bibfnamefont {E.}~\bibnamefont
  {Komatsu}}, \bibinfo {author} {\bibfnamefont {J.}~\bibnamefont {Dunkley}},
  \bibinfo {author} {\bibfnamefont {M.~R.}\ \bibnamefont {Nolta}}, \ and\
  \bibinfo {author} {\bibnamefont {et~al.}},\ }\href@noop {} {\bibfield
  {journal} {\bibinfo  {journal} {ApJS}\ }\textbf {\bibinfo {volume} {180}},\
  \bibinfo {pages} {330} (\bibinfo {year} {2009})}\BibitemShut {NoStop}%
\bibitem [{\citenamefont {Jarosik}\ \emph {et~al.}(2011)\citenamefont
  {Jarosik}, \citenamefont {Bennett}, \citenamefont {Dunkley}, \citenamefont
  {Gold}, \citenamefont {Greason}, \citenamefont {Halpern}, \citenamefont
  {Hill}, \citenamefont {Hinshaw}, \citenamefont {Kogut}, \citenamefont
  {Komatsu},\ and\ \citenamefont {et~al.}}]{Jarosik:2010iu}%
  \BibitemOpen
  \bibfield  {author} {\bibinfo {author} {\bibfnamefont {N.}~\bibnamefont
  {Jarosik}}, \bibinfo {author} {\bibfnamefont {C.~L.}\ \bibnamefont
  {Bennett}}, \bibinfo {author} {\bibfnamefont {J.}~\bibnamefont {Dunkley}},
  \bibinfo {author} {\bibfnamefont {B.}~\bibnamefont {Gold}}, \bibinfo {author}
  {\bibfnamefont {M.~R.}\ \bibnamefont {Greason}}, \bibinfo {author}
  {\bibfnamefont {M.}~\bibnamefont {Halpern}}, \bibinfo {author} {\bibfnamefont
  {R.~S.}\ \bibnamefont {Hill}}, \bibinfo {author} {\bibfnamefont
  {G.}~\bibnamefont {Hinshaw}}, \bibinfo {author} {\bibfnamefont
  {A.}~\bibnamefont {Kogut}}, \bibinfo {author} {\bibfnamefont
  {E.}~\bibnamefont {Komatsu}}, \ and\ \bibinfo {author} {\bibnamefont
  {et~al.}},\ }\href@noop {} {\bibfield  {journal} {\bibinfo  {journal}
  {\apjs}\ }\textbf {\bibinfo {volume} {192}},\ \bibinfo {pages} {14} (\bibinfo
  {year} {2011})}\BibitemShut {NoStop}%
\bibitem [{\citenamefont {Komatsu}\ \emph {et~al.}(2011)\citenamefont
  {Komatsu}, \citenamefont {Smith}, \citenamefont {Dunkley},\ and\
  \citenamefont {et~al.}}]{Komatsu2011}%
  \BibitemOpen
  \bibfield  {author} {\bibinfo {author} {\bibfnamefont {E.}~\bibnamefont
  {Komatsu}}, \bibinfo {author} {\bibfnamefont {K.~M.}\ \bibnamefont {Smith}},
  \bibinfo {author} {\bibfnamefont {J.}~\bibnamefont {Dunkley}}, \ and\
  \bibinfo {author} {\bibnamefont {et~al.}},\ }\href@noop {} {\bibfield
  {journal} {\bibinfo  {journal} {\apjs}\ }\textbf {\bibinfo {volume} {192}},\
  \bibinfo {pages} {18} (\bibinfo {year} {2011})}\BibitemShut {NoStop}%
\bibitem [{\citenamefont {{Planck Collaboration XIV}}(2016)}]{Ade:2015yua}%
  \BibitemOpen
  \bibfield  {author} {\bibinfo {author} {\bibnamefont {{Planck Collaboration
  XIV}}} (\bibinfo {collaboration} {Planck Collaboration}),\ }\href@noop {}
  {\bibfield  {journal} {\bibinfo  {journal} {Astron.Astrophys.}\ }\textbf
  {\bibinfo {volume} {594}},\ \bibinfo {pages} {A14} (\bibinfo {year}
  {2016})}\BibitemShut {NoStop}%
\bibitem [{\citenamefont {Tegmark}\ \emph {et~al.}(2004)\citenamefont {Tegmark}
  \emph {et~al.}}]{Tegmark:2003ud}%
  \BibitemOpen
  \bibfield  {author} {\bibinfo {author} {\bibfnamefont {M.}~\bibnamefont
  {Tegmark}} \emph {et~al.} (\bibinfo {collaboration} {SDSS Collaboration}),\
  }\href {\doibase 10.1103/PhysRevD.69.103501} {\bibfield  {journal} {\bibinfo
  {journal} {Phys. Rev. D}\ }\textbf {\bibinfo {volume} {69}},\ \bibinfo
  {pages} {103501} (\bibinfo {year} {2004})}\BibitemShut {NoStop}%
\bibitem [{\citenamefont {Cole}\ \emph {et~al.}(2005)\citenamefont {Cole} \emph
  {et~al.}}]{Cole:2005sx}%
  \BibitemOpen
  \bibfield  {author} {\bibinfo {author} {\bibfnamefont {S.}~\bibnamefont
  {Cole}} \emph {et~al.} (\bibinfo {collaboration} {2dFGRS Collaboration}),\
  }\href {\doibase 10.1111/j.1365-2966.2005.09318.x} {\bibfield  {journal}
  {\bibinfo  {journal} {MNRAS}\ }\textbf {\bibinfo {volume} {362}},\ \bibinfo
  {pages} {505} (\bibinfo {year} {2005})}\BibitemShut {NoStop}%
\bibitem [{\citenamefont {Eisenstein}\ \emph {et~al.}(2005)\citenamefont
  {Eisenstein} \emph {et~al.}}]{Eisenstein:2005su}%
  \BibitemOpen
  \bibfield  {author} {\bibinfo {author} {\bibfnamefont {D.~J.}\ \bibnamefont
  {Eisenstein}} \emph {et~al.} (\bibinfo {collaboration} {SDSS
  Collaboration}),\ }\href {\doibase 10.1086/466512} {\bibfield  {journal}
  {\bibinfo  {journal} {ApJ}\ }\textbf {\bibinfo {volume} {633}},\ \bibinfo
  {pages} {560} (\bibinfo {year} {2005})}\BibitemShut {NoStop}%
\bibitem [{\citenamefont {Percival}\ \emph {et~al.}(2010)\citenamefont
  {Percival}, \citenamefont {Reid}, \citenamefont {Eisenstein},\ and\
  \citenamefont {et~al.}}]{Percival2010}%
  \BibitemOpen
  \bibfield  {author} {\bibinfo {author} {\bibfnamefont {W.~J.}\ \bibnamefont
  {Percival}}, \bibinfo {author} {\bibfnamefont {B.~A.}\ \bibnamefont {Reid}},
  \bibinfo {author} {\bibfnamefont {D.~J.}\ \bibnamefont {Eisenstein}}, \ and\
  \bibinfo {author} {\bibnamefont {et~al.}},\ }\href@noop {} {\bibfield
  {journal} {\bibinfo  {journal} {\mnras}\ }\textbf {\bibinfo {volume} {401}},\
  \bibinfo {pages} {2148} (\bibinfo {year} {2010})}\BibitemShut {NoStop}%
\bibitem [{\citenamefont {Blake}\ \emph {et~al.}(2011)\citenamefont {Blake}
  \emph {et~al.}}]{Blake:2011rj}%
  \BibitemOpen
  \bibfield  {author} {\bibinfo {author} {\bibfnamefont {C.}~\bibnamefont
  {Blake}} \emph {et~al.},\ }\href {\doibase 10.1111/j.1365-2966.2011.18903.x}
  {\bibfield  {journal} {\bibinfo  {journal} {Mon. Not. Roy. Astron. Soc.}\
  }\textbf {\bibinfo {volume} {415}},\ \bibinfo {pages} {2876} (\bibinfo {year}
  {2011})},\ \Eprint {http://arxiv.org/abs/1104.2948} {arXiv:1104.2948
  [astro-ph.CO]} \BibitemShut {NoStop}%
\bibitem [{\citenamefont {Reid}\ \emph {et~al.}(2012)\citenamefont {Reid},
  \citenamefont {Samushia}, \citenamefont {White}, \citenamefont {Percival},
  \citenamefont {Manera} \emph {et~al.}}]{Reid:2012sw}%
  \BibitemOpen
  \bibfield  {author} {\bibinfo {author} {\bibfnamefont {B.~A.}\ \bibnamefont
  {Reid}}, \bibinfo {author} {\bibfnamefont {L.}~\bibnamefont {Samushia}},
  \bibinfo {author} {\bibfnamefont {M.}~\bibnamefont {White}}, \bibinfo
  {author} {\bibfnamefont {W.~J.}\ \bibnamefont {Percival}}, \bibinfo {author}
  {\bibfnamefont {M.}~\bibnamefont {Manera}},  \emph {et~al.},\ }\href
  {\doibase 10.1111/j.1365-2966.2012.21779.x} {\bibfield  {journal} {\bibinfo
  {journal} {MNRAS}\ }\textbf {\bibinfo {volume} {426}},\ \bibinfo {pages}
  {2719} (\bibinfo {year} {2012})}\BibitemShut {NoStop}%
\bibitem [{\citenamefont {Alcaniz}(2004)}]{Alcaniz:2003qy}%
  \BibitemOpen
  \bibfield  {author} {\bibinfo {author} {\bibfnamefont {J.~S.}\ \bibnamefont
  {Alcaniz}},\ }\href {\doibase 10.1103/PhysRevD.69.083521} {\bibfield
  {journal} {\bibinfo  {journal} {Phys. Rev.}\ }\textbf {\bibinfo {volume}
  {D69}},\ \bibinfo {pages} {083521} (\bibinfo {year} {2004})},\ \Eprint
  {http://arxiv.org/abs/astro-ph/0312424} {arXiv:astro-ph/0312424 [astro-ph]}
  \BibitemShut {NoStop}%
\bibitem [{\citenamefont {Wang}\ and\ \citenamefont
  {Steinhardt}(1998)}]{Wang1998}%
  \BibitemOpen
  \bibfield  {author} {\bibinfo {author} {\bibfnamefont {L.}~\bibnamefont
  {Wang}}\ and\ \bibinfo {author} {\bibfnamefont {P.~J.}\ \bibnamefont
  {Steinhardt}},\ }\href@noop {} {\bibfield  {journal} {\bibinfo  {journal}
  {\apj}\ }\textbf {\bibinfo {volume} {508}},\ \bibinfo {pages} {483} (\bibinfo
  {year} {1998})}\BibitemShut {NoStop}%
\bibitem [{\citenamefont {Allen}\ \emph {et~al.}(2004)\citenamefont {Allen},
  \citenamefont {Schmidt}, \citenamefont {Ebeling}, \citenamefont {Fabian},\
  and\ \citenamefont {van Speybroeck}}]{Allen:2004cd}%
  \BibitemOpen
  \bibfield  {author} {\bibinfo {author} {\bibfnamefont {S.~W.}\ \bibnamefont
  {Allen}}, \bibinfo {author} {\bibfnamefont {R.~W.}\ \bibnamefont {Schmidt}},
  \bibinfo {author} {\bibfnamefont {H.}~\bibnamefont {Ebeling}}, \bibinfo
  {author} {\bibfnamefont {A.~C.}\ \bibnamefont {Fabian}}, \ and\ \bibinfo
  {author} {\bibfnamefont {L.}~\bibnamefont {van Speybroeck}},\ }\href
  {\doibase 10.1111/j.1365-2966.2004.08080.x} {\bibfield  {journal} {\bibinfo
  {journal} {Mon. Not. Roy. Astron. Soc.}\ }\textbf {\bibinfo {volume} {353}},\
  \bibinfo {pages} {457} (\bibinfo {year} {2004})},\ \Eprint
  {http://arxiv.org/abs/astro-ph/0405340} {arXiv:astro-ph/0405340 [astro-ph]}
  \BibitemShut {NoStop}%
\bibitem [{\citenamefont {Benjamin}\ \emph {et~al.}(2007)\citenamefont
  {Benjamin}, \citenamefont {Heymans}, \citenamefont {Semboloni}, \citenamefont
  {Van~Waerbeke}, \citenamefont {Hoekstra}, \citenamefont {Erben},
  \citenamefont {Gladders}, \citenamefont {Hetterscheidt}, \citenamefont
  {Mellier},\ and\ \citenamefont {Yee}}]{Benjamin:2007ys}%
  \BibitemOpen
  \bibfield  {author} {\bibinfo {author} {\bibfnamefont {J.}~\bibnamefont
  {Benjamin}}, \bibinfo {author} {\bibfnamefont {C.}~\bibnamefont {Heymans}},
  \bibinfo {author} {\bibfnamefont {E.}~\bibnamefont {Semboloni}}, \bibinfo
  {author} {\bibfnamefont {L.}~\bibnamefont {Van~Waerbeke}}, \bibinfo {author}
  {\bibfnamefont {H.}~\bibnamefont {Hoekstra}}, \bibinfo {author}
  {\bibfnamefont {T.}~\bibnamefont {Erben}}, \bibinfo {author} {\bibfnamefont
  {M.~D.}\ \bibnamefont {Gladders}}, \bibinfo {author} {\bibfnamefont
  {M.}~\bibnamefont {Hetterscheidt}}, \bibinfo {author} {\bibfnamefont
  {Y.}~\bibnamefont {Mellier}}, \ and\ \bibinfo {author} {\bibfnamefont
  {H.~K.~C.}\ \bibnamefont {Yee}},\ }\href {\doibase
  10.1111/j.1365-2966.2007.12202.x} {\bibfield  {journal} {\bibinfo  {journal}
  {Mon. Not. Roy. Astron. Soc.}\ }\textbf {\bibinfo {volume} {381}},\ \bibinfo
  {pages} {702} (\bibinfo {year} {2007})},\ \Eprint
  {http://arxiv.org/abs/astro-ph/0703570} {arXiv:astro-ph/0703570 [astro-ph]}
  \BibitemShut {NoStop}%
\bibitem [{\citenamefont {Amendola}\ \emph {et~al.}(2008)\citenamefont
  {Amendola}, \citenamefont {Kunz},\ and\ \citenamefont
  {Sapone}}]{Amendola:2007rr}%
  \BibitemOpen
  \bibfield  {author} {\bibinfo {author} {\bibfnamefont {L.}~\bibnamefont
  {Amendola}}, \bibinfo {author} {\bibfnamefont {M.}~\bibnamefont {Kunz}}, \
  and\ \bibinfo {author} {\bibfnamefont {D.}~\bibnamefont {Sapone}},\ }\href
  {\doibase 10.1088/1475-7516/2008/04/013} {\bibfield  {journal} {\bibinfo
  {journal} {JCAP}\ }\textbf {\bibinfo {volume} {0804}},\ \bibinfo {pages}
  {013} (\bibinfo {year} {2008})},\ \Eprint {http://arxiv.org/abs/0704.2421}
  {arXiv:0704.2421 [astro-ph]} \BibitemShut {NoStop}%
\bibitem [{\citenamefont {Fu}\ \emph {et~al.}(2008)\citenamefont {Fu} \emph
  {et~al.}}]{Fu:2007qq}%
  \BibitemOpen
  \bibfield  {author} {\bibinfo {author} {\bibfnamefont {L.}~\bibnamefont {Fu}}
  \emph {et~al.},\ }\href {\doibase 10.1051/0004-6361:20078522} {\bibfield
  {journal} {\bibinfo  {journal} {Astron. Astrophys.}\ }\textbf {\bibinfo
  {volume} {479}},\ \bibinfo {pages} {9} (\bibinfo {year} {2008})},\ \Eprint
  {http://arxiv.org/abs/0712.0884} {arXiv:0712.0884 [astro-ph]} \BibitemShut
  {NoStop}%
\bibitem [{\citenamefont {Peebles}\ and\ \citenamefont
  {Ratra}(2003)}]{Peebles2003}%
  \BibitemOpen
  \bibfield  {author} {\bibinfo {author} {\bibfnamefont {P.~J.}\ \bibnamefont
  {Peebles}}\ and\ \bibinfo {author} {\bibfnamefont {B.}~\bibnamefont
  {Ratra}},\ }\href@noop {} {\bibfield  {journal} {\bibinfo  {journal} {Reviews
  of Modern Physics}\ }\textbf {\bibinfo {volume} {75}},\ \bibinfo {pages}
  {559} (\bibinfo {year} {2003})}\BibitemShut {NoStop}%
\bibitem [{\citenamefont {Weinberg}(1989)}]{Weinberg1989}%
  \BibitemOpen
  \bibfield  {author} {\bibinfo {author} {\bibfnamefont {S.}~\bibnamefont
  {Weinberg}},\ }\href@noop {} {\bibfield  {journal} {\bibinfo  {journal}
  {Reviews of Modern Physics}\ }\textbf {\bibinfo {volume} {61}},\ \bibinfo
  {pages} {1} (\bibinfo {year} {1989})}\BibitemShut {NoStop}%
\bibitem [{\citenamefont {Sahni}\ and\ \citenamefont
  {Starobinsky}(2000)}]{Sahni:1999gb}%
  \BibitemOpen
  \bibfield  {author} {\bibinfo {author} {\bibfnamefont {V.}~\bibnamefont
  {Sahni}}\ and\ \bibinfo {author} {\bibfnamefont {A.~A.}\ \bibnamefont
  {Starobinsky}},\ }\href@noop {} {\bibfield  {journal} {\bibinfo  {journal}
  {IJMPD}\ }\textbf {\bibinfo {volume} {9}},\ \bibinfo {pages} {373} (\bibinfo
  {year} {2000})}\BibitemShut {NoStop}%
\bibitem [{\citenamefont {Sol\`a}(2013)}]{Sola:2013gha}%
  \BibitemOpen
  \bibfield  {author} {\bibinfo {author} {\bibfnamefont {J.}~\bibnamefont
  {Sol\`a}},\ }\href {\doibase 10.1088/1742-6596/453/1/012015} {\bibfield
  {journal} {\bibinfo  {journal} {J. Phys. Conf. Ser.}\ }\textbf {\bibinfo
  {volume} {453}},\ \bibinfo {pages} {012015} (\bibinfo {year} {2013})},\
  \Eprint {http://arxiv.org/abs/1306.1527} {arXiv:1306.1527 [gr-qc]}
  \BibitemShut {NoStop}%
\bibitem [{\citenamefont {Horava}\ and\ \citenamefont
  {Minic}(2000)}]{Horava:2000tb}%
  \BibitemOpen
  \bibfield  {author} {\bibinfo {author} {\bibfnamefont {P.}~\bibnamefont
  {Horava}}\ and\ \bibinfo {author} {\bibfnamefont {D.}~\bibnamefont {Minic}},\
  }\href {\doibase 10.1103/PhysRevLett.85.1610} {\bibfield  {journal} {\bibinfo
   {journal} {Phys. Rev. Lett.}\ }\textbf {\bibinfo {volume} {85}},\ \bibinfo
  {pages} {1610} (\bibinfo {year} {2000})},\ \Eprint
  {http://arxiv.org/abs/hep-th/0001145} {arXiv:hep-th/0001145 [hep-th]}
  \BibitemShut {NoStop}%
\bibitem [{\citenamefont {Thomas}(2002)}]{Thomas:2002pq}%
  \BibitemOpen
  \bibfield  {author} {\bibinfo {author} {\bibfnamefont {S.~D.}\ \bibnamefont
  {Thomas}},\ }\href {\doibase 10.1103/PhysRevLett.89.081301} {\bibfield
  {journal} {\bibinfo  {journal} {Phys. Rev. Lett.}\ }\textbf {\bibinfo
  {volume} {89}},\ \bibinfo {pages} {081301} (\bibinfo {year}
  {2002})}\BibitemShut {NoStop}%
\bibitem [{\citenamefont {Wei}\ and\ \citenamefont {Cai}(2008)}]{Wei:2007ty}%
  \BibitemOpen
  \bibfield  {author} {\bibinfo {author} {\bibfnamefont {H.}~\bibnamefont
  {Wei}}\ and\ \bibinfo {author} {\bibfnamefont {R.-G.}\ \bibnamefont {Cai}},\
  }\href {\doibase 10.1016/j.physletb.2007.12.030} {\bibfield  {journal}
  {\bibinfo  {journal} {Phys. Lett.}\ }\textbf {\bibinfo {volume} {B660}},\
  \bibinfo {pages} {113} (\bibinfo {year} {2008})},\ \Eprint
  {http://arxiv.org/abs/0708.0884} {arXiv:0708.0884 [astro-ph]} \BibitemShut
  {NoStop}%
\bibitem [{\citenamefont {Akhlaghi}\ \emph {et~al.}(2018)\citenamefont
  {Akhlaghi}, \citenamefont {Malekjani}, \citenamefont {Basilakos},\ and\
  \citenamefont {Haghi}}]{Akhlaghi:2018knk}%
  \BibitemOpen
  \bibfield  {author} {\bibinfo {author} {\bibfnamefont {I.~A.}\ \bibnamefont
  {Akhlaghi}}, \bibinfo {author} {\bibfnamefont {M.}~\bibnamefont {Malekjani}},
  \bibinfo {author} {\bibfnamefont {S.}~\bibnamefont {Basilakos}}, \ and\
  \bibinfo {author} {\bibfnamefont {H.}~\bibnamefont {Haghi}},\ }\href
  {\doibase 10.1093/mnras/sty903} {\bibfield  {journal} {\bibinfo  {journal}
  {Mon. Not. Roy. Astron. Soc.}\ }\textbf {\bibinfo {volume} {477}},\ \bibinfo
  {pages} {3659} (\bibinfo {year} {2018})},\ \Eprint
  {http://arxiv.org/abs/1804.02989} {arXiv:1804.02989 [gr-qc]} \BibitemShut
  {NoStop}%
\bibitem [{\citenamefont {Malekjani}\ \emph {et~al.}(2018)\citenamefont
  {Malekjani}, \citenamefont {Rezaei},\ and\ \citenamefont
  {Akhlaghi}}]{Malekjani:2018qcz}%
  \BibitemOpen
  \bibfield  {author} {\bibinfo {author} {\bibfnamefont {M.}~\bibnamefont
  {Malekjani}}, \bibinfo {author} {\bibfnamefont {M.}~\bibnamefont {Rezaei}}, \
  and\ \bibinfo {author} {\bibfnamefont {I.~A.}\ \bibnamefont {Akhlaghi}},\
  }\href {\doibase 10.1103/PhysRevD.98.063533} {\  (\bibinfo {year} {2018}),\
  10.1103/PhysRevD.98.063533},\ \Eprint {http://arxiv.org/abs/1809.08792}
  {arXiv:1809.08792 [gr-qc]} \BibitemShut {NoStop}%
\bibitem [{\citenamefont {Basilakos}\ \emph {et~al.}(2012)\citenamefont
  {Basilakos}, \citenamefont {Polarski},\ and\ \citenamefont
  {Sol\`a}}]{Basilakos:2012ra}%
  \BibitemOpen
  \bibfield  {author} {\bibinfo {author} {\bibfnamefont {S.}~\bibnamefont
  {Basilakos}}, \bibinfo {author} {\bibfnamefont {D.}~\bibnamefont {Polarski}},
  \ and\ \bibinfo {author} {\bibfnamefont {J.}~\bibnamefont {Sol\`a}},\ }\href
  {\doibase 10.1103/PhysRevD.86.043010} {\bibfield  {journal} {\bibinfo
  {journal} {Phys. Rev.}\ }\textbf {\bibinfo {volume} {D86}},\ \bibinfo {pages}
  {043010} (\bibinfo {year} {2012})},\ \Eprint {http://arxiv.org/abs/1204.4806}
  {arXiv:1204.4806 [gr-qc]} \BibitemShut {NoStop}%
\bibitem [{\citenamefont {Basilakos}\ and\ \citenamefont
  {Sol\`a}(2014)}]{Basilakos:2014tha}%
  \BibitemOpen
  \bibfield  {author} {\bibinfo {author} {\bibfnamefont {S.}~\bibnamefont
  {Basilakos}}\ and\ \bibinfo {author} {\bibfnamefont {J.}~\bibnamefont
  {Sol\`a}},\ }\href {\doibase 10.1103/PhysRevD.90.023008} {\bibfield
  {journal} {\bibinfo  {journal} {Phys. Rev.}\ }\textbf {\bibinfo {volume}
  {D90}},\ \bibinfo {pages} {023008} (\bibinfo {year} {2014})},\ \Eprint
  {http://arxiv.org/abs/1402.6594} {arXiv:1402.6594 [astro-ph.CO]} \BibitemShut
  {NoStop}%
\bibitem [{\citenamefont {Lima}\ \emph {et~al.}(2013)\citenamefont {Lima},
  \citenamefont {Basilakos},\ and\ \citenamefont {Sol\`a}}]{Lima:2012mu}%
  \BibitemOpen
  \bibfield  {author} {\bibinfo {author} {\bibfnamefont {J.~A.~S.}\
  \bibnamefont {Lima}}, \bibinfo {author} {\bibfnamefont {S.}~\bibnamefont
  {Basilakos}}, \ and\ \bibinfo {author} {\bibfnamefont {J.}~\bibnamefont
  {Sol\`a}},\ }\href {\doibase 10.1093/mnras/stt220} {\bibfield  {journal}
  {\bibinfo  {journal} {Mon. Not. Roy. Astron. Soc.}\ }\textbf {\bibinfo
  {volume} {431}},\ \bibinfo {pages} {923} (\bibinfo {year} {2013})},\ \Eprint
  {http://arxiv.org/abs/1209.2802} {arXiv:1209.2802 [gr-qc]} \BibitemShut
  {NoStop}%
\bibitem [{\citenamefont {Perico}\ \emph {et~al.}(2013)\citenamefont {Perico},
  \citenamefont {Lima}, \citenamefont {Basilakos},\ and\ \citenamefont
  {Sol\`a}}]{Perico:2013mna}%
  \BibitemOpen
  \bibfield  {author} {\bibinfo {author} {\bibfnamefont {E.~L.~D.}\
  \bibnamefont {Perico}}, \bibinfo {author} {\bibfnamefont {J.~A.~S.}\
  \bibnamefont {Lima}}, \bibinfo {author} {\bibfnamefont {S.}~\bibnamefont
  {Basilakos}}, \ and\ \bibinfo {author} {\bibfnamefont {J.}~\bibnamefont
  {Sol\`a}},\ }\href {\doibase 10.1103/PhysRevD.88.063531} {\bibfield
  {journal} {\bibinfo  {journal} {Phys. Rev.}\ }\textbf {\bibinfo {volume}
  {D88}},\ \bibinfo {pages} {063531} (\bibinfo {year} {2013})},\ \Eprint
  {http://arxiv.org/abs/1306.0591} {arXiv:1306.0591 [astro-ph.CO]} \BibitemShut
  {NoStop}%
\bibitem [{\citenamefont {Sol\`a}\ and\ \citenamefont
  {G\'omez-Valent}(2015)}]{Sola:2015rra}%
  \BibitemOpen
  \bibfield  {author} {\bibinfo {author} {\bibfnamefont {J.}~\bibnamefont
  {Sol\`a}}\ and\ \bibinfo {author} {\bibfnamefont {A.}~\bibnamefont
  {G\'omez-Valent}},\ }\href {\doibase 10.1142/S0218271815410035} {\bibfield
  {journal} {\bibinfo  {journal} {Int. J. Mod. Phys.}\ }\textbf {\bibinfo
  {volume} {D24}},\ \bibinfo {pages} {1541003} (\bibinfo {year} {2015})},\
  \Eprint {http://arxiv.org/abs/1501.03832} {arXiv:1501.03832 [gr-qc]}
  \BibitemShut {NoStop}%
\bibitem [{\citenamefont {Sol\`a}(2015)}]{Sola:2015csa}%
  \BibitemOpen
  \bibfield  {author} {\bibinfo {author} {\bibfnamefont {J.}~\bibnamefont
  {Sol\`a}},\ }\href {\doibase 10.1142/S0218271815440277} {\bibfield  {journal}
  {\bibinfo  {journal} {Int. J. Mod. Phys.}\ }\textbf {\bibinfo {volume}
  {D24}},\ \bibinfo {pages} {1544027} (\bibinfo {year} {2015})},\ \Eprint
  {http://arxiv.org/abs/1505.05863} {arXiv:1505.05863 [gr-qc]} \BibitemShut
  {NoStop}%
\bibitem [{\citenamefont {Schutzhold}(2002)}]{Schutzhold:2002pr}%
  \BibitemOpen
  \bibfield  {author} {\bibinfo {author} {\bibfnamefont {R.}~\bibnamefont
  {Schutzhold}},\ }\href {\doibase 10.1103/PhysRevLett.89.081302} {\bibfield
  {journal} {\bibinfo  {journal} {Phys. Rev. Lett.}\ }\textbf {\bibinfo
  {volume} {89}},\ \bibinfo {pages} {081302} (\bibinfo {year} {2002})},\
  \Eprint {http://arxiv.org/abs/gr-qc/0204018} {arXiv:gr-qc/0204018 [gr-qc]}
  \BibitemShut {NoStop}%
\bibitem [{\citenamefont {Urban}\ and\ \citenamefont
  {Zhitnitsky}(2010)}]{Urban2010a}%
  \BibitemOpen
  \bibfield  {author} {\bibinfo {author} {\bibfnamefont {F.}~\bibnamefont
  {Urban}}\ and\ \bibinfo {author} {\bibfnamefont {A.}~\bibnamefont
  {Zhitnitsky}},\ }\href@noop {} {\bibfield  {journal} {\bibinfo  {journal}
  {Phys. Lett. B}\ }\textbf {\bibinfo {volume} {688}},\ \bibinfo {pages} {9}
  (\bibinfo {year} {2010})}\BibitemShut {NoStop}%
\bibitem [{\citenamefont {Witten}(1979)}]{Witten1979}%
  \BibitemOpen
  \bibfield  {author} {\bibinfo {author} {\bibfnamefont {E.}~\bibnamefont
  {Witten}},\ }\href@noop {} {\bibfield  {journal} {\bibinfo  {journal}
  {Nucl.Phys. B}\ }\textbf {\bibinfo {volume} {156}},\ \bibinfo {pages} {269}
  (\bibinfo {year} {1979})}\BibitemShut {NoStop}%
\bibitem [{\citenamefont {Veneziano}(1979)}]{Veneziano1979}%
  \BibitemOpen
  \bibfield  {author} {\bibinfo {author} {\bibfnamefont {G.}~\bibnamefont
  {Veneziano}},\ }\href@noop {} {\bibfield  {journal} {\bibinfo  {journal}
  {Nucl.Phys. B}\ }\textbf {\bibinfo {volume} {159}},\ \bibinfo {pages} {213}
  (\bibinfo {year} {1979})}\BibitemShut {NoStop}%
\bibitem [{\citenamefont {Rosenzweig}\ \emph {et~al.}(1980)\citenamefont
  {Rosenzweig}, \citenamefont {Schechter},\ and\ \citenamefont
  {Trahern}}]{Rosenzweig1980}%
  \BibitemOpen
  \bibfield  {author} {\bibinfo {author} {\bibfnamefont {C.}~\bibnamefont
  {Rosenzweig}}, \bibinfo {author} {\bibfnamefont {J.}~\bibnamefont
  {Schechter}}, \ and\ \bibinfo {author} {\bibfnamefont {C.}~\bibnamefont
  {Trahern}},\ }\href@noop {} {\bibfield  {journal} {\bibinfo  {journal} {Phys.
  Rev. D}\ }\textbf {\bibinfo {volume} {21}},\ \bibinfo {pages} {3388}
  (\bibinfo {year} {1980})}\BibitemShut {NoStop}%
\bibitem [{\citenamefont {Verlinde}(2011)}]{Verlinde:2010hp}%
  \BibitemOpen
  \bibfield  {author} {\bibinfo {author} {\bibfnamefont {E.~P.}\ \bibnamefont
  {Verlinde}},\ }\href {\doibase 10.1007/JHEP04(2011)029} {\bibfield  {journal}
  {\bibinfo  {journal} {JHEP}\ }\textbf {\bibinfo {volume} {04}},\ \bibinfo
  {pages} {029} (\bibinfo {year} {2011})},\ \Eprint
  {http://arxiv.org/abs/1001.0785} {arXiv:1001.0785 [hep-th]} \BibitemShut
  {NoStop}%
\bibitem [{\citenamefont {Easson}\ \emph {et~al.}(2011)\citenamefont {Easson},
  \citenamefont {Frampton},\ and\ \citenamefont {Smoot}}]{Easson:2010av}%
  \BibitemOpen
  \bibfield  {author} {\bibinfo {author} {\bibfnamefont {D.~A.}\ \bibnamefont
  {Easson}}, \bibinfo {author} {\bibfnamefont {P.~H.}\ \bibnamefont
  {Frampton}}, \ and\ \bibinfo {author} {\bibfnamefont {G.~F.}\ \bibnamefont
  {Smoot}},\ }\href {\doibase 10.1016/j.physletb.2010.12.025} {\bibfield
  {journal} {\bibinfo  {journal} {Phys. Lett.}\ }\textbf {\bibinfo {volume}
  {B696}},\ \bibinfo {pages} {273} (\bibinfo {year} {2011})},\ \Eprint
  {http://arxiv.org/abs/1002.4278} {arXiv:1002.4278 [hep-th]} \BibitemShut
  {NoStop}%
\bibitem [{\citenamefont {Komatsu}(2019)}]{Komatsu:2018meb}%
  \BibitemOpen
  \bibfield  {author} {\bibinfo {author} {\bibfnamefont {N.}~\bibnamefont
  {Komatsu}},\ }\href {\doibase 10.1103/PhysRevD.99.043523} {\bibfield
  {journal} {\bibinfo  {journal} {Phys. Rev.}\ }\textbf {\bibinfo {volume}
  {D99}},\ \bibinfo {pages} {043523} (\bibinfo {year} {2019})},\ \Eprint
  {http://arxiv.org/abs/1810.11138} {arXiv:1810.11138 [gr-qc]} \BibitemShut
  {NoStop}%
\bibitem [{\citenamefont {G\'omez-Valent}\ \emph {et~al.}(2015)\citenamefont
  {G\'omez-Valent}, \citenamefont {Karimkhani},\ and\ \citenamefont
  {Sol\`a}}]{Gomez-Valent:2015pia}%
  \BibitemOpen
  \bibfield  {author} {\bibinfo {author} {\bibfnamefont {A.}~\bibnamefont
  {G\'omez-Valent}}, \bibinfo {author} {\bibfnamefont {E.}~\bibnamefont
  {Karimkhani}}, \ and\ \bibinfo {author} {\bibfnamefont {J.}~\bibnamefont
  {Sol\`a}},\ }\href {\doibase 10.1088/1475-7516/2015/12/048} {\bibfield
  {journal} {\bibinfo  {journal} {JCAP}\ }\textbf {\bibinfo {volume} {1512}},\
  \bibinfo {pages} {048} (\bibinfo {year} {2015})},\ \Eprint
  {http://arxiv.org/abs/1509.03298} {arXiv:1509.03298 [gr-qc]} \BibitemShut
  {NoStop}%
\bibitem [{\citenamefont {G\'omez-Valent}\ and\ \citenamefont
  {Basilakos}(2015)}]{Gomez-Valent:2014rxa}%
  \BibitemOpen
  \bibfield  {author} {\bibinfo {author} {\bibfnamefont {S.~J.}\ \bibnamefont
  {G\'omez-Valent}, \bibfnamefont {A.}}\ and\ \bibinfo {author} {\bibfnamefont
  {S.}~\bibnamefont {Basilakos}},\ }\href {\doibase
  10.1088/1475-7516/2015/01/004} {\bibfield  {journal} {\bibinfo  {journal}
  {JCAP}\ }\textbf {\bibinfo {volume} {1501}},\ \bibinfo {pages} {004}
  (\bibinfo {year} {2015})},\ \Eprint {http://arxiv.org/abs/1409.7048}
  {arXiv:1409.7048 [astro-ph.CO]} \BibitemShut {NoStop}%
\bibitem [{\citenamefont {G\'omez-Valent}\ and\ \citenamefont
  {Sol\`a}(2015)}]{Gomez-Valent:2014fda}%
  \BibitemOpen
  \bibfield  {author} {\bibinfo {author} {\bibfnamefont {A.}~\bibnamefont
  {G\'omez-Valent}}\ and\ \bibinfo {author} {\bibfnamefont {J.}~\bibnamefont
  {Sol\`a}},\ }\href {\doibase 10.1093/mnras/stv209} {\bibfield  {journal}
  {\bibinfo  {journal} {Mon. Not. Roy. Astron. Soc.}\ }\textbf {\bibinfo
  {volume} {448}},\ \bibinfo {pages} {2810} (\bibinfo {year} {2015})},\ \Eprint
  {http://arxiv.org/abs/1412.3785} {arXiv:1412.3785 [astro-ph.CO]} \BibitemShut
  {NoStop}%
\bibitem [{\citenamefont {Sol\`a}\ \emph {et~al.}(2015)\citenamefont {Sol\`a},
  \citenamefont {G\'omez-Valent},\ and\ \citenamefont
  {de~Cruz~P\'erez}}]{Sola:2015wwa}%
  \BibitemOpen
  \bibfield  {author} {\bibinfo {author} {\bibfnamefont {J.}~\bibnamefont
  {Sol\`a}}, \bibinfo {author} {\bibfnamefont {A.}~\bibnamefont
  {G\'omez-Valent}}, \ and\ \bibinfo {author} {\bibfnamefont {J.}~\bibnamefont
  {de~Cruz~P\'erez}},\ }\href {\doibase 10.1088/2041-8205/811/1/L14} {\bibfield
   {journal} {\bibinfo  {journal} {Astrophys. J.}\ }\textbf {\bibinfo {volume}
  {811}},\ \bibinfo {pages} {L14} (\bibinfo {year} {2015})},\ \Eprint
  {http://arxiv.org/abs/1506.05793} {arXiv:1506.05793 [gr-qc]} \BibitemShut
  {NoStop}%
\bibitem [{\citenamefont {Sol\`a}\ \emph {et~al.}(2017)\citenamefont {Sol\`a},
  \citenamefont {G\'omez-Valent},\ and\ \citenamefont
  {de~Cruz~P\'erez}}]{Sola:2016jky}%
  \BibitemOpen
  \bibfield  {author} {\bibinfo {author} {\bibfnamefont {J.}~\bibnamefont
  {Sol\`a}}, \bibinfo {author} {\bibfnamefont {A.}~\bibnamefont
  {G\'omez-Valent}}, \ and\ \bibinfo {author} {\bibfnamefont {J.}~\bibnamefont
  {de~Cruz~P\'erez}},\ }\href {\doibase 10.3847/1538-4357/836/1/43} {\bibfield
  {journal} {\bibinfo  {journal} {Astrophys. J.}\ }\textbf {\bibinfo {volume}
  {836}},\ \bibinfo {pages} {43} (\bibinfo {year} {2017})},\ \Eprint
  {http://arxiv.org/abs/1602.02103} {arXiv:1602.02103 [astro-ph.CO]}
  \BibitemShut {NoStop}%
\bibitem [{\citenamefont {Sol\`a}\ \emph
  {et~al.}(2018{\natexlab{a}})\citenamefont {Sol\`a}, \citenamefont
  {de~Cruz~P\'erez},\ and\ \citenamefont {G\'omez-Valent}}]{Sola:2016ecz}%
  \BibitemOpen
  \bibfield  {author} {\bibinfo {author} {\bibfnamefont {J.}~\bibnamefont
  {Sol\`a}}, \bibinfo {author} {\bibfnamefont {J.}~\bibnamefont
  {de~Cruz~P\'erez}}, \ and\ \bibinfo {author} {\bibfnamefont {A.}~\bibnamefont
  {G\'omez-Valent}},\ }\href {\doibase 10.1209/0295-5075/121/39001} {\bibfield
  {journal} {\bibinfo  {journal} {EPL}\ }\textbf {\bibinfo {volume} {121}},\
  \bibinfo {pages} {39001} (\bibinfo {year} {2018}{\natexlab{a}})},\ \Eprint
  {http://arxiv.org/abs/1606.00450} {arXiv:1606.00450 [gr-qc]} \BibitemShut
  {NoStop}%
\bibitem [{\citenamefont {Sol\`a}\ \emph
  {et~al.}(2018{\natexlab{b}})\citenamefont {Sol\`a}, \citenamefont {P\'erez},\
  and\ \citenamefont {G\'omez-Valent}}]{Sola:2017jbl}%
  \BibitemOpen
  \bibfield  {author} {\bibinfo {author} {\bibfnamefont {J.}~\bibnamefont
  {Sol\`a}}, \bibinfo {author} {\bibfnamefont {J.~d.~C.}\ \bibnamefont
  {P\'erez}}, \ and\ \bibinfo {author} {\bibfnamefont {A.}~\bibnamefont
  {G\'omez-Valent}},\ }\href {\doibase 10.1093/mnras/sty1253} {\bibfield
  {journal} {\bibinfo  {journal} {Mon. Not. Roy. Astron. Soc.}\ }\textbf
  {\bibinfo {volume} {478}},\ \bibinfo {pages} {4357} (\bibinfo {year}
  {2018}{\natexlab{b}})},\ \Eprint {http://arxiv.org/abs/1703.08218}
  {arXiv:1703.08218 [astro-ph.CO]} \BibitemShut {NoStop}%
\bibitem [{\citenamefont {Solà}\ \emph {et~al.}(2017)\citenamefont {Solà},
  \citenamefont {Gómez-Valent},\ and\ \citenamefont
  {de~Cruz~Pérez}}]{Sola:2017znb}%
  \BibitemOpen
  \bibfield  {author} {\bibinfo {author} {\bibfnamefont {J.}~\bibnamefont
  {Solà}}, \bibinfo {author} {\bibfnamefont {A.}~\bibnamefont
  {Gómez-Valent}}, \ and\ \bibinfo {author} {\bibfnamefont {J.}~\bibnamefont
  {de~Cruz~Pérez}},\ }\href {\doibase 10.1016/j.physletb.2017.09.073}
  {\bibfield  {journal} {\bibinfo  {journal} {Phys. Lett.}\ }\textbf {\bibinfo
  {volume} {B774}},\ \bibinfo {pages} {317} (\bibinfo {year} {2017})},\ \Eprint
  {http://arxiv.org/abs/1705.06723} {arXiv:1705.06723 [astro-ph.CO]}
  \BibitemShut {NoStop}%
\bibitem [{\citenamefont {Cai}\ \emph {et~al.}(2011)\citenamefont {Cai},
  \citenamefont {Tuo}, \citenamefont {Zhang},\ and\ \citenamefont
  {Su}}]{Cai:2010uf}%
  \BibitemOpen
  \bibfield  {author} {\bibinfo {author} {\bibfnamefont {R.-G.}\ \bibnamefont
  {Cai}}, \bibinfo {author} {\bibfnamefont {Z.-L.}\ \bibnamefont {Tuo}},
  \bibinfo {author} {\bibfnamefont {H.-B.}\ \bibnamefont {Zhang}}, \ and\
  \bibinfo {author} {\bibfnamefont {Q.}~\bibnamefont {Su}},\ }\href {\doibase
  10.1103/PhysRevD.84.123501} {\bibfield  {journal} {\bibinfo  {journal} {Phys.
  Rev.}\ }\textbf {\bibinfo {volume} {D84}},\ \bibinfo {pages} {123501}
  (\bibinfo {year} {2011})},\ \Eprint {http://arxiv.org/abs/1011.3212}
  {arXiv:1011.3212 [astro-ph.CO]} \BibitemShut {NoStop}%
\bibitem [{\citenamefont {Feng}\ \emph {et~al.}(2013)\citenamefont {Feng},
  \citenamefont {Li},\ and\ \citenamefont {Shen}}]{Feng:2012gr}%
  \BibitemOpen
  \bibfield  {author} {\bibinfo {author} {\bibfnamefont {C.-J.}\ \bibnamefont
  {Feng}}, \bibinfo {author} {\bibfnamefont {X.-Z.}\ \bibnamefont {Li}}, \ and\
  \bibinfo {author} {\bibfnamefont {X.-Y.}\ \bibnamefont {Shen}},\ }\href
  {\doibase 10.1103/PhysRevD.87.023006} {\bibfield  {journal} {\bibinfo
  {journal} {Phys. Rev.}\ }\textbf {\bibinfo {volume} {D87}},\ \bibinfo {pages}
  {023006} (\bibinfo {year} {2013})},\ \Eprint {http://arxiv.org/abs/1202.0058}
  {arXiv:1202.0058 [astro-ph.CO]} \BibitemShut {NoStop}%
\bibitem [{\citenamefont {Khurshudyan}\ \emph {et~al.}(2015)\citenamefont
  {Khurshudyan}, \citenamefont {Khurshudyan},\ and\ \citenamefont
  {Myrzakulov}}]{Khurshudyan:2013oba}%
  \BibitemOpen
  \bibfield  {author} {\bibinfo {author} {\bibfnamefont {M.}~\bibnamefont
  {Khurshudyan}}, \bibinfo {author} {\bibfnamefont {A.}~\bibnamefont
  {Khurshudyan}}, \ and\ \bibinfo {author} {\bibfnamefont {R.}~\bibnamefont
  {Myrzakulov}},\ }\href {\doibase 10.1007/s10509-015-2341-4} {\bibfield
  {journal} {\bibinfo  {journal} {Astrophys. Space Sci.}\ }\textbf {\bibinfo
  {volume} {357}},\ \bibinfo {pages} {113} (\bibinfo {year} {2015})},\ \Eprint
  {http://arxiv.org/abs/1307.7859} {arXiv:1307.7859 [gr-qc]} \BibitemShut
  {NoStop}%
\bibitem [{\citenamefont {Alavirad}\ and\ \citenamefont
  {Sheykhi}(2014)}]{Alavirad:2014kqa}%
  \BibitemOpen
  \bibfield  {author} {\bibinfo {author} {\bibfnamefont {H.}~\bibnamefont
  {Alavirad}}\ and\ \bibinfo {author} {\bibfnamefont {A.}~\bibnamefont
  {Sheykhi}},\ }\href {\doibase 10.1016/j.physletb.2014.05.023} {\bibfield
  {journal} {\bibinfo  {journal} {Phys. Lett.}\ }\textbf {\bibinfo {volume}
  {B734}},\ \bibinfo {pages} {148} (\bibinfo {year} {2014})},\ \Eprint
  {http://arxiv.org/abs/1405.2515} {arXiv:1405.2515 [astro-ph.CO]} \BibitemShut
  {NoStop}%
\bibitem [{\citenamefont {{Macaulay}}\ \emph {et~al.}(2013)\citenamefont
  {{Macaulay}}, \citenamefont {{Wehus}},\ and\ \citenamefont
  {{Eriksen}}}]{Macaulay:2013swa}%
  \BibitemOpen
  \bibfield  {author} {\bibinfo {author} {\bibfnamefont {E.}~\bibnamefont
  {{Macaulay}}}, \bibinfo {author} {\bibfnamefont {I.~K.}\ \bibnamefont
  {{Wehus}}}, \ and\ \bibinfo {author} {\bibfnamefont {H.~K.}\ \bibnamefont
  {{Eriksen}}},\ }\href@noop {} {\bibfield  {journal} {\bibinfo  {journal}
  {Physical Review Letters}\ }\textbf {\bibinfo {volume} {111}},\ \bibinfo
  {eid} {161301} (\bibinfo {year} {2013})},\ \Eprint
  {http://arxiv.org/abs/1303.6583} {arXiv:1303.6583 [astro-ph.CO]} \BibitemShut
  {NoStop}%
\bibitem [{\citenamefont {{Riess}}\ \emph {et~al.}(2018)\citenamefont
  {{Riess}}, \citenamefont {{Casertano}}, \citenamefont {{Yuan}}, \citenamefont
  {{Macri}}, \citenamefont {{Anderson}}, \citenamefont {{MacKenty}},
  \citenamefont {{Bowers}}, \citenamefont {{Clubb}}, \citenamefont
  {{Filippenko}}, \citenamefont {{Jones}},\ and\ \citenamefont
  {{Tucker}}}]{Riess:2018uxu}%
  \BibitemOpen
  \bibfield  {author} {\bibinfo {author} {\bibfnamefont {A.~G.}\ \bibnamefont
  {{Riess}}}, \bibinfo {author} {\bibfnamefont {S.}~\bibnamefont
  {{Casertano}}}, \bibinfo {author} {\bibfnamefont {W.}~\bibnamefont {{Yuan}}},
  \bibinfo {author} {\bibfnamefont {L.}~\bibnamefont {{Macri}}}, \bibinfo
  {author} {\bibfnamefont {J.}~\bibnamefont {{Anderson}}}, \bibinfo {author}
  {\bibfnamefont {J.~W.}\ \bibnamefont {{MacKenty}}}, \bibinfo {author}
  {\bibfnamefont {J.~B.}\ \bibnamefont {{Bowers}}}, \bibinfo {author}
  {\bibfnamefont {K.~I.}\ \bibnamefont {{Clubb}}}, \bibinfo {author}
  {\bibfnamefont {A.~V.}\ \bibnamefont {{Filippenko}}}, \bibinfo {author}
  {\bibfnamefont {D.~O.}\ \bibnamefont {{Jones}}}, \ and\ \bibinfo {author}
  {\bibfnamefont {B.~E.}\ \bibnamefont {{Tucker}}},\ }\href {\doibase
  10.3847/1538-4357/aaadb7} {\bibfield  {journal} {\bibinfo  {journal} {\apj}\
  }\textbf {\bibinfo {volume} {855}},\ \bibinfo {eid} {136} (\bibinfo {year}
  {2018})},\ \Eprint {http://arxiv.org/abs/1801.01120} {arXiv:1801.01120
  [astro-ph.SR]} \BibitemShut {NoStop}%
\bibitem [{\citenamefont {Aghanim}\ \emph {et~al.}(2018)\citenamefont {Aghanim}
  \emph {et~al.}}]{Aghanim:2018eyx}%
  \BibitemOpen
  \bibfield  {author} {\bibinfo {author} {\bibfnamefont {N.}~\bibnamefont
  {Aghanim}} \emph {et~al.} (\bibinfo {collaboration} {Planck}),\ }\href@noop
  {} {\  (\bibinfo {year} {2018})},\ \Eprint {http://arxiv.org/abs/1807.06209}
  {arXiv:1807.06209 [astro-ph.CO]} \BibitemShut {NoStop}%
\bibitem [{\citenamefont {Erickson}\ \emph {et~al.}(2002)\citenamefont
  {Erickson}, \citenamefont {Caldwell}, \citenamefont {Steinhardt},
  \citenamefont {Armendariz-Picon},\ and\ \citenamefont
  {Mukhanov}}]{Erickson:2001bq}%
  \BibitemOpen
  \bibfield  {author} {\bibinfo {author} {\bibfnamefont {J.~K.}\ \bibnamefont
  {Erickson}}, \bibinfo {author} {\bibfnamefont {R.~R.}\ \bibnamefont
  {Caldwell}}, \bibinfo {author} {\bibfnamefont {P.~J.}\ \bibnamefont
  {Steinhardt}}, \bibinfo {author} {\bibfnamefont {C.}~\bibnamefont
  {Armendariz-Picon}}, \ and\ \bibinfo {author} {\bibfnamefont {V.~F.}\
  \bibnamefont {Mukhanov}},\ }\href {\doibase 10.1103/PhysRevLett.88.121301}
  {\bibfield  {journal} {\bibinfo  {journal} {Phys. Rev. Lett.}\ }\textbf
  {\bibinfo {volume} {88}},\ \bibinfo {pages} {121301} (\bibinfo {year}
  {2002})}\BibitemShut {NoStop}%
\bibitem [{\citenamefont {Bean}\ and\ \citenamefont
  {Dor{\'e}}(2004)}]{Bean:2003fb}%
  \BibitemOpen
  \bibfield  {author} {\bibinfo {author} {\bibfnamefont {R.}~\bibnamefont
  {Bean}}\ and\ \bibinfo {author} {\bibfnamefont {O.}~\bibnamefont
  {Dor{\'e}}},\ }\href {\doibase 10.1103/PhysRevD.69.083503} {\bibfield
  {journal} {\bibinfo  {journal} {Phys. Rev. D}\ }\textbf {\bibinfo {volume}
  {69}},\ \bibinfo {pages} {083503} (\bibinfo {year} {2004})}\BibitemShut
  {NoStop}%
\bibitem [{\citenamefont {Mehrabi}\ \emph
  {et~al.}(2015{\natexlab{a}})\citenamefont {Mehrabi}, \citenamefont
  {Basilakos}, \citenamefont {Malekjani},\ and\ \citenamefont
  {Davari}}]{Mehrabi:2015kta}%
  \BibitemOpen
  \bibfield  {author} {\bibinfo {author} {\bibfnamefont {A.}~\bibnamefont
  {Mehrabi}}, \bibinfo {author} {\bibfnamefont {S.}~\bibnamefont {Basilakos}},
  \bibinfo {author} {\bibfnamefont {M.}~\bibnamefont {Malekjani}}, \ and\
  \bibinfo {author} {\bibfnamefont {Z.}~\bibnamefont {Davari}},\ }\href
  {\doibase 10.1103/PhysRevD.92.123513} {\bibfield  {journal} {\bibinfo
  {journal} {Phys. Rev.}\ }\textbf {\bibinfo {volume} {D92}},\ \bibinfo {pages}
  {123513} (\bibinfo {year} {2015}{\natexlab{a}})},\ \Eprint
  {http://arxiv.org/abs/1510.03996} {arXiv:1510.03996 [astro-ph.CO]}
  \BibitemShut {NoStop}%
\bibitem [{\citenamefont {Malekjani}\ \emph {et~al.}(2017)\citenamefont
  {Malekjani}, \citenamefont {Basilakos}, \citenamefont {Davari}, \citenamefont
  {Mehrabi},\ and\ \citenamefont {Rezaei}}]{Malekjani:2016edh}%
  \BibitemOpen
  \bibfield  {author} {\bibinfo {author} {\bibfnamefont {M.}~\bibnamefont
  {Malekjani}}, \bibinfo {author} {\bibfnamefont {S.}~\bibnamefont
  {Basilakos}}, \bibinfo {author} {\bibfnamefont {Z.}~\bibnamefont {Davari}},
  \bibinfo {author} {\bibfnamefont {A.}~\bibnamefont {Mehrabi}}, \ and\
  \bibinfo {author} {\bibfnamefont {M.}~\bibnamefont {Rezaei}},\ }\href@noop {}
  {\bibfield  {journal} {\bibinfo  {journal} {Mon. Not. Roy. Astron. Soc.}\
  }\textbf {\bibinfo {volume} {464}},\ \bibinfo {pages} {1192} (\bibinfo {year}
  {2017})},\ \Eprint {http://arxiv.org/abs/1609.01998} {arXiv:1609.01998
  [astro-ph.CO]} \BibitemShut {NoStop}%
\bibitem [{\citenamefont {Abramo}\ \emph {et~al.}(2009)\citenamefont {Abramo},
  \citenamefont {Batista}, \citenamefont {Liberato},\ and\ \citenamefont
  {Rosenfeld}}]{Abramo:2008ip}%
  \BibitemOpen
  \bibfield  {author} {\bibinfo {author} {\bibfnamefont {L.~R.}\ \bibnamefont
  {Abramo}}, \bibinfo {author} {\bibfnamefont {R.~C.}\ \bibnamefont {Batista}},
  \bibinfo {author} {\bibfnamefont {L.}~\bibnamefont {Liberato}}, \ and\
  \bibinfo {author} {\bibfnamefont {R.}~\bibnamefont {Rosenfeld}},\ }\href
  {\doibase 10.1103/PhysRevD.79.023516} {\bibfield  {journal} {\bibinfo
  {journal} {Phys. Rev.}\ }\textbf {\bibinfo {volume} {D79}},\ \bibinfo {pages}
  {023516} (\bibinfo {year} {2009})},\ \Eprint {http://arxiv.org/abs/0806.3461}
  {arXiv:0806.3461 [astro-ph]} \BibitemShut {NoStop}%
\bibitem [{\citenamefont {Grande}\ \emph {et~al.}(2009)\citenamefont {Grande},
  \citenamefont {Pelinson},\ and\ \citenamefont {Sol\`a}}]{Grande:2008re}%
  \BibitemOpen
  \bibfield  {author} {\bibinfo {author} {\bibfnamefont {J.}~\bibnamefont
  {Grande}}, \bibinfo {author} {\bibfnamefont {A.}~\bibnamefont {Pelinson}}, \
  and\ \bibinfo {author} {\bibfnamefont {J.}~\bibnamefont {Sol\`a}},\ }\href
  {\doibase 10.1103/PhysRevD.79.043006} {\bibfield  {journal} {\bibinfo
  {journal} {Phys. Rev.}\ }\textbf {\bibinfo {volume} {D79}},\ \bibinfo {pages}
  {043006} (\bibinfo {year} {2009})},\ \Eprint {http://arxiv.org/abs/0809.3462}
  {arXiv:0809.3462 [astro-ph]} \BibitemShut {NoStop}%
\bibitem [{\citenamefont {Llinares}\ and\ \citenamefont {Mota}(2013)}]{mota6}%
  \BibitemOpen
  \bibfield  {author} {\bibinfo {author} {\bibfnamefont {C.}~\bibnamefont
  {Llinares}}\ and\ \bibinfo {author} {\bibfnamefont {D.~F.}\ \bibnamefont
  {Mota}},\ }\href {\doibase 10.1103/PhysRevLett.110.161101} {\bibfield
  {journal} {\bibinfo  {journal} {Phys. Rev. Lett.}\ }\textbf {\bibinfo
  {volume} {110}},\ \bibinfo {pages} {161101} (\bibinfo {year} {2013})},\
  \Eprint {http://arxiv.org/abs/1302.1774} {arXiv:1302.1774 [astro-ph.CO]}
  \BibitemShut {NoStop}%
\bibitem [{\citenamefont {Rezaei}\ \emph {et~al.}(2017)\citenamefont {Rezaei},
  \citenamefont {Malekjani}, \citenamefont {Basilakos}, \citenamefont
  {Mehrabi},\ and\ \citenamefont {Mota}}]{Rezaei:2017yyj}%
  \BibitemOpen
  \bibfield  {author} {\bibinfo {author} {\bibfnamefont {M.}~\bibnamefont
  {Rezaei}}, \bibinfo {author} {\bibfnamefont {M.}~\bibnamefont {Malekjani}},
  \bibinfo {author} {\bibfnamefont {S.}~\bibnamefont {Basilakos}}, \bibinfo
  {author} {\bibfnamefont {A.}~\bibnamefont {Mehrabi}}, \ and\ \bibinfo
  {author} {\bibfnamefont {D.~F.}\ \bibnamefont {Mota}},\ }\href {\doibase
  10.3847/1538-4357/aa7898} {\bibfield  {journal} {\bibinfo  {journal}
  {Astrophys. J.}\ }\textbf {\bibinfo {volume} {843}},\ \bibinfo {pages} {65}
  (\bibinfo {year} {2017})},\ \Eprint {http://arxiv.org/abs/1706.02537}
  {arXiv:1706.02537 [astro-ph.CO]} \BibitemShut {NoStop}%
\bibitem [{\citenamefont {Rezaei}(2019{\natexlab{a}})}]{Rezaei:2019roe}%
  \BibitemOpen
  \bibfield  {author} {\bibinfo {author} {\bibfnamefont {M.}~\bibnamefont
  {Rezaei}},\ }\href {\doibase 10.1093/mnras/stz394} {\bibfield  {journal}
  {\bibinfo  {journal} {Mon. Not. Roy. Astron. Soc.}\ }\textbf {\bibinfo
  {volume} {485}},\ \bibinfo {pages} {550} (\bibinfo {year}
  {2019}{\natexlab{a}})},\ \Eprint {http://arxiv.org/abs/1902.04776}
  {arXiv:1902.04776 [gr-qc]} \BibitemShut {NoStop}%
\bibitem [{\citenamefont {Sol\`a}\ \emph {et~al.}(2019)\citenamefont {Sol\`a},
  \citenamefont {G\'omez-Valent},\ and\ \citenamefont
  {P\'erez}}]{Sola:2018sjf}%
  \BibitemOpen
  \bibfield  {author} {\bibinfo {author} {\bibfnamefont {J.}~\bibnamefont
  {Sol\`a}}, \bibinfo {author} {\bibfnamefont {A.}~\bibnamefont
  {G\'omez-Valent}}, \ and\ \bibinfo {author} {\bibfnamefont {J.~d.~C.}\
  \bibnamefont {P\'erez}},\ }\href@noop {} {\bibfield  {journal} {\bibinfo
  {journal} {Phys. Dark. Univ.}\ }\textbf {\bibinfo {volume} {25}},\ \bibinfo
  {pages} {100311} (\bibinfo {year} {2019})},\ \Eprint
  {http://arxiv.org/abs/1811.03505} {arXiv:1811.03505 [astro-ph.CO]}
  \BibitemShut {NoStop}%
\bibitem [{\citenamefont {Sola}\ \emph {et~al.}(2019)\citenamefont {Sola},
  \citenamefont {Gomez-Valent},\ and\ \citenamefont {Perez}}]{Sola:2019lnw}%
  \BibitemOpen
  \bibfield  {author} {\bibinfo {author} {\bibfnamefont {J.}~\bibnamefont
  {Sola}}, \bibinfo {author} {\bibfnamefont {A.}~\bibnamefont {Gomez-Valent}},
  \ and\ \bibinfo {author} {\bibfnamefont {J.~d.~C.}\ \bibnamefont {Perez}},\
  }\href@noop {} {\  (\bibinfo {year} {2019})},\ \Eprint
  {http://arxiv.org/abs/1904.11470} {arXiv:1904.11470 [astro-ph.CO]}
  \BibitemShut {NoStop}%
\bibitem [{\citenamefont {Zhitnitsky}(2011)}]{Zhitnitsky:2011tr}%
  \BibitemOpen
  \bibfield  {author} {\bibinfo {author} {\bibfnamefont {A.~R.}\ \bibnamefont
  {Zhitnitsky}},\ }\href {\doibase 10.1103/PhysRevD.84.124008} {\bibfield
  {journal} {\bibinfo  {journal} {Phys. Rev.}\ }\textbf {\bibinfo {volume}
  {D84}},\ \bibinfo {pages} {124008} (\bibinfo {year} {2011})},\ \Eprint
  {http://arxiv.org/abs/1105.6088} {arXiv:1105.6088 [hep-th]} \BibitemShut
  {NoStop}%
\bibitem [{\citenamefont {Basilakos}\ \emph {et~al.}(2009)\citenamefont
  {Basilakos}, \citenamefont {Plionis},\ and\ \citenamefont
  {Sol\`a}}]{Basilakos:2009wi}%
  \BibitemOpen
  \bibfield  {author} {\bibinfo {author} {\bibfnamefont {S.}~\bibnamefont
  {Basilakos}}, \bibinfo {author} {\bibfnamefont {M.}~\bibnamefont {Plionis}},
  \ and\ \bibinfo {author} {\bibfnamefont {J.}~\bibnamefont {Sol\`a}},\ }\href
  {\doibase 10.1103/PhysRevD.80.083511} {\bibfield  {journal} {\bibinfo
  {journal} {Phys. Rev.}\ }\textbf {\bibinfo {volume} {D80}},\ \bibinfo {pages}
  {083511} (\bibinfo {year} {2009})},\ \Eprint {http://arxiv.org/abs/0907.4555}
  {arXiv:0907.4555 [astro-ph.CO]} \BibitemShut {NoStop}%
\bibitem [{\citenamefont {Malekjani}\ \emph {et~al.}(2015)\citenamefont
  {Malekjani}, \citenamefont {Naderi},\ and\ \citenamefont
  {Pace}}]{Malekjani:2015pza}%
  \BibitemOpen
  \bibfield  {author} {\bibinfo {author} {\bibfnamefont {M.}~\bibnamefont
  {Malekjani}}, \bibinfo {author} {\bibfnamefont {T.}~\bibnamefont {Naderi}}, \
  and\ \bibinfo {author} {\bibfnamefont {F.}~\bibnamefont {Pace}},\ }\href
  {\doibase 10.1093/mnras/stv1909} {\bibfield  {journal} {\bibinfo  {journal}
  {Mon. Not. Roy. Astron. Soc.}\ }\textbf {\bibinfo {volume} {453}},\ \bibinfo
  {pages} {4148} (\bibinfo {year} {2015})},\ \Eprint
  {http://arxiv.org/abs/1508.04697} {arXiv:1508.04697 [gr-qc]} \BibitemShut
  {NoStop}%
\bibitem [{\citenamefont {Hinshaw}\ \emph {et~al.}(2013)\citenamefont {Hinshaw}
  \emph {et~al.}}]{Hinshaw:2012aka}%
  \BibitemOpen
  \bibfield  {author} {\bibinfo {author} {\bibfnamefont {G.}~\bibnamefont
  {Hinshaw}} \emph {et~al.} (\bibinfo {collaboration} {WMAP}),\ }\href
  {\doibase 10.1088/0067-0049/208/2/19} {\bibfield  {journal} {\bibinfo
  {journal} {ApJS}\ }\textbf {\bibinfo {volume} {208}},\ \bibinfo {pages} {19}
  (\bibinfo {year} {2013})}\BibitemShut {NoStop}%
\bibitem [{\citenamefont {Armendariz-Picon}\ \emph {et~al.}(2000)\citenamefont
  {Armendariz-Picon}, \citenamefont {Mukhanov},\ and\ \citenamefont
  {Steinhardt}}]{ArmendarizPicon:2000dh}%
  \BibitemOpen
  \bibfield  {author} {\bibinfo {author} {\bibfnamefont {C.}~\bibnamefont
  {Armendariz-Picon}}, \bibinfo {author} {\bibfnamefont {V.~F.}\ \bibnamefont
  {Mukhanov}}, \ and\ \bibinfo {author} {\bibfnamefont {P.~J.}\ \bibnamefont
  {Steinhardt}},\ }\href {\doibase 10.1103/PhysRevLett.85.4438} {\bibfield
  {journal} {\bibinfo  {journal} {Phys. Rev. Lett.}\ }\textbf {\bibinfo
  {volume} {85}},\ \bibinfo {pages} {4438} (\bibinfo {year}
  {2000})}\BibitemShut {NoStop}%
\bibitem [{\citenamefont {Abramo}\ \emph {et~al.}(2007)\citenamefont {Abramo},
  \citenamefont {Batista}, \citenamefont {Liberato},\ and\ \citenamefont
  {Rosenfeld}}]{Abramo2007}%
  \BibitemOpen
  \bibfield  {author} {\bibinfo {author} {\bibfnamefont {L.~R.}\ \bibnamefont
  {Abramo}}, \bibinfo {author} {\bibfnamefont {R.~C.}\ \bibnamefont {Batista}},
  \bibinfo {author} {\bibfnamefont {L.}~\bibnamefont {Liberato}}, \ and\
  \bibinfo {author} {\bibfnamefont {R.}~\bibnamefont {Rosenfeld}},\ }\href@noop
  {} {\bibfield  {journal} {\bibinfo  {journal} {JCAP}\ }\textbf {\bibinfo
  {volume} {11}},\ \bibinfo {pages} {12} (\bibinfo {year} {2007})}\BibitemShut
  {NoStop}%
\bibitem [{\citenamefont {Ballesteros}\ and\ \citenamefont
  {Riotto}(2008)}]{Ballesteros:2008qk}%
  \BibitemOpen
  \bibfield  {author} {\bibinfo {author} {\bibfnamefont {G.}~\bibnamefont
  {Ballesteros}}\ and\ \bibinfo {author} {\bibfnamefont {A.}~\bibnamefont
  {Riotto}},\ }\href {\doibase 10.1016/j.physletb.2008.08.035} {\bibfield
  {journal} {\bibinfo  {journal} {Phys. Lett. B}\ }\textbf {\bibinfo {volume}
  {668}},\ \bibinfo {pages} {171} (\bibinfo {year} {2008})}\BibitemShut
  {NoStop}%
\bibitem [{\citenamefont {Pace}\ \emph {et~al.}(2010)\citenamefont {Pace},
  \citenamefont {Waizmann},\ and\ \citenamefont {Bartelmann}}]{Pace2010}%
  \BibitemOpen
  \bibfield  {author} {\bibinfo {author} {\bibfnamefont {F.}~\bibnamefont
  {Pace}}, \bibinfo {author} {\bibfnamefont {J.~C.}\ \bibnamefont {Waizmann}},
  \ and\ \bibinfo {author} {\bibfnamefont {M.}~\bibnamefont {Bartelmann}},\
  }\href@noop {} {\bibfield  {journal} {\bibinfo  {journal} {\mnras}\ }\textbf
  {\bibinfo {volume} {406}},\ \bibinfo {pages} {1865} (\bibinfo {year}
  {2010})}\BibitemShut {NoStop}%
\bibitem [{\citenamefont {Basilakos}\ \emph
  {et~al.}(2010{\natexlab{a}})\citenamefont {Basilakos}, \citenamefont
  {Plionis},\ and\ \citenamefont {Sol{\`a}}}]{Basilakos2010}%
  \BibitemOpen
  \bibfield  {author} {\bibinfo {author} {\bibfnamefont {S.}~\bibnamefont
  {Basilakos}}, \bibinfo {author} {\bibfnamefont {M.}~\bibnamefont {Plionis}},
  \ and\ \bibinfo {author} {\bibfnamefont {J.}~\bibnamefont {Sol{\`a}}},\
  }\href@noop {} {\bibfield  {journal} {\bibinfo  {journal} {\prd}\ }\textbf
  {\bibinfo {volume} {82}},\ \bibinfo {pages} {083512} (\bibinfo {year}
  {2010}{\natexlab{a}})}\BibitemShut {NoStop}%
\bibitem [{\citenamefont {Basilakos}\ \emph
  {et~al.}(2010{\natexlab{b}})\citenamefont {Basilakos}, \citenamefont
  {Plionis},\ and\ \citenamefont {Sol\`a}}]{Basilakos:2010rs}%
  \BibitemOpen
  \bibfield  {author} {\bibinfo {author} {\bibfnamefont {S.}~\bibnamefont
  {Basilakos}}, \bibinfo {author} {\bibfnamefont {M.}~\bibnamefont {Plionis}},
  \ and\ \bibinfo {author} {\bibfnamefont {J.}~\bibnamefont {Sol\`a}},\ }\href
  {\doibase 10.1103/PhysRevD.82.083512} {\bibfield  {journal} {\bibinfo
  {journal} {Phys. Rev.}\ }\textbf {\bibinfo {volume} {D82}},\ \bibinfo {pages}
  {083512} (\bibinfo {year} {2010}{\natexlab{b}})},\ \Eprint
  {http://arxiv.org/abs/1005.5592} {arXiv:1005.5592 [astro-ph.CO]} \BibitemShut
  {NoStop}%
\bibitem [{\citenamefont {Batista}\ and\ \citenamefont
  {Pace}(2013)}]{Batista:2013oca}%
  \BibitemOpen
  \bibfield  {author} {\bibinfo {author} {\bibfnamefont {R.}~\bibnamefont
  {Batista}}\ and\ \bibinfo {author} {\bibfnamefont {F.}~\bibnamefont {Pace}},\
  }\href {\doibase 10.1088/1475-7516/2013/06/044} {\bibfield  {journal}
  {\bibinfo  {journal} {JCAP}\ }\textbf {\bibinfo {volume} {1306}},\ \bibinfo
  {pages} {044} (\bibinfo {year} {2013})}\BibitemShut {NoStop}%
\bibitem [{\citenamefont {Pace}\ \emph {et~al.}(2014)\citenamefont {Pace},
  \citenamefont {Batista},\ and\ \citenamefont {Del~Popolo}}]{Pace:2014taa}%
  \BibitemOpen
  \bibfield  {author} {\bibinfo {author} {\bibfnamefont {F.}~\bibnamefont
  {Pace}}, \bibinfo {author} {\bibfnamefont {R.~C.}\ \bibnamefont {Batista}}, \
  and\ \bibinfo {author} {\bibfnamefont {A.}~\bibnamefont {Del~Popolo}},\
  }\href {\doibase 10.1093/mnras/stu1782} {\bibfield  {journal} {\bibinfo
  {journal} {MNRAS}\ }\textbf {\bibinfo {volume} {445}},\ \bibinfo {pages}
  {648} (\bibinfo {year} {2014})}\BibitemShut {NoStop}%
\bibitem [{\citenamefont {Mehrabi}\ \emph
  {et~al.}(2015{\natexlab{b}})\citenamefont {Mehrabi}, \citenamefont
  {Basilakos},\ and\ \citenamefont {Pace}}]{Mehrabi:2015hva}%
  \BibitemOpen
  \bibfield  {author} {\bibinfo {author} {\bibfnamefont {A.}~\bibnamefont
  {Mehrabi}}, \bibinfo {author} {\bibfnamefont {S.}~\bibnamefont {Basilakos}},
  \ and\ \bibinfo {author} {\bibfnamefont {F.}~\bibnamefont {Pace}},\ }\href
  {\doibase 10.1093/mnras/stv1478} {\bibfield  {journal} {\bibinfo  {journal}
  {MNRAS}\ }\textbf {\bibinfo {volume} {452}},\ \bibinfo {pages} {2930}
  (\bibinfo {year} {2015}{\natexlab{b}})},\ \Eprint
  {http://arxiv.org/abs/1504.01262} {arXiv:1504.01262 [astro-ph.CO]}
  \BibitemShut {NoStop}%
\bibitem [{\citenamefont {Rezaei}\ and\ \citenamefont
  {Malekjani}(2017)}]{Rezaei:2017hon}%
  \BibitemOpen
  \bibfield  {author} {\bibinfo {author} {\bibfnamefont {M.}~\bibnamefont
  {Rezaei}}\ and\ \bibinfo {author} {\bibfnamefont {M.}~\bibnamefont
  {Malekjani}},\ }\href {\doibase 10.1103/PhysRevD.96.063519} {\bibfield
  {journal} {\bibinfo  {journal} {Phys. Rev. D}\ }\textbf {\bibinfo {volume}
  {96}},\ \bibinfo {pages} {063519} (\bibinfo {year} {2017})}\BibitemShut
  {NoStop}%
\bibitem [{\citenamefont {Rezaei}(2019{\natexlab{b}})}]{Rezaei:2019roz}%
  \BibitemOpen
  \bibfield  {author} {\bibinfo {author} {\bibfnamefont {M.}~\bibnamefont
  {Rezaei}},\ }\href {\doibase 10.1093/mnras/stz733} {\bibfield  {journal}
  {\bibinfo  {journal} {Mon. Not. Roy. Astron. Soc.}\ }\textbf {\bibinfo
  {volume} {485}},\ \bibinfo {pages} {4841} (\bibinfo {year}
  {2019}{\natexlab{b}})}\BibitemShut {NoStop}%
\bibitem [{\citenamefont {G\'omez-Valent}\ and\ \citenamefont
  {Sol\`a}(2018)}]{Gomez-Valent:2018nib}%
  \BibitemOpen
  \bibfield  {author} {\bibinfo {author} {\bibfnamefont {A.}~\bibnamefont
  {G\'omez-Valent}}\ and\ \bibinfo {author} {\bibfnamefont {J.}~\bibnamefont
  {Sol\`a}},\ }\href {\doibase 10.1093/mnras/sty1028} {\bibfield  {journal}
  {\bibinfo  {journal} {Mon. Not. Roy. Astron. Soc.}\ }\textbf {\bibinfo
  {volume} {478}},\ \bibinfo {pages} {126} (\bibinfo {year} {2018})},\ \Eprint
  {http://arxiv.org/abs/1801.08501} {arXiv:1801.08501 [astro-ph.CO]}
  \BibitemShut {NoStop}%
\bibitem [{\citenamefont {Bardeen}\ \emph {et~al.}(1986)\citenamefont
  {Bardeen}, \citenamefont {Bond}, \citenamefont {Kaiser},\ and\ \citenamefont
  {Szalay}}]{Bardeen:1985tr}%
  \BibitemOpen
  \bibfield  {author} {\bibinfo {author} {\bibfnamefont {J.~M.}\ \bibnamefont
  {Bardeen}}, \bibinfo {author} {\bibfnamefont {J.~R.}\ \bibnamefont {Bond}},
  \bibinfo {author} {\bibfnamefont {N.}~\bibnamefont {Kaiser}}, \ and\ \bibinfo
  {author} {\bibfnamefont {A.~S.}\ \bibnamefont {Szalay}},\ }\href {\doibase
  10.1086/164143} {\bibfield  {journal} {\bibinfo  {journal} {Astrophys. J.}\
  }\textbf {\bibinfo {volume} {304}},\ \bibinfo {pages} {15} (\bibinfo {year}
  {1986})}\BibitemShut {NoStop}%
\bibitem [{\citenamefont {Liddle}\ and\ \citenamefont
  {Lyth}(2000)}]{Liddle:2000cg}%
  \BibitemOpen
  \bibfield  {author} {\bibinfo {author} {\bibfnamefont {A.~R.}\ \bibnamefont
  {Liddle}}\ and\ \bibinfo {author} {\bibfnamefont {D.~H.}\ \bibnamefont
  {Lyth}},\ }\href@noop {} {\emph {\bibinfo {title} {{Cosmological inflation
  and large scale structure}}}}\ (\bibinfo {year} {2000})\BibitemShut {NoStop}%
\bibitem [{\citenamefont {Scolnic}\ \emph {et~al.}(2018)\citenamefont {Scolnic}
  \emph {et~al.}}]{Scolnic:2017caz}%
  \BibitemOpen
  \bibfield  {author} {\bibinfo {author} {\bibfnamefont {D.~M.}\ \bibnamefont
  {Scolnic}} \emph {et~al.},\ }\href {\doibase 10.3847/1538-4357/aab9bb}
  {\bibfield  {journal} {\bibinfo  {journal} {Astrophys. J.}\ }\textbf
  {\bibinfo {volume} {859}},\ \bibinfo {pages} {101} (\bibinfo {year}
  {2018})},\ \Eprint {http://arxiv.org/abs/1710.00845} {arXiv:1710.00845
  [astro-ph.CO]} \BibitemShut {NoStop}%
\bibitem [{\citenamefont {Bond}\ \emph {et~al.}(1997)\citenamefont {Bond},
  \citenamefont {Efstathiou},\ and\ \citenamefont {Tegmark}}]{Bond:1997wr}%
  \BibitemOpen
  \bibfield  {author} {\bibinfo {author} {\bibfnamefont {J.~R.}\ \bibnamefont
  {Bond}}, \bibinfo {author} {\bibfnamefont {G.}~\bibnamefont {Efstathiou}}, \
  and\ \bibinfo {author} {\bibfnamefont {M.}~\bibnamefont {Tegmark}},\ }\href
  {\doibase 10.1093/mnras/291.1.L33} {\bibfield  {journal} {\bibinfo  {journal}
  {Mon. Not. Roy. Astron. Soc.}\ }\textbf {\bibinfo {volume} {291}},\ \bibinfo
  {pages} {L33} (\bibinfo {year} {1997})},\ \Eprint
  {http://arxiv.org/abs/astro-ph/9702100} {arXiv:astro-ph/9702100 [astro-ph]}
  \BibitemShut {NoStop}%
\bibitem [{\citenamefont {Efstathiou}\ and\ \citenamefont
  {Bond}(1999)}]{Efstathiou:1998xx}%
  \BibitemOpen
  \bibfield  {author} {\bibinfo {author} {\bibfnamefont {G.}~\bibnamefont
  {Efstathiou}}\ and\ \bibinfo {author} {\bibfnamefont {J.~R.}\ \bibnamefont
  {Bond}},\ }\href {\doibase 10.1046/j.1365-8711.1999.02274.x} {\bibfield
  {journal} {\bibinfo  {journal} {Mon. Not. Roy. Astron. Soc.}\ }\textbf
  {\bibinfo {volume} {304}},\ \bibinfo {pages} {75} (\bibinfo {year} {1999})},\
  \Eprint {http://arxiv.org/abs/astro-ph/9807103} {arXiv:astro-ph/9807103
  [astro-ph]} \BibitemShut {NoStop}%
\bibitem [{\citenamefont {Wang}\ and\ \citenamefont
  {Mukherjee}(2007)}]{Wang:2007mza}%
  \BibitemOpen
  \bibfield  {author} {\bibinfo {author} {\bibfnamefont {Y.}~\bibnamefont
  {Wang}}\ and\ \bibinfo {author} {\bibfnamefont {P.}~\bibnamefont
  {Mukherjee}},\ }\href {\doibase 10.1103/PhysRevD.76.103533} {\bibfield
  {journal} {\bibinfo  {journal} {Phys. Rev.}\ }\textbf {\bibinfo {volume}
  {D76}},\ \bibinfo {pages} {103533} (\bibinfo {year} {2007})},\ \Eprint
  {http://arxiv.org/abs/astro-ph/0703780} {arXiv:astro-ph/0703780 [astro-ph]}
  \BibitemShut {NoStop}%
\bibitem [{\citenamefont {Chen}\ \emph {et~al.}(2018)\citenamefont {Chen},
  \citenamefont {Huang},\ and\ \citenamefont {Wang}}]{Chen:2018dbv}%
  \BibitemOpen
  \bibfield  {author} {\bibinfo {author} {\bibfnamefont {L.}~\bibnamefont
  {Chen}}, \bibinfo {author} {\bibfnamefont {Q.-G.}\ \bibnamefont {Huang}}, \
  and\ \bibinfo {author} {\bibfnamefont {K.}~\bibnamefont {Wang}},\ }\href@noop
  {} {\  (\bibinfo {year} {2018})},\ \Eprint {http://arxiv.org/abs/1808.05724}
  {arXiv:1808.05724 [astro-ph.CO]} \BibitemShut {NoStop}%
\bibitem [{\citenamefont {Shafer}\ and\ \citenamefont
  {Huterer}(2014)}]{Shafer:2013pxa}%
  \BibitemOpen
  \bibfield  {author} {\bibinfo {author} {\bibfnamefont {D.~L.}\ \bibnamefont
  {Shafer}}\ and\ \bibinfo {author} {\bibfnamefont {D.}~\bibnamefont
  {Huterer}},\ }\href {\doibase 10.1103/PhysRevD.89.063510} {\bibfield
  {journal} {\bibinfo  {journal} {Phys. Rev.}\ }\textbf {\bibinfo {volume}
  {D89}},\ \bibinfo {pages} {063510} (\bibinfo {year} {2014})},\ \Eprint
  {http://arxiv.org/abs/1312.1688} {arXiv:1312.1688 [astro-ph.CO]} \BibitemShut
  {NoStop}%
\bibitem [{\citenamefont {Beutler}\ \emph {et~al.}(2011)\citenamefont
  {Beutler}, \citenamefont {Blake}, \citenamefont {Colless}, \citenamefont
  {Jones}, \citenamefont {Staveley-Smith} \emph {et~al.}}]{Beutler:2011hx}%
  \BibitemOpen
  \bibfield  {author} {\bibinfo {author} {\bibfnamefont {F.}~\bibnamefont
  {Beutler}}, \bibinfo {author} {\bibfnamefont {C.}~\bibnamefont {Blake}},
  \bibinfo {author} {\bibfnamefont {M.}~\bibnamefont {Colless}}, \bibinfo
  {author} {\bibfnamefont {D.~H.}\ \bibnamefont {Jones}}, \bibinfo {author}
  {\bibfnamefont {L.}~\bibnamefont {Staveley-Smith}},  \emph {et~al.},\ }\href
  {\doibase 10.1111/j.1365-2966.2011.19250.x} {\bibfield  {journal} {\bibinfo
  {journal} {MNRAS}\ }\textbf {\bibinfo {volume} {416}},\ \bibinfo {pages}
  {3017} (\bibinfo {year} {2011})}\BibitemShut {NoStop}%
\bibitem [{\citenamefont {Font-Ribera}\ \emph {et~al.}(2014)\citenamefont
  {Font-Ribera} \emph {et~al.}}]{Font-Ribera:2013wce}%
  \BibitemOpen
  \bibfield  {author} {\bibinfo {author} {\bibfnamefont {A.}~\bibnamefont
  {Font-Ribera}} \emph {et~al.} (\bibinfo {collaboration} {BOSS}),\ }\href
  {\doibase 10.1088/1475-7516/2014/05/027} {\bibfield  {journal} {\bibinfo
  {journal} {JCAP}\ }\textbf {\bibinfo {volume} {1405}},\ \bibinfo {pages}
  {027} (\bibinfo {year} {2014})},\ \Eprint {http://arxiv.org/abs/1311.1767}
  {arXiv:1311.1767 [astro-ph.CO]} \BibitemShut {NoStop}%
\bibitem [{\citenamefont {Ross}\ \emph {et~al.}(2015)\citenamefont {Ross},
  \citenamefont {Samushia}, \citenamefont {Howlett}, \citenamefont {Percival},
  \citenamefont {Burden},\ and\ \citenamefont {Manera}}]{Ross:2014qpa}%
  \BibitemOpen
  \bibfield  {author} {\bibinfo {author} {\bibfnamefont {A.~J.}\ \bibnamefont
  {Ross}}, \bibinfo {author} {\bibfnamefont {L.}~\bibnamefont {Samushia}},
  \bibinfo {author} {\bibfnamefont {C.}~\bibnamefont {Howlett}}, \bibinfo
  {author} {\bibfnamefont {W.~J.}\ \bibnamefont {Percival}}, \bibinfo {author}
  {\bibfnamefont {A.}~\bibnamefont {Burden}}, \ and\ \bibinfo {author}
  {\bibfnamefont {M.}~\bibnamefont {Manera}},\ }\href {\doibase
  10.1093/mnras/stv154} {\bibfield  {journal} {\bibinfo  {journal} {Mon. Not.
  Roy. Astron. Soc.}\ }\textbf {\bibinfo {volume} {449}},\ \bibinfo {pages}
  {835} (\bibinfo {year} {2015})},\ \Eprint {http://arxiv.org/abs/1409.3242}
  {arXiv:1409.3242 [astro-ph.CO]} \BibitemShut {NoStop}%
\bibitem [{\citenamefont {Alam}\ \emph {et~al.}(2017)\citenamefont {Alam} \emph
  {et~al.}}]{Alam:2016hwk}%
  \BibitemOpen
  \bibfield  {author} {\bibinfo {author} {\bibfnamefont {S.}~\bibnamefont
  {Alam}} \emph {et~al.} (\bibinfo {collaboration} {BOSS}),\ }\href {\doibase
  10.1093/mnras/stx721} {\bibfield  {journal} {\bibinfo  {journal} {Mon. Not.
  Roy. Astron. Soc.}\ }\textbf {\bibinfo {volume} {470}},\ \bibinfo {pages}
  {2617} (\bibinfo {year} {2017})},\ \Eprint {http://arxiv.org/abs/1607.03155}
  {arXiv:1607.03155 [astro-ph.CO]} \BibitemShut {NoStop}%
\bibitem [{\citenamefont {Bautista}\ \emph {et~al.}(2017)\citenamefont
  {Bautista} \emph {et~al.}}]{Bautista:2017zgn}%
  \BibitemOpen
  \bibfield  {author} {\bibinfo {author} {\bibfnamefont {J.~E.}\ \bibnamefont
  {Bautista}} \emph {et~al.},\ }\href {\doibase 10.1051/0004-6361/201730533}
  {\bibfield  {journal} {\bibinfo  {journal} {Astron. Astrophys.}\ }\textbf
  {\bibinfo {volume} {603}},\ \bibinfo {pages} {A12} (\bibinfo {year}
  {2017})},\ \Eprint {http://arxiv.org/abs/1702.00176} {arXiv:1702.00176
  [astro-ph.CO]} \BibitemShut {NoStop}%
\bibitem [{\citenamefont {Ata}\ \emph {et~al.}(2018)\citenamefont {Ata} \emph
  {et~al.}}]{Ata:2017dya}%
  \BibitemOpen
  \bibfield  {author} {\bibinfo {author} {\bibfnamefont {M.}~\bibnamefont
  {Ata}} \emph {et~al.},\ }\href {\doibase 10.1093/mnras/stx2630} {\bibfield
  {journal} {\bibinfo  {journal} {Mon. Not. Roy. Astron. Soc.}\ }\textbf
  {\bibinfo {volume} {473}},\ \bibinfo {pages} {4773} (\bibinfo {year}
  {2018})},\ \Eprint {http://arxiv.org/abs/1705.06373} {arXiv:1705.06373
  [astro-ph.CO]} \BibitemShut {NoStop}%
\bibitem [{\citenamefont {Park}\ and\ \citenamefont
  {Ratra}(2018{\natexlab{a}})}]{Park:2018tgj}%
  \BibitemOpen
  \bibfield  {author} {\bibinfo {author} {\bibfnamefont {C.-G.}\ \bibnamefont
  {Park}}\ and\ \bibinfo {author} {\bibfnamefont {B.}~\bibnamefont {Ratra}},\
  }\href@noop {} {\  (\bibinfo {year} {2018}{\natexlab{a}})},\ \Eprint
  {http://arxiv.org/abs/1809.03598} {arXiv:1809.03598 [astro-ph.CO]}
  \BibitemShut {NoStop}%
\bibitem [{\citenamefont {Serra}\ \emph {et~al.}(2009)\citenamefont {Serra},
  \citenamefont {Cooray}, \citenamefont {Holz}, \citenamefont {Melchiorri},
  \citenamefont {Pandolfi},\ and\ \citenamefont {Sarkar}}]{Serra:2009yp}%
  \BibitemOpen
  \bibfield  {author} {\bibinfo {author} {\bibfnamefont {P.}~\bibnamefont
  {Serra}}, \bibinfo {author} {\bibfnamefont {A.}~\bibnamefont {Cooray}},
  \bibinfo {author} {\bibfnamefont {D.~E.}\ \bibnamefont {Holz}}, \bibinfo
  {author} {\bibfnamefont {A.}~\bibnamefont {Melchiorri}}, \bibinfo {author}
  {\bibfnamefont {S.}~\bibnamefont {Pandolfi}}, \ and\ \bibinfo {author}
  {\bibfnamefont {D.}~\bibnamefont {Sarkar}},\ }\href {\doibase
  10.1103/PhysRevD.80.121302} {\bibfield  {journal} {\bibinfo  {journal} {Phys.
  Rev. D}\ }\textbf {\bibinfo {volume} {80}},\ \bibinfo {pages} {121302}
  (\bibinfo {year} {2009})}\BibitemShut {NoStop}%
\bibitem [{\citenamefont {Burles}\ \emph {et~al.}(2001)\citenamefont {Burles},
  \citenamefont {Nollett},\ and\ \citenamefont {Turner}}]{Burles:2000zk}%
  \BibitemOpen
  \bibfield  {author} {\bibinfo {author} {\bibfnamefont {S.}~\bibnamefont
  {Burles}}, \bibinfo {author} {\bibfnamefont {K.~M.}\ \bibnamefont {Nollett}},
  \ and\ \bibinfo {author} {\bibfnamefont {M.~S.}\ \bibnamefont {Turner}},\
  }\href {\doibase 10.1086/320251} {\bibfield  {journal} {\bibinfo  {journal}
  {ApJ}\ }\textbf {\bibinfo {volume} {552}},\ \bibinfo {pages} {L1} (\bibinfo
  {year} {2001})}\BibitemShut {NoStop}%
\bibitem [{\citenamefont {Farooq}\ and\ \citenamefont
  {Ratra}(2013)}]{Farooq:2013hq}%
  \BibitemOpen
  \bibfield  {author} {\bibinfo {author} {\bibfnamefont {O.}~\bibnamefont
  {Farooq}}\ and\ \bibinfo {author} {\bibfnamefont {B.}~\bibnamefont {Ratra}},\
  }\href {\doibase 10.1088/2041-8205/766/1/L7} {\bibfield  {journal} {\bibinfo
  {journal} {Astrophys. J.}\ }\textbf {\bibinfo {volume} {766}},\ \bibinfo
  {pages} {L7} (\bibinfo {year} {2013})},\ \Eprint
  {http://arxiv.org/abs/1301.5243} {arXiv:1301.5243 [astro-ph.CO]} \BibitemShut
  {NoStop}%
\bibitem [{\citenamefont {Blake}\ \emph {et~al.}(2012)\citenamefont {Blake}
  \emph {et~al.}}]{Blake:2012pj}%
  \BibitemOpen
  \bibfield  {author} {\bibinfo {author} {\bibfnamefont {C.}~\bibnamefont
  {Blake}} \emph {et~al.},\ }\href {\doibase 10.1111/j.1365-2966.2012.21473.x}
  {\bibfield  {journal} {\bibinfo  {journal} {Mon. Not. Roy. Astron. Soc.}\
  }\textbf {\bibinfo {volume} {425}},\ \bibinfo {pages} {405} (\bibinfo {year}
  {2012})},\ \Eprint {http://arxiv.org/abs/1204.3674} {arXiv:1204.3674
  [astro-ph.CO]} \BibitemShut {NoStop}%
\bibitem [{\citenamefont {Chuang}\ and\ \citenamefont
  {Wang}(2012)}]{Chuang:2011fy}%
  \BibitemOpen
  \bibfield  {author} {\bibinfo {author} {\bibfnamefont {C.-H.}\ \bibnamefont
  {Chuang}}\ and\ \bibinfo {author} {\bibfnamefont {Y.}~\bibnamefont {Wang}},\
  }\href {\doibase 10.1111/j.1365-2966.2012.21565.x} {\bibfield  {journal}
  {\bibinfo  {journal} {Mon. Not. Roy. Astron. Soc.}\ }\textbf {\bibinfo
  {volume} {426}},\ \bibinfo {pages} {226} (\bibinfo {year} {2012})},\ \Eprint
  {http://arxiv.org/abs/1102.2251} {arXiv:1102.2251 [astro-ph.CO]} \BibitemShut
  {NoStop}%
\bibitem [{\citenamefont {Busca}\ \emph {et~al.}(2013)\citenamefont {Busca}
  \emph {et~al.}}]{Busca:2012bu}%
  \BibitemOpen
  \bibfield  {author} {\bibinfo {author} {\bibfnamefont {N.~G.}\ \bibnamefont
  {Busca}} \emph {et~al.},\ }\href {\doibase 10.1051/0004-6361/201220724}
  {\bibfield  {journal} {\bibinfo  {journal} {Astron. Astrophys.}\ }\textbf
  {\bibinfo {volume} {552}},\ \bibinfo {pages} {A96} (\bibinfo {year}
  {2013})},\ \Eprint {http://arxiv.org/abs/1211.2616} {arXiv:1211.2616
  [astro-ph.CO]} \BibitemShut {NoStop}%
\bibitem [{\citenamefont {Nesseris}\ \emph {et~al.}(2017)\citenamefont
  {Nesseris}, \citenamefont {Pantazis},\ and\ \citenamefont
  {Perivolaropoulos}}]{Nesseris:2017vor}%
  \BibitemOpen
  \bibfield  {author} {\bibinfo {author} {\bibfnamefont {S.}~\bibnamefont
  {Nesseris}}, \bibinfo {author} {\bibfnamefont {G.}~\bibnamefont {Pantazis}},
  \ and\ \bibinfo {author} {\bibfnamefont {L.}~\bibnamefont
  {Perivolaropoulos}},\ }\href {\doibase 10.1103/PhysRevD.96.023542} {\bibfield
   {journal} {\bibinfo  {journal} {Phys. Rev.}\ }\textbf {\bibinfo {volume}
  {D96}},\ \bibinfo {pages} {023542} (\bibinfo {year} {2017})},\ \Eprint
  {http://arxiv.org/abs/1703.10538} {arXiv:1703.10538 [astro-ph.CO]}
  \BibitemShut {NoStop}%
\bibitem [{\citenamefont {Song}\ and\ \citenamefont
  {Percival}(2009)}]{Song:2008qt}%
  \BibitemOpen
  \bibfield  {author} {\bibinfo {author} {\bibfnamefont {Y.-S.}\ \bibnamefont
  {Song}}\ and\ \bibinfo {author} {\bibfnamefont {W.~J.}\ \bibnamefont
  {Percival}},\ }\href {\doibase 10.1088/1475-7516/2009/10/004} {\bibfield
  {journal} {\bibinfo  {journal} {JCAP}\ }\textbf {\bibinfo {volume} {0910}},\
  \bibinfo {pages} {004} (\bibinfo {year} {2009})}\BibitemShut {NoStop}%
\bibitem [{\citenamefont {Samushia}\ \emph {et~al.}(2012)\citenamefont
  {Samushia}, \citenamefont {Percival},\ and\ \citenamefont
  {Raccanelli}}]{Samushia:2011cs}%
  \BibitemOpen
  \bibfield  {author} {\bibinfo {author} {\bibfnamefont {L.}~\bibnamefont
  {Samushia}}, \bibinfo {author} {\bibfnamefont {W.~J.}\ \bibnamefont
  {Percival}}, \ and\ \bibinfo {author} {\bibfnamefont {A.}~\bibnamefont
  {Raccanelli}},\ }\href {\doibase 10.1111/j.1365-2966.2011.20169.x} {\bibfield
   {journal} {\bibinfo  {journal} {MNRAS}\ }\textbf {\bibinfo {volume} {420}},\
  \bibinfo {pages} {2102} (\bibinfo {year} {2012})}\BibitemShut {NoStop}%
\bibitem [{\citenamefont {Hudson}\ and\ \citenamefont
  {Turnbull}(2013)}]{Hudson:2012gt}%
  \BibitemOpen
  \bibfield  {author} {\bibinfo {author} {\bibfnamefont {M.~J.}\ \bibnamefont
  {Hudson}}\ and\ \bibinfo {author} {\bibfnamefont {S.~J.}\ \bibnamefont
  {Turnbull}},\ }\href {\doibase 10.1088/2041-8205/751/2/L30} {\bibfield
  {journal} {\bibinfo  {journal} {ApJ}\ }\textbf {\bibinfo {volume} {751}},\
  \bibinfo {pages} {L30} (\bibinfo {year} {2013})}\BibitemShut {NoStop}%
\bibitem [{\citenamefont {Blake}\ \emph {et~al.}(2013)\citenamefont {Blake}
  \emph {et~al.}}]{Blake:2013nif}%
  \BibitemOpen
  \bibfield  {author} {\bibinfo {author} {\bibfnamefont {C.}~\bibnamefont
  {Blake}} \emph {et~al.},\ }\href {\doibase 10.1093/mnras/stt1791} {\bibfield
  {journal} {\bibinfo  {journal} {Mon. Not. Roy. Astron. Soc.}\ }\textbf
  {\bibinfo {volume} {436}},\ \bibinfo {pages} {3089} (\bibinfo {year}
  {2013})},\ \Eprint {http://arxiv.org/abs/1309.5556} {arXiv:1309.5556
  [astro-ph.CO]} \BibitemShut {NoStop}%
\bibitem [{\citenamefont {Sanchez}\ \emph {et~al.}(2014)\citenamefont {Sanchez}
  \emph {et~al.}}]{Sanchez:2013tga}%
  \BibitemOpen
  \bibfield  {author} {\bibinfo {author} {\bibfnamefont {A.~G.}\ \bibnamefont
  {Sanchez}} \emph {et~al.},\ }\href {\doibase 10.1093/mnras/stu342} {\bibfield
   {journal} {\bibinfo  {journal} {Mon. Not. Roy. Astron. Soc.}\ }\textbf
  {\bibinfo {volume} {440}},\ \bibinfo {pages} {2692} (\bibinfo {year}
  {2014})},\ \Eprint {http://arxiv.org/abs/1312.4854} {arXiv:1312.4854
  [astro-ph.CO]} \BibitemShut {NoStop}%
\bibitem [{\citenamefont {Chuang}\ \emph {et~al.}(2016)\citenamefont {Chuang}
  \emph {et~al.}}]{Chuang:2013wga}%
  \BibitemOpen
  \bibfield  {author} {\bibinfo {author} {\bibfnamefont {C.-H.}\ \bibnamefont
  {Chuang}} \emph {et~al.},\ }\href {\doibase 10.1093/mnras/stw1535} {\bibfield
   {journal} {\bibinfo  {journal} {Mon. Not. Roy. Astron. Soc.}\ }\textbf
  {\bibinfo {volume} {461}},\ \bibinfo {pages} {3781} (\bibinfo {year}
  {2016})},\ \Eprint {http://arxiv.org/abs/1312.4889} {arXiv:1312.4889
  [astro-ph.CO]} \BibitemShut {NoStop}%
\bibitem [{\citenamefont {Howlett}\ \emph {et~al.}(2015)\citenamefont
  {Howlett}, \citenamefont {Ross}, \citenamefont {Samushia}, \citenamefont
  {Percival},\ and\ \citenamefont {Manera}}]{Howlett:2014opa}%
  \BibitemOpen
  \bibfield  {author} {\bibinfo {author} {\bibfnamefont {C.}~\bibnamefont
  {Howlett}}, \bibinfo {author} {\bibfnamefont {A.}~\bibnamefont {Ross}},
  \bibinfo {author} {\bibfnamefont {L.}~\bibnamefont {Samushia}}, \bibinfo
  {author} {\bibfnamefont {W.}~\bibnamefont {Percival}}, \ and\ \bibinfo
  {author} {\bibfnamefont {M.}~\bibnamefont {Manera}},\ }\href {\doibase
  10.1093/mnras/stu2693} {\bibfield  {journal} {\bibinfo  {journal} {Mon. Not.
  Roy. Astron. Soc.}\ }\textbf {\bibinfo {volume} {449}},\ \bibinfo {pages}
  {848} (\bibinfo {year} {2015})},\ \Eprint {http://arxiv.org/abs/1409.3238}
  {arXiv:1409.3238 [astro-ph.CO]} \BibitemShut {NoStop}%
\bibitem [{\citenamefont {Feix}\ \emph {et~al.}(2015)\citenamefont {Feix},
  \citenamefont {Nusser},\ and\ \citenamefont {Branchini}}]{Feix:2015dla}%
  \BibitemOpen
  \bibfield  {author} {\bibinfo {author} {\bibfnamefont {M.}~\bibnamefont
  {Feix}}, \bibinfo {author} {\bibfnamefont {A.}~\bibnamefont {Nusser}}, \ and\
  \bibinfo {author} {\bibfnamefont {E.}~\bibnamefont {Branchini}},\ }\href@noop
  {} {\bibfield  {journal} {\bibinfo  {journal} {ArXiv e-prints, 1503.05945}\ }
  (\bibinfo {year} {2015})}\BibitemShut {NoStop}%
\bibitem [{\citenamefont {Okumura}\ \emph {et~al.}(2016)\citenamefont {Okumura}
  \emph {et~al.}}]{Okumura:2015lvp}%
  \BibitemOpen
  \bibfield  {author} {\bibinfo {author} {\bibfnamefont {T.}~\bibnamefont
  {Okumura}} \emph {et~al.},\ }\href {\doibase 10.1093/pasj/psw029} {\bibfield
  {journal} {\bibinfo  {journal} {Publ. Astron. Soc. Jap.}\ }\textbf {\bibinfo
  {volume} {68}},\ \bibinfo {pages} {38} (\bibinfo {year} {2016})},\ \Eprint
  {http://arxiv.org/abs/1511.08083} {arXiv:1511.08083 [astro-ph.CO]}
  \BibitemShut {NoStop}%
\bibitem [{\citenamefont {Pezzotta}\ \emph {et~al.}(2017)\citenamefont
  {Pezzotta} \emph {et~al.}}]{Pezzotta:2016gbo}%
  \BibitemOpen
  \bibfield  {author} {\bibinfo {author} {\bibfnamefont {A.}~\bibnamefont
  {Pezzotta}} \emph {et~al.},\ }\href {\doibase 10.1051/0004-6361/201630295}
  {\bibfield  {journal} {\bibinfo  {journal} {Astron. Astrophys.}\ }\textbf
  {\bibinfo {volume} {604}},\ \bibinfo {pages} {A33} (\bibinfo {year}
  {2017})},\ \Eprint {http://arxiv.org/abs/1612.05645} {arXiv:1612.05645
  [astro-ph.CO]} \BibitemShut {NoStop}%
\bibitem [{\citenamefont {Huterer}\ \emph {et~al.}(2017)\citenamefont
  {Huterer}, \citenamefont {Shafer}, \citenamefont {Scolnic},\ and\
  \citenamefont {Schmidt}}]{Huterer:2016uyq}%
  \BibitemOpen
  \bibfield  {author} {\bibinfo {author} {\bibfnamefont {D.}~\bibnamefont
  {Huterer}}, \bibinfo {author} {\bibfnamefont {D.}~\bibnamefont {Shafer}},
  \bibinfo {author} {\bibfnamefont {D.}~\bibnamefont {Scolnic}}, \ and\
  \bibinfo {author} {\bibfnamefont {F.}~\bibnamefont {Schmidt}},\ }\href
  {\doibase 10.1088/1475-7516/2017/05/015} {\bibfield  {journal} {\bibinfo
  {journal} {JCAP}\ }\textbf {\bibinfo {volume} {1705}},\ \bibinfo {pages}
  {015} (\bibinfo {year} {2017})},\ \Eprint {http://arxiv.org/abs/1611.09862}
  {arXiv:1611.09862 [astro-ph.CO]} \BibitemShut {NoStop}%
\bibitem [{\citenamefont {Park}\ and\ \citenamefont
  {Ratra}(2018{\natexlab{b}})}]{Park:2018fxx}%
  \BibitemOpen
  \bibfield  {author} {\bibinfo {author} {\bibfnamefont {C.-G.}\ \bibnamefont
  {Park}}\ and\ \bibinfo {author} {\bibfnamefont {B.}~\bibnamefont {Ratra}},\
  }\href {\doibase 10.3847/1538-4357/aae82d} {\bibfield  {journal} {\bibinfo
  {journal} {Astrophys. J.}\ }\textbf {\bibinfo {volume} {868}},\ \bibinfo
  {pages} {83} (\bibinfo {year} {2018}{\natexlab{b}})},\ \Eprint
  {http://arxiv.org/abs/1807.07421} {arXiv:1807.07421 [astro-ph.CO]}
  \BibitemShut {NoStop}%
\bibitem [{\citenamefont {Martinelli}\ \emph {et~al.}(2019)\citenamefont
  {Martinelli}, \citenamefont {Hogg}, \citenamefont {Peirone}, \citenamefont
  {Bruni},\ and\ \citenamefont {Wands}}]{Martinelli:2019dau}%
  \BibitemOpen
  \bibfield  {author} {\bibinfo {author} {\bibfnamefont {M.}~\bibnamefont
  {Martinelli}}, \bibinfo {author} {\bibfnamefont {N.~B.}\ \bibnamefont
  {Hogg}}, \bibinfo {author} {\bibfnamefont {S.}~\bibnamefont {Peirone}},
  \bibinfo {author} {\bibfnamefont {M.}~\bibnamefont {Bruni}}, \ and\ \bibinfo
  {author} {\bibfnamefont {D.}~\bibnamefont {Wands}},\ }\href@noop {} {\
  (\bibinfo {year} {2019})},\ \Eprint {http://arxiv.org/abs/1902.10694}
  {arXiv:1902.10694 [astro-ph.CO]} \BibitemShut {NoStop}%
\bibitem [{\citenamefont {Capozziello}\ \emph {et~al.}(2011)\citenamefont
  {Capozziello}, \citenamefont {Lazkoz},\ and\ \citenamefont
  {Salzano}}]{Capozziello:2011tj}%
  \BibitemOpen
  \bibfield  {author} {\bibinfo {author} {\bibfnamefont {S.}~\bibnamefont
  {Capozziello}}, \bibinfo {author} {\bibfnamefont {R.}~\bibnamefont {Lazkoz}},
  \ and\ \bibinfo {author} {\bibfnamefont {V.}~\bibnamefont {Salzano}},\ }\href
  {\doibase 10.1103/PhysRevD.84.124061} {\bibfield  {journal} {\bibinfo
  {journal} {Phys. Rev.}\ }\textbf {\bibinfo {volume} {D84}},\ \bibinfo {pages}
  {124061} (\bibinfo {year} {2011})},\ \Eprint {http://arxiv.org/abs/1104.3096}
  {arXiv:1104.3096 [astro-ph.CO]} \BibitemShut {NoStop}%
\bibitem [{\citenamefont {Kass}\ and\ \citenamefont
  {Raftery}(1995)}]{Kass:1995loi}%
  \BibitemOpen
  \bibfield  {author} {\bibinfo {author} {\bibfnamefont {R.~E.}\ \bibnamefont
  {Kass}}\ and\ \bibinfo {author} {\bibfnamefont {A.~E.}\ \bibnamefont
  {Raftery}},\ }\href {\doibase 10.1080/01621459.1995.10476572} {\bibfield
  {journal} {\bibinfo  {journal} {J. Am. Statist. Assoc.}\ }\textbf {\bibinfo
  {volume} {90}},\ \bibinfo {pages} {773} (\bibinfo {year} {1995})}\BibitemShut
  {NoStop}%
\bibitem [{\citenamefont {Burnham}\ and\ \citenamefont
  {Anderson}(2002)}]{Burnham}%
  \BibitemOpen
  \bibfield  {author} {\bibinfo {author} {\bibfnamefont {K.~P.}\ \bibnamefont
  {Burnham}}\ and\ \bibinfo {author} {\bibfnamefont {D.~R.}\ \bibnamefont
  {Anderson}},\ }\href@noop {} {\emph {\bibinfo {title} {Pade Approximants}}}\
  (\bibinfo  {publisher} {Springer-Verlag New York},\ \bibinfo {year}
  {2002})\BibitemShut {NoStop}%
\bibitem [{\citenamefont {Farooq}\ \emph {et~al.}(2017)\citenamefont {Farooq},
  \citenamefont {Madiyar}, \citenamefont {Crandall},\ and\ \citenamefont
  {Ratra}}]{Farooq:2016zwm}%
  \BibitemOpen
  \bibfield  {author} {\bibinfo {author} {\bibfnamefont {O.}~\bibnamefont
  {Farooq}}, \bibinfo {author} {\bibfnamefont {F.~R.}\ \bibnamefont {Madiyar}},
  \bibinfo {author} {\bibfnamefont {S.}~\bibnamefont {Crandall}}, \ and\
  \bibinfo {author} {\bibfnamefont {B.}~\bibnamefont {Ratra}},\ }\href
  {\doibase 10.3847/1538-4357/835/1/26} {\bibfield  {journal} {\bibinfo
  {journal} {Astrophys. J.}\ }\textbf {\bibinfo {volume} {835}},\ \bibinfo
  {pages} {26} (\bibinfo {year} {2017})},\ \Eprint
  {http://arxiv.org/abs/1607.03537} {arXiv:1607.03537 [astro-ph.CO]}
  \BibitemShut {NoStop}%
\bibitem [{\citenamefont {Taddei}\ and\ \citenamefont
  {Amendola}(2015)}]{Taddei:2014wqa}%
  \BibitemOpen
  \bibfield  {author} {\bibinfo {author} {\bibfnamefont {L.}~\bibnamefont
  {Taddei}}\ and\ \bibinfo {author} {\bibfnamefont {L.}~\bibnamefont
  {Amendola}},\ }\href {\doibase 10.1088/1475-7516/2015/02/001} {\bibfield
  {journal} {\bibinfo  {journal} {JCAP}\ }\textbf {\bibinfo {volume} {1502}},\
  \bibinfo {pages} {001} (\bibinfo {year} {2015})},\ \Eprint
  {http://arxiv.org/abs/1408.3520} {arXiv:1408.3520 [astro-ph.CO]} \BibitemShut
  {NoStop}%
\bibitem [{\citenamefont {Riess}\ \emph {et~al.}(2019)\citenamefont {Riess},
  \citenamefont {Casertano}, \citenamefont {Yuan}, \citenamefont {Macri},\ and\
  \citenamefont {Scolnic}}]{Riess:2019cxk}%
  \BibitemOpen
  \bibfield  {author} {\bibinfo {author} {\bibfnamefont {A.~G.}\ \bibnamefont
  {Riess}}, \bibinfo {author} {\bibfnamefont {S.}~\bibnamefont {Casertano}},
  \bibinfo {author} {\bibfnamefont {W.}~\bibnamefont {Yuan}}, \bibinfo {author}
  {\bibfnamefont {L.~M.}\ \bibnamefont {Macri}}, \ and\ \bibinfo {author}
  {\bibfnamefont {D.}~\bibnamefont {Scolnic}},\ }\href {\doibase
  10.3847/1538-4357/ab1422} {\bibfield  {journal} {\bibinfo  {journal}
  {Astrophys. J.}\ }\textbf {\bibinfo {volume} {876}},\ \bibinfo {pages} {85}
  (\bibinfo {year} {2019})},\ \Eprint {http://arxiv.org/abs/1903.07603}
  {arXiv:1903.07603 [astro-ph.CO]} \BibitemShut {NoStop}%
\end{thebibliography}%

\end{document}